\documentclass[11pt]{article}
\usepackage{amsmath, amssymb, amsthm}
\usepackage{mathtools}
\usepackage{hyperref}
\usepackage{enumitem}
\usepackage{geometry}
\usepackage{authblk}
\usepackage{graphicx}

\theoremstyle{definition}

\title{\textbf{Groupoid Representation of Lorentz Transformations Including Superluminal Observers}}

\author{Marco Zaopo\thanks{Email: \texttt{marco.zaopo@gmail.com}}}

\begin{document}
	
	\maketitle

		\begin{abstract}
			We propose a Minkowski-pointed Poincare groupoid as an extension of special
			relativity including frames that are superluminal with respect to the
			rest frame.  We first construct a real 4-dimensional matrix representation of
			superluminal Lorentz transformations and show that every such transformation
			can be factorized into an ordinary subluminal Lorentz boost, an infinite-speed
			boost obtained as limit of a superluminal one, and, possibly, a spatial rotation.  For each spatial direction
			\(\hat n\), the infinite-speed limit of a superluminal Lorentz boost defines a matrix
			\(\Lambda_\infty(\hat n)\), which exchanges the temporal direction of the
			rest frame with the spatial direction \(\hat n\).
			We then show that the matrix formed as the product of the one representing \(+\infty\) limit and the one representing \(-\infty\) limit, defines an
			involution \(C_{\hat n}\) that stabilizes the Poincare group for each $\hat n$ and generates
			direction-dependent semidirect-product extensions. These extended
			Poincare groups appear as isotropy groups of metric objects constituting the groupoid, 
			while infinite-speed boosts \(\Lambda_\infty(\hat n)\) are arrows between different
			objects.
			To each object we attach an observable algebra generated by the Poincare
			generators. Groupoid arrows induce \(*\)-isomorphisms between the observable algebras,
			implementing the covariant transport of four-vector and tensor observables
			between metric objects.  The Minkowski-pointed readout is then obtained by
			contracting indices of products of transported observables with the Minkowski metric.  A
			physical wavefunction is defined as a joint eigenfunction of the two
			Minkowski-pointed Casimir observables built from the algebra of a given object.
			This classification contains, besides the ordinary massive and massless
			wavefunctions, tachyonic \(SO(2,1)\) wavefunctions: states whose
			four-momentum is spacelike in the rest frame coordinates.
			As a kinematic application, we compare this construction with the Wu
			\(^{60}\mathrm{Co}\) experiment under the working hypothesis that beta
			radiation belongs to the tachyonic \(SO(2,1)\) sector.  Groupoid transport
			associates the oriented nuclear preparation with a preferred momentum axis
			opposite to the nuclear polarization.  Reversal of the magnetic preparation
			exchanges the two oriented arrows and reverses the transported momentum,
			consistently with the observed reversal of the detector-counting asymmetry.
			The measured dipolar angular law is compatible with this covariance and
			therefore does not falsify the proposed source--arrow correspondence.  The
			present result establishes a kinematic and representation-theoretic
			correspondence and initiates a program to investigate whether
			groupoid-covariant dynamics can provide a viable route beyond the Standard
			Model.
		\end{abstract}

	\clearpage
	
	\tableofcontents
	
	\clearpage
	\section{Introduction}
	
	The possibility of extending Lorentz transformations beyond the subluminal
	domain has a long history.  A first attempt to understand superluminal
	transformations from a group-theoretical point of view was given, to the
	author's knowledge, in \cite{Recami1973}, where subluminal and superluminal
	transformations were represented by means of complex space-time coordinates.
	A later analysis in real space-time coordinates was developed in
	\cite{ChandolaRajput1989}; however, that work mainly classified real
	representations of superluminal boosts with respect to a static observer, and
	did not provide a complete representation of transformations between
	subluminal and superluminal reference frames.  More recently,
	\cite{DraganEtAl2023} introduced a real-coordinate transformation between a
	subluminal and a superluminal observer, called a superflip.
	
	The present work develops this line of thought in a different direction.  We
	construct a real four-dimensional matrix representation of Lorentz
	transformations including frames that are superluminal with respect to rest frame.  The starting point is the velocity-space inversion defined in
\eqref{eq:srf-velocity-inversion-v-to-V},
which maps subluminal velocities \(|\mathbf v|<c\) to superluminal velocities
	\(|\mathbf V|>c\), and conversely.  This inversion is the three-dimensional
	inversion of the velocity vector through the sphere of radius \(c\).  Exploiting
	the induced symmetry of the Lorentz boost matrices, we obtain a real
	representation of boosts for velocity parameters in
	\(\mathbb R^3\setminus\{|\mathbf v|=c\}\), rather than only in the ordinary
	subluminal ball.
	
	A key role is played by the infinite-speed limits of the superluminal boosts.
	For every spatial direction \(\hat n\), parametrized in
\eqref{eq:srf-unit-vector}, we introduce the two infinite-speed limits
\(\Lambda_{\pm\infty}(\hat n)\) defined in
\eqref{eq:srf-lambda-plus-definition} and
\eqref{eq:srf-lambda-minus-definition}.

These matrices are involutive, real, and have determinant \(-1\).  In a basis
	adapted to \(\hat n\), \(\Lambda_\infty(\hat n)\) exchanges the temporal
	direction of the rest frame with the spatial direction \(\hat n\),
	while leaving the transverse spatial directions unchanged.  Every
	superluminal boost can then be factorized into an ordinary subluminal boost,
	an infinite-speed boost, and, possibly, a spatial rotation.  Thus the
	infinite-speed matrices are not isolated singular objects, but structural
	building blocks of the superluminal sector.
	
	The product of the two oriented limits defines the proper involution
\(C_{\hat n}\) in \eqref{eq:srf-Cn-definition}, with the properties
collected in \eqref{eq:srf-Cn-involution}.  It normalizes the proper
orthochronous Lorentz group and induces an involutive automorphism of the
connected Poincare group \eqref{eq:srf-connected-poincare-group}.
Consequently, each direction determines the semidirect-product extension
\eqref{eq:srf-extended-poincare-group}.

This construction preserves the standard locally compact semidirect-product
	structure needed for the application of Mackey theory and harmonic-analysis
	methods \cite{HewittRoss1970,Folland2016}.
	
	The main conceptual step of this paper is to reinterpret the above
	direction-dependent extensions
	as isotropy groups of a single Poincare groupoid.  To every spatial direction \(\hat n\) we associate the directional metric
object of \eqref{eq:gpd-projective-directional-object}, using the transported
metric \eqref{eq:srf-directional-metric-two-signs}.  The ordinary Minkowski
object is defined in \eqref{eq:gpd-object-eta}.  The infinite-speed matrices
then represent the parallel arrows of \eqref{eq:gpd-parallel-oriented-arrows}
between metric objects.  By contrast, \(C_{\hat n}\) is a derived isotropy
	loop.  In this sense, the groupoid separates two structures: infinite-speed boosts are arrows between objects, while their product
	\(C_{\hat n}\) belongs to the isotropy structure of the objects and generates
	the corresponding extended Poincare groups.
	
	This groupoid formulation also clarifies the physical role of the Minkowski
	metric.  The extension developed here is covariant among metric objects, but
	an experiment is always performed through a pointed readout object.  In the
	present work the pointed object is the Minkowski laboratory frame
	\(\mathcal E_\eta\).  This means that the empirical meaning of spectral
	quadruples is assigned using the laboratory time--space partition:
	one component is read as temporal/energy-like and the remaining three as
	spatial/momentum-like.  Thus the Minkowski metric is not introduced as an
	absolute background for all observers but assume the role of selecting a given space-time partition as part of the definition of rest (or laboratory) frame.
	
	To each object \(\mathcal E_\alpha\) we attach an observable algebra generated
by its Poincare translation and Lorentz generators, together with the
available discrete isotropy loop.  The algebra generated by the infinitesimal
observables is made precise in \eqref{eq:gpd-algebraic-observable-algebra}.
Every groupoid arrow induces a \(*\)-isomorphism between these algebras; its
action on translation and Lorentz generators is given in
\eqref{eq:gpd-enveloping-isomorphism-momentum} and
\eqref{eq:gpd-enveloping-isomorphism-lorentz}.

	Wavefunctions are then classified by the spectral data of the Casimir
	operators built from these observable algebras.  However, the physically
	meaningful classification is performed through the Minkowski-pointed readout.
	For an object \(\mathcal E_\alpha\), the readout operators are defined in
\eqref{eq:gpd-first-pointed-classification-operator} and
\eqref{eq:gpd-second-pointed-classification-operator}, using the pointed
Pauli--Lubanski observable \eqref{eq:gpd-pointed-pauli-lubanski}.

The ordinary massive and massless wavefunctions arise from the Minkowski
	object itself.  Additional tachyonic sectors arise when states of directional
	objects are classified with the Minkowski metric.  For example, the massive representative
\eqref{eq:uir-massive-positive-representative} is transported to one of the
opposite spatial momenta in \eqref{eq:gpd-positive-arrow-massive-transport}
and \eqref{eq:gpd-negative-arrow-massive-transport}.  Intrinsically, it
remains massive with respect to the directional metric, as expressed by
\eqref{eq:gpd-directional-intrinsic-massive-value}; the Minkowski metric
instead gives the spacelike value
\eqref{eq:gpd-directional-pointed-spacelike-value}.

Its readout little group is therefore \(SO(2,1)\), and the corresponding
	tachyonic wave equations are the joint eigenvalue equations of the two
	Minkowski-pointed Casimir observables.  In this sense, tachyonic
	wavefunctions are not introduced as additional particles, but arise as
	laboratory readouts of directional groupoid states.
	
	Since the 1960s, quantum
	mechanical descriptions of tachyonic particles have been investigated in
	several forms.  Early attempts include \cite{Feinberg1967,DharSudarshan1968}.
	A crucial point, emphasized in later discussions such as \cite{Schwartz2017}
	and already implicit in Wigner's classification \cite{Wigner1939}, is that
	tachyonic wavefunctions should be associated with unitary representations of
	\(SO(2,1)\), not with the \(SO(3)\) little group of ordinary massive
	particles.  The present framework gives this observation a geometrical
	origin: the \(SO(2,1)\) sector is the Minkowski readout of directional
	groupoid states.
	
	The final part of the paper applies this construction to the Wu
	\(^{60}\mathrm{Co}\) beta-decay experiment. In the standard electroweak interpretation, the Wu asymmetry is explained
	dynamically by the chiral \(V-A\) structure of the weak interaction.  The
	framework developed here provides a geometrical representation-theoretic interpretation:
	groupoid transport associates the oriented nuclear preparation with a preferred momentum axis
	opposite to the nuclear polarization and reversal of the magnetic preparation
	exchanges the two oriented arrows and reverses the transported momentum,
	consistently with the observed detector-counting asymmetry of the original experiment.
	The measured dipolar angular law is compatible with this covariance and
	therefore does not falsify the proposed source--arrow correspondence. This explanation will be compared with the standard one invoking chirality as a derived phenomenological property peculiar of electroweak processes which has no geometric origin. The
	present result establishes a kinematic and representation-theoretic
	correspondence and initiates a program to investigate whether
	groupoid-covariant dynamics can provide a viable route beyond the Standard
	Model interpreting electroweak radiation as superluminal.

\section{Space-time Representations of Superluminal Reference Frames and Poincare Group Extensions}
\label{sec:superluminal-spacetime-poincare-extensions}

In this section we construct the local group-theoretical ingredients from which
the Poincare groupoid will be built.  We first recall the ordinary proper
orthochronous Lorentz group.  We then introduce real matrix representatives for
reference frames which are superluminal with respect to rest
frame, following the line of investigation initiated in
\cite{Recami1973,ChandolaRajput1989} and recently developed in real
space-time coordinates in \cite{DraganEtAl2023}.  The key point is that the
superluminal sector admits finite infinite-speed limits.  The corresponding
matrices \(\Lambda_{\pm\infty}(\hat n)\) are not ordinary automorphisms of one
Minkowski object; however, their closed product defines a proper involution
\(C_{\hat n}\), which stabilizes the Poincare group and generates a
direction-dependent semidirect-product extension.

Throughout this section we use the Minkowski metric
\begin{equation}
	\eta
	=
	\operatorname{diag}(-1,+1,+1,+1).
	\label{eq:srf-minkowski-metric}
\end{equation}
Four-vectors are written as
\begin{equation*}
	x
	=
	(x^0,\mathbf x),
	\qquad
	x^0=ct,
	\qquad
	\mathbf x=(x^1,x^2,x^3).
\end{equation*}
A spatial unit vector is denoted by
\begin{equation}
	\hat n
	=
	(n_x,n_y,n_z)=(\sin\theta\cos\phi,\sin\theta\sin\phi,\cos\theta).
	\label{eq:srf-unit-vector}
\end{equation}

\subsection{The Proper Orthochronous Lorentz Group}
\label{subsec:proper-orthochronous-lorentz-group}

The proper orthochronous Lorentz group is the connected component with the
identity in the Lorentz group.  It is defined by
\begin{equation*}
	SO^+(1,3)
	=
	\left\{
	L\in GL(4,\mathbb R):
	L^T\eta L=\eta,\;
	\det L=1,\;
	L^0{}_{0}>0
	\right\}.
\end{equation*}
This is the group of real linear transformations preserving the Minkowski
quadratic form and the time orientation. In ordinary special relativity these transformations are generated by subluminal boosts and spatial rotations \cite{Einstein1905}.

Let
\begin{equation*}
	\mathbf v=(v_x,v_y,v_z),
	\qquad
	|\mathbf v|<c.
\end{equation*}
The standard Lorentz factor is
\begin{equation*}
	\gamma_s
	=
	\frac{1}{\sqrt{1-\frac{|\mathbf v|^2}{c^2}}},
	\qquad
	K_s
	=
	\gamma_s-1.
\end{equation*}
The boost associated with \(\mathbf v\) is
\begin{equation}
	\Lambda_s(\mathbf v)=
	\begin{pmatrix}
		\gamma_s
		&
		-\gamma_s \dfrac{v_x}{c}
		&
		-\gamma_s \dfrac{v_y}{c}
		&
		-\gamma_s \dfrac{v_z}{c}
		\\[6pt]
		-\gamma_s \dfrac{v_x}{c}
		&
		1+K_s\dfrac{v_x^2}{|\mathbf v|^2}
		&
		K_s\dfrac{v_xv_y}{|\mathbf v|^2}
		&
		K_s\dfrac{v_xv_z}{|\mathbf v|^2}
		\\[6pt]
		-\gamma_s \dfrac{v_y}{c}
		&
		K_s\dfrac{v_xv_y}{|\mathbf v|^2}
		&
		1+K_s\dfrac{v_y^2}{|\mathbf v|^2}
		&
		K_s\dfrac{v_yv_z}{|\mathbf v|^2}
		\\[6pt]
		-\gamma_s \dfrac{v_z}{c}
		&
		K_s\dfrac{v_xv_z}{|\mathbf v|^2}
		&
		K_s\dfrac{v_yv_z}{|\mathbf v|^2}
		&
		1+K_s\dfrac{v_z^2}{|\mathbf v|^2}
	\end{pmatrix}.
	\label{eq:srf-subluminal-boost-matrix}
\end{equation}
It satisfies
\begin{equation*}
	\Lambda_s(\mathbf v)^T\eta\Lambda_s(\mathbf v)=\eta,
	\qquad
	\det\Lambda_s(\mathbf v)=1,
	\qquad
	\Lambda_s(\mathbf v)^0{}_{0}>0.
\end{equation*}
A spatial rotation is represented by
\begin{equation*}
	R(\Theta)
	=
	\begin{pmatrix}
		1&0\\
		0&\mathcal R(\Theta)
	\end{pmatrix},
	\qquad
	\mathcal R(\Theta)\in SO(3).
\end{equation*}
Thus a generic proper orthochronous Lorentz transformation may be written as
\begin{equation*}
	L_s(\mathbf v,\Theta)
	=
	\Lambda_s(\mathbf v)R(\Theta).
\end{equation*}
The defining identities are
\begin{equation*}
	L_s(\mathbf v,\Theta)^T\eta L_s(\mathbf v,\Theta)=\eta,
	\qquad
	\det L_s(\mathbf v,\Theta)=1,
	\qquad
	L_s(\mathbf v,\Theta)^0{}_{0}>0.
\end{equation*}
The group law is the ordinary composition of Lorentz transformations.  For
non-collinear boosts, the product of two boosts is a boost composed with a
spatial rotation, namely the Thomas precession \cite{Jackson1998}.

\subsection{Superluminal Reference Frames Representation and Infinite-Speed Boosts}
\label{subsec:superluminal-representations-infinite-speed}

We now extend the boost representation to reference frames whose velocity
parameter is superluminal with respect to the chosen reference frame.  The
starting point is the inversion of the velocity vector through the sphere of
radius \(c\):
\begin{equation}
 \mathbf V=\frac{c^2}{|\mathbf v|^2}\mathbf v,
 \qquad
 \mathbf v=\frac{c^2}{|\mathbf V|^2}\mathbf V.
 \label{eq:srf-velocity-inversion-v-to-V}
\end{equation}

Therefore
\begin{equation*}
	|\mathbf v|<c
	\quad\Longleftrightarrow\quad
	|\mathbf V|>c.
\end{equation*}
The spatial direction of the boost is common to both velocity parameters:
\begin{equation*}
	\hat n
	=
	\frac{\mathbf v}{|\mathbf v|}
	=
	\frac{\mathbf V}{|\mathbf V|}.
\end{equation*}

For a superluminal velocity parameter \(\mathbf V\), define
\begin{equation*}
	\gamma_S
	=
	\frac{1}{\sqrt{\frac{|\mathbf V|^2}{c^2}-1}},
	\qquad
	K_S
	=
	\gamma_S-1.
\end{equation*}
The corresponding real superluminal boost representative is
\begin{equation}
	\Lambda_S(\mathbf V)=
	\begin{pmatrix}
		\gamma_S
		&
		-\gamma_S \dfrac{V_x}{c}
		&
		-\gamma_S \dfrac{V_y}{c}
		&
		-\gamma_S \dfrac{V_z}{c}
		\\[6pt]
		-\gamma_S \dfrac{V_x}{c}
		&
		1+K_S\dfrac{V_x^2}{|\mathbf V|^2}
		&
		K_S\dfrac{V_xV_y}{|\mathbf V|^2}
		&
		K_S\dfrac{V_xV_z}{|\mathbf V|^2}
		\\[6pt]
		-\gamma_S \dfrac{V_y}{c}
		&
		K_S\dfrac{V_xV_y}{|\mathbf V|^2}
		&
		1+K_S\dfrac{V_y^2}{|\mathbf V|^2}
		&
		K_S\dfrac{V_yV_z}{|\mathbf V|^2}
		\\[6pt]
		-\gamma_S \dfrac{V_z}{c}
		&
		K_S\dfrac{V_xV_z}{|\mathbf V|^2}
		&
		K_S\dfrac{V_yV_z}{|\mathbf V|^2}
		&
		1+K_S\dfrac{V_z^2}{|\mathbf V|^2}
	\end{pmatrix}.
	\label{eq:srf-superluminal-boost-matrix}
\end{equation}
Using the velocity inversion, one has
\begin{equation}
	\frac{V_iV_j}{|\mathbf V|^2}
	=
	\frac{v_iv_j}{|\mathbf v|^2},
	\qquad
	i,j=1,2,3.
	\label{eq:srf-directional-projector-invariance}
\end{equation}
Moreover,
\begin{equation}
	\gamma_S
	=
	\gamma_s\frac{|\mathbf v|}{c}
	=
	\gamma_s\frac{c}{|\mathbf V|}.
	\label{eq:srf-gamma-S-gamma-s-relation}
\end{equation}
These identities imply that the superluminal boost is obtained from an ordinary
boost by multiplication with a finite infinite-speed matrix (see appendix \ref{app:superluminal-boost-factorization} for a proof):
\begin{equation}
	\Lambda_S(\mathbf V)
	=
	\Lambda_s(\mathbf v)\Lambda_+(\hat n).
	\label{eq:srf-superluminal-factorization}
\end{equation}
Here
\begin{equation}
	\Lambda_+(\hat n)
	:=
	\Lambda_{\infty}(\hat n)
	=
	\lim_{|\mathbf V|\to\infty}\Lambda_S(\mathbf V).
	\label{eq:srf-lambda-plus-definition}
\end{equation}
Using \eqref{eq:srf-unit-vector}, one obtains
\begin{equation}
	\Lambda_+(\hat n)
	=
	\begin{pmatrix}
		0&-n_x&-n_y&-n_z\\
		-n_x&1-n_x^2&-n_xn_y&-n_xn_z\\
		-n_y&-n_xn_y&1-n_y^2&-n_yn_z\\
		-n_z&-n_xn_z&-n_yn_z&1-n_z^2
	\end{pmatrix}.
	\label{eq:srf-lambda-plus-matrix}
\end{equation}

The opposite infinite-speed limit is defined by
\begin{equation}
	\Lambda_-(\hat n)
	:=
	\Lambda_{-\infty}(\hat n)
	=
	\lim_{|\mathbf V|\to\infty}\Lambda_S(-\mathbf V).
	\label{eq:srf-lambda-minus-definition}
\end{equation}
Its explicit form is
\begin{equation}
	\Lambda_-(\hat n)
	=
	\begin{pmatrix}
		0&n_x&n_y&n_z\\
		n_x&1-n_x^2&-n_xn_y&-n_xn_z\\
		n_y&-n_xn_y&1-n_y^2&-n_yn_z\\
		n_z&-n_xn_z&-n_yn_z&1-n_z^2
	\end{pmatrix}.
	\label{eq:srf-lambda-minus-matrix}
\end{equation}
The two infinite-speed matrices satisfy
\begin{equation}
	\Lambda_+(\hat n)^2=I,
	\qquad
	\Lambda_-(\hat n)^2=I.
	\label{eq:srf-lambda-plus-minus-involutions}
\end{equation}
Their determinants are
\begin{equation*}
	\det\Lambda_+(\hat n)
	=
	\det\Lambda_-(\hat n)
	=
	-1.
\end{equation*}
Their action on the time axis and on the longitudinal spatial axis is
\begin{equation*}
	\Lambda_+(\hat n)e_0
	=
	-(0,\hat n),
	\qquad
	\Lambda_-(\hat n)e_0
	=
	(0,\hat n).
\end{equation*}
Furthermore,
\begin{equation*}
	\Lambda_+(\hat n)(0,\hat n)
	=
	-e_0,
	\qquad
	\Lambda_-(\hat n)(0,\hat n)
	=
	e_0.
\end{equation*}
Let
\[
\mathcal B_{\hat n}
=
(e_0,e_{\hat n},e_{\perp1},e_{\perp2}),
\qquad
e_{\hat n}=(0,\hat n),
\]
be a basis adapted to the direction \(\hat n\). In this basis,
\[
[\Lambda_{+\infty}(\hat n)]_{\mathcal B_{\hat n}}
=
\begin{pmatrix}
	0&-1&0&0\\
	-1&0&0&0\\
	0&0&1&0\\
	0&0&0&1
\end{pmatrix}.
\]
Hence \(\Lambda_{+\infty}(\hat n)\) interchanges the temporal and
longitudinal basis directions with an orientation reversal,
\[
e_0\mapsto-e_{\hat n},
\qquad
e_{\hat n}\mapsto-e_0,
\]
while fixing the two-dimensional transverse spatial subspace pointwise.

For transformations in arbitrary spatial direction and with arbitrary spatial
axis orientation, we write
\begin{equation*}
	L_S(\mathbf V,\Theta)
	=
	\Lambda_s(\mathbf v)\Lambda_+(\hat n)R(\Theta).
\end{equation*}
This representation is real in four-dimensional space-time coordinates and constitutes matrices containing transformation laws for coordinates of two observers moving with relative speed greater than $c$.

\subsection{Proper Orthochronous and Poincare Group Extensions}
\label{subsec:proper-orthochronous-poincare-extensions}

Define
\begin{equation}
	C_{\hat n}
	:=
	\Lambda_+(\hat n)\Lambda_-(\hat n)	=
	\Lambda_-(\hat n)\Lambda_+(\hat n).
	\label{eq:srf-Cn-definition}
\end{equation}

A direct computation gives
\begin{equation*}
	C_{\hat n}
	=
	\begin{pmatrix}
		-1&0&0&0\\
		0&1-2n_x^2&-2n_xn_y&-2n_xn_z\\
		0&-2n_xn_y&1-2n_y^2&-2n_yn_z\\
		0&-2n_xn_z&-2n_yn_z&1-2n_z^2
	\end{pmatrix}.
\end{equation*}
In block form this is
\begin{equation}
	C_{\hat n}
	=
	\begin{pmatrix}
		-1&0\\
		0&I_3-2\hat n\hat n^T
	\end{pmatrix}.
	\label{eq:srf-Cn-block-form}
\end{equation}
The matrix \(C_{\hat n}\) is a metric-preserving proper involution:
\begin{equation}
 C_{\hat n}^2=I,\qquad C_{\hat n}^{T}\eta C_{\hat n}=\eta,
 \qquad \det C_{\hat n}=1.
 \label{eq:srf-Cn-involution}
\end{equation}

Therefore \(C_{\hat n}\) is a proper but non-orthochronous Lorentz
transformation.  It normalizes the proper orthochronous Lorentz group:
\begin{equation*}
	C_{\hat n}SO^+(1,3)C_{\hat n}^{-1}
	=
	SO^+(1,3).
\end{equation*}
Hence conjugation by \(C_{\hat n}\) defines an involutive automorphism
\begin{equation*}
	\alpha_{C_{\hat n}}:
	SO^+(1,3)\longrightarrow SO^+(1,3),
	\qquad
	\alpha_{C_{\hat n}}(L)
	=
	C_{\hat n}LC_{\hat n}^{-1}.
\end{equation*}
This gives the direction-dependent Lorentz extension
\begin{equation}
	\mathcal L_{\eta,C_{\hat n}}
	:=
	SO^+(1,3)
	\rtimes_{\operatorname{Ad}_{C_{\hat n}}}
	\langle C_{\hat n}\rangle.
	\label{eq:srf-directional-lorentz-extension}
\end{equation}
Since
\begin{equation*}
	\langle C_{\hat n}\rangle\simeq\mathbb Z_2,
\end{equation*}
one may also write
\begin{equation*}
	\mathcal L_{\eta,C_{\hat n}}
	\simeq
	SO^+(1,3)
	\rtimes_{\operatorname{Ad}_{C_{\hat n}}}
	\mathbb Z_2.
\end{equation*}
The elements of the extension are
\begin{equation*}
	L C_{\hat n}^{\epsilon},
	\qquad
	L\in SO^+(1,3),
	\qquad
	\epsilon\in\{0,1\}.
\end{equation*}
The multiplication law is
\begin{equation*}
	(L,C_{\hat n}^{\epsilon})
	(L',C_{\hat n}^{\epsilon'})
	=
	\left(
	L C_{\hat n}^{\epsilon}L'C_{\hat n}^{-\epsilon},
	C_{\hat n}^{\epsilon+\epsilon'}
	\right).
\end{equation*}
This is a finite semidirect-product extension of the proper orthochronous
Lorentz group, in the standard sense of locally compact group extensions
\cite{Folland2016,HewittRoss1970}.

Let
\begin{equation}\label{definition-translation-group}
	\mathbb V\simeq\mathbb R^{1,3}
\end{equation}
be the translation group.  The connected Poincare group is
\begin{equation}
	\mathcal P_0
	:=
	\mathbb V\rtimes SO^+(1,3).
	\label{eq:srf-connected-poincare-group}
\end{equation}
Its product is
\begin{equation*}
	(a,L)(a',L')
	=
	(a+La',LL').
\end{equation*}
Since \(C_{\hat n}\) preserves the Minkowski metric and normalizes
\(SO^+(1,3)\), it also normalizes \(\mathcal P_0\).  The induced automorphism is
\begin{equation*}
	\alpha_{C_{\hat n}}(a,L)
	=
	(C_{\hat n}a,C_{\hat n}LC_{\hat n}^{-1}).
\end{equation*}
Thus we define the direction-dependent extended Poincare group by
\begin{equation}
	\mathcal P_{\eta,C_{\hat n}}
	:=
	\mathcal P_0
	\rtimes_{\operatorname{Ad}_{C_{\hat n}}}
	\mathbb Z_2.
	\label{eq:srf-extended-poincare-group}
\end{equation}
Equivalently,
\begin{equation}
	\mathcal P_{\eta,C_{\hat n}}
	=
	\mathbb V
	\rtimes
	\mathcal L_{\eta,C_{\hat n}}.
	\label{eq:srf-extended-poincare-equivalent}
\end{equation}
This is the isotropy group that will later be attached to the corresponding
metric object of the Poincare groupoid.

\subsection{Unitary Irreducible Representations of the Extended Poincare Group}
\label{subsec:uirs-extended-poincare-group}

We now classify the unitary irreducible representations of the two-component
Poincare extension generated by the isotropy loop \(C_{\hat n}\).  This is the
local representation-theoretic classification associated with a fixed metric
object. The groupoid interpretation will be introduced in the next section.

From the discussion in the previous section the extended Poincare group is therefore \eqref{eq:srf-extended-poincare-equivalent}, or, equivalently:
\begin{equation*}
	\mathcal P_{\eta,C_{\hat n}}
	=
	\mathcal P_0
	\cup
	\mathcal P_0C_{\hat n}.
\end{equation*}
Its identity component is \(\mathcal P_0\), while the second component is
obtained by composing ordinary Poincare group elements
with \(C_{\hat n}\).

We first recall the ordinary Wigner classification of \(\mathcal P_0\)
\cite{Wigner1939,Weinberg}.  A
character of the translation group is
\begin{equation*}
	\chi_p(a)
	=
	\exp\left(-\frac{i}{\hbar}p(a)\right),
	\qquad
	p\in\mathbb V^*.
\end{equation*}
For a representative momentum \(p\), the little group is
\begin{equation*}
	H_p
	=
	\operatorname{Stab}_{SO^+(1,3)}(p).
\end{equation*}
Given a unitary irreducible representation \(\sigma\) of \(H_p\), the ordinary
Wigner representation is
\begin{equation}
	\pi_{p,\sigma}
	=
	\operatorname{Ind}_{\mathbb V\rtimes H_p}^{\mathcal P_0}
	\left(
	\chi_p\otimes\sigma
	\right).
	\label{eq:uir-wigner-representation}
\end{equation}
With the convention \eqref{eq:srf-minkowski-metric} the three ordinary orbit types are
\begin{equation}
	p^T\eta p=-M^2,
	\qquad
	p^0>0,
	\qquad
	H_p\simeq SO(3),
	\label{eq:uir-massive-orbit-type}
\end{equation}
\begin{equation}
	p^T\eta p=0,
	\qquad
	p^0>0,
	\qquad
	H_p\simeq ISO(2),
	\label{eq:uir-massless-orbit-type}
\end{equation}
and
\begin{equation}
	p^T\eta p=+M^2,
	\qquad
	H_p\simeq SO(2,1).
	\label{eq:uir-tachyonic-orbit-type}
\end{equation}

The automorphism \(\alpha_{C_{\hat n}}\) acts on ordinary Wigner
representations by conjugation:
\begin{equation}
	\pi^{C_{\hat n}}(g)
	=
	\pi\left(\alpha_{C_{\hat n}}^{-1}(g)\right),
	\qquad
	g\in\mathcal P_0.
	\label{eq:uir-Cn-action-on-representations}
\end{equation}
Because the automorphism is involutive, its inverse in
\eqref{eq:uir-Cn-action-on-representations} can equally be replaced by
\(\alpha_{C_{\hat n}}\).

At the level of momentum labels, the action is
\begin{equation*}
	p
	\longmapsto
	C_{\hat n}^{-1 T}p.
\end{equation*}
Using
\begin{equation*}
	C_{\hat n}^{-1}=C_{\hat n},
	\qquad
	C_{\hat n}^{T}=C_{\hat n},
\end{equation*}
this becomes
\begin{equation}
	p
	\longmapsto
	C_{\hat n}p.
	\label{eq:uir-Cn-momentum-action}
\end{equation}
Therefore
\begin{equation*}
	\left(\pi_{p,\sigma}\right)^{C_{\hat n}}
	\simeq
	\pi_{C_{\hat n}p,\sigma^{C_{\hat n}}},
\end{equation*}
where
\begin{equation*}
	\sigma^{C_{\hat n}}(h')
	=
	\sigma(C_{\hat n}^{-1}h'C_{\hat n}),
	\qquad
	h'\in H_{C_{\hat n}p}.
\end{equation*}
The little groups are related by
\begin{equation*}
	H_{C_{\hat n}p}
	=
	C_{\hat n}H_pC_{\hat n}^{-1}.
\end{equation*}

The theory of unitary irreducible representations of semidirect-product group
extensions was developed by Clifford and Mackey
\cite{Clifford1937,Mackey1958,Mackey}.  We detail the split index-two case
relevant to \eqref{eq:srf-extended-poincare-equivalent} in
Appendix~\ref{app:clifford-mackey-index-two}; see also
\cite{Serre1977} for the corresponding index-two form of Clifford theory.
Clifford theory constrains the restriction of an irreducible representation
of a group \(G\) to a normal subgroup \(N\), whereas Mackey induction provides
a construction of irreducible representations of \(G\) from representation
data associated with \(N\).  Combining these results allows us to classify
the UIRs of the extended Poincare group from the Wigner UIRs of its connected
Poincare subgroup.
Starting with an ordinary Wigner UIR
\(\pi\) of \(\mathcal P_0\), the theory predicts there are two cases.

We use the conjugate representation defined in
\eqref{eq:uir-Cn-action-on-representations}.

\paragraph{Case A}

Assume that
\[
	\pi^{C_{\hat n}}
	\not\simeq
	\pi
\]
thus $\pi^{C_{\hat n}}$ and $\pi$ are inequivalent as unitary representations of the connected Poincare group
\(\mathcal P_0\).

The orbit of $\pi$, a Wigner representation, under
\[
	\mathcal P_{\eta,C_{\hat n}}/\mathcal P_0
	\simeq
	\mathbb Z_2
\]
therefore contains the two elements:
\begin{equation*}
	\mathcal O_{\pi}
	=
	\left\{
	\pi,
	\pi^{C_{\hat n}}
	\right\}.
\end{equation*}

\(\pi\) and \(\pi^{C_{\hat n}}\) may be realized on two tagged
copies of the same Hilbert space.  We therefore use the canonical carrier
\begin{equation*}
	\mathcal H_{\Pi}^{\mathrm{ext}}
	=
	\mathcal H_\pi
	\oplus
	\mathcal H_\pi.
\end{equation*}
The two sector subspaces are
\begin{align*}
	\mathcal H_+
	&:=
	\mathcal H_\pi\oplus\{0\},
	\\
	\mathcal H_-
	&:=
	\{0\}\oplus\mathcal H_\pi.
\end{align*}
A general vector is written as
\begin{equation*}
	\Psi
	=
	(\psi_+,\psi_-),
	\qquad
	\psi_+,\psi_-\in\mathcal H_\pi.
\end{equation*}

The connected
subgroup acts according to
\begin{equation}
	\Pi(g)
	=
	\begin{pmatrix}
		\pi(g)&0\\
		0&\pi^{C_{\hat n}}(g)
	\end{pmatrix}
	=
	\begin{pmatrix}
		\pi(g)&0\\
		0&
		\pi\!\left(
		\alpha_{C_{\hat n}}(g)
		\right)
	\end{pmatrix},
	\qquad
	g\in\mathcal P_0.
	\label{eq:uir-case-A-connected-action}
\end{equation}
Thus the first copy carries \(\pi\), whereas the second copy carries the
inequivalent conjugate representation \(\pi^{C_{\hat n}}\).

The nontrivial element of the quotient group is represented by the exchange
operator because of the action \eqref{eq:uir-Cn-momentum-action} on momentum space 
\begin{equation}
	\Pi(C_{\hat n})
	=
	S_C
	:=
	\begin{pmatrix}
		0&I_{\mathcal H_\pi}\\
		I_{\mathcal H_\pi}&0
	\end{pmatrix}.
	\label{eq:uir-case-A-discrete-exchange-operator}
\end{equation}
The operator \eqref{eq:uir-case-A-discrete-exchange-operator} exchanges
\(\psi_+\) and \(\psi_-\), and therefore maps each sector subspace onto
the other.

Since \(C_{\hat n}^2=I\), one has
\begin{equation*}
	S_C^2
	=
	\begin{pmatrix}
		I_{\mathcal H_\pi}&0\\
		0&I_{\mathcal H_\pi}
	\end{pmatrix}
	=
	I_{\mathcal H_{\Pi}^{\mathrm{ext}}}.
\end{equation*}
Moreover,
\begin{equation*}
	\begin{aligned}
		S_C\Pi(g)S_C^{-1}
		&=
		\begin{pmatrix}
			\pi^{C_{\hat n}}(g)&0\\
			0&\pi(g)
		\end{pmatrix}
		\\
		&=
		\begin{pmatrix}
			\pi\!\left(
			\alpha_{C_{\hat n}}(g)
			\right)&0\\
			0&
			\pi\!\left(
			\alpha_{C_{\hat n}}^2(g)
			\right)
		\end{pmatrix}.
	\end{aligned}
\end{equation*}
Because \(C_{\hat n}^2=I\), the induced automorphism is involutive:
\[
	\alpha_{C_{\hat n}}^2
	=
	\operatorname{id}_{\mathcal P_0}
\]
Therefore
\begin{equation}
	S_C\Pi(g)S_C^{-1}
	=
	\Pi\!\left(
	\alpha_{C_{\hat n}}(g)
	\right),
	\qquad
	g\in\mathcal P_0.
	\label{eq:uir-case-A-semidir-product-relation}
\end{equation}
This is the covariance relation required by the semidirect-product
structure.

The exchange operator \(S_C\) is part of the representation of the
extended group.  It transforms the action of \(g\) into the action
of its conjugate
\(\alpha_{C_{\hat n}}(g)\).

The irreducibility of the extended representation can be seen directly.
Since \(\pi\) and \(\pi^{C_{\hat n}}\) are inequivalent irreducible
representations, Schur's lemma \cite{Shur} implies that every bounded operator commuting
with the restriction of \(\Pi\) to \(\mathcal P_0\) has the form
\begin{equation*}
	X
	=
	\begin{pmatrix}
		aI_{\mathcal H_\pi}&0\\
		0&bI_{\mathcal H_\pi}
	\end{pmatrix}.
\end{equation*}
Requiring \(X\) also to commute with \(S_C\) gives $a=b$.
The commutant of the extended representation therefore consists only of
scalar multiples of the identity.  Hence the induced representation of the extended Poincarè group
\begin{equation}
	\Pi
	=
	\operatorname{Ind}_{\mathcal P_0}^{
		\mathcal P_{\eta,C_{\hat n}}
	}
	\pi
	\label{eq:uir-case-A-induced}
\end{equation}
is irreducible.

The alternative convention
\begin{equation*}
	\Pi(C_{\hat n})
	=
	-S_C
\end{equation*}
does not define an inequivalent representation in Case A.  Indeed, the
unitary operator
\begin{equation*}
	V
	=
	\begin{pmatrix}
		I_{\mathcal H_\pi}&0\\
		0&-I_{\mathcal H_\pi}
	\end{pmatrix}
\end{equation*}
satisfies
\begin{equation*}
	V\Pi(g)V^{-1}
	=
	\Pi(g),
	\qquad
	VS_CV^{-1}
	=
	-S_C.
\end{equation*}
The two sign conventions are therefore unitarily equivalent.  

The restriction of the extended representation to the connected
Poincare subgroup is
\begin{equation}
	\operatorname{Res}_{\mathcal P_0}^{
		\mathcal P_{\eta,C_{\hat n}}
	}
	\Pi
	\simeq
	\pi
	\oplus
	\pi^{C_{\hat n}}.
	\label{eq:uir-case-A-restriction}
\end{equation}

In summary, in case A, since $\pi$ and $\pi^{C_{\hat n}}$ are inequivalent and must live in the same hilbert space and the automorphism is involutive and unitary, the only thing that the extension element can do is exchange the two inequivalent sectors of the irreducible representation up to a sign.

\paragraph{Case B}
In the second case,
\[
	\pi^{C_{\hat n}}\simeq\pi.
\]
thus $\pi^{C_{\hat n}}$ and $\pi$ are equivalent as Poincarè group representations.
Then there exists a unitary intertwiner
\begin{equation*}
	A_C:\mathcal H_\pi\longrightarrow\mathcal H_\pi
\end{equation*}
such that
\begin{equation}
	A_C\pi(g)A_C^{-1}
	=
	\pi(\alpha_{C_{\hat n}}(g)).
	\label{eq:uir-case-B-intertwining-relation}
\end{equation}
Since \(C_{\hat n}^2=I\), one may normalize \(A_C\) so that $	A_C^2=I.$

Then \(\pi\) has two inequivalent extensions to
\(\mathcal P_{\eta,C_{\hat n}}\), distinguished by the sign with which
\(C_{\hat n}\) is represented:
\begin{equation}
	\Pi_+(C_{\hat n})=+A_C,
	\qquad
	\Pi_-(C_{\hat n})=-A_C.
	\label{eq:uir-case-B-two-extensions}
\end{equation}

The classification of the UIRs lying above the orbit of $\pi$ is therefore:
\begin{equation*}
	\begin{array}{c|c|c}
		\text{condition}
		&
		\text{extended-group UIR}
		&
		\text{restriction to }\mathcal P_0
		\\[3pt]
		\hline
		\pi^{C_{\hat n}}\not\simeq\pi
		&
		\operatorname{Ind}_{\mathcal P_0}^{\mathcal P_{\eta,C_{\hat n}}}\pi
		&
		\pi\oplus\pi^{C_{\hat n}}
		\\[6pt]
		\pi^{C_{\hat n}}\simeq\pi
		&
		\Pi_+,\ \Pi_-
		&
		\pi
	\end{array}
\end{equation*}
In the first line, \(C_{\hat n}\) exchanges two inequivalent ordinary Wigner
sectors.  In the second line, it acts internally on a single Wigner sector, and
the two extensions are distinguished by a discrete label $\pm 1$.

We now specialize the above considerations for the different types of Wigner UIRs of Poincarè group, which is the normal subgroup of the extended group of which we know the UIRs \cite{Wigner1939} and that we will employ to find all possible UIRs of the extended group.

\subsubsection{Massive sector}
\label{subsubsec:uir-massive-sector}

Consider the positive-energy massive representative
\begin{equation}
	p_+
	=
	(M,0,0,0)^T,
	\qquad
	M>0.
	\label{eq:uir-massive-positive-representative}
\end{equation}
It satisfies
\begin{equation}
	p_+^T\eta p_+
	=
	-M^2,
	\qquad
	p_+^0>0.
	\label{eq:uir-massive-positive-invariant}
\end{equation}
The little group is $	H_{p_+}\simeq SO(3).$
The corresponding ordinary massive Wigner representation is
\begin{equation*}
	\pi_{M,s}^{+}
	=
	\operatorname{Ind}_{\mathbb V\rtimes SO(3)}^{\mathcal P_0}
	\left(
	\chi_{p_+}\otimes\sigma_s
	\right),
\end{equation*}
where $\chi_{p_+}$ is the character of the representative momentum and \(\sigma_s\) is the spin-\(s\) representation of \(SO(3)\).

Define \(p_-:=(-M,0,0,0)^T\).  The action of \(C_{\hat n}\) is
\begin{equation}
 C_{\hat n}p_+=p_-.
 \label{eq:uir-Cn-massive-branch-exchange}
\end{equation}

The two connected mass hyperboloids are exchanged:
\begin{equation*}
	\mathcal O_M^+
	\longleftrightarrow
	\mathcal O_M^-.
\end{equation*}
At the representation level,
\begin{equation*}
	\left(\pi_{M,s}^{+}\right)^{C_{\hat n}}
	\simeq
	\pi_{M,s}^{-}.
\end{equation*}
As representations of the connected Poincare group,
\begin{equation*}
	\pi_{M,s}^{+}
	\not\simeq
	\pi_{M,s}^{-},
\end{equation*}
because their spectra lie on different connected energy branches.  Hence the
massive sector is an instance of Case A.

The corresponding extended massive UIR is
\begin{equation*}
	\Pi_{M,s}^{\rm ext}
	=
	\operatorname{Ind}_{\mathcal P_0}^{\mathcal P_{\eta,C_{\hat n}}}
	\pi_{M,s}^{+}.
\end{equation*}
Its restriction to the connected Poincare group is
\begin{equation}
	\operatorname{Res}_{\mathcal P_0}^{\mathcal P_{\eta,C_{\hat n}}}
	\Pi_{M,s}^{\rm ext}
	\simeq
	\pi_{M,s}^{+}
	\oplus
	\pi_{M,s}^{-}.
	\label{eq:uir-extended-massive-restriction}
\end{equation}
Thus the extended massive representation contains both massive energy branches.

The two massive Wigner representations may be realized on
\begin{align*}
	\mathcal H_{M,s}^{+}
	&=
	L^2\!\left(\mathcal O_M^+,d\mu_M;\mathcal V_s\right),
	\\
	\mathcal H_{M,s}^{-}
	&=
	L^2\!\left(\mathcal O_M^-,d\mu_M;\mathcal V_s\right),
\end{align*}
where \(\mathcal V_s\) carries the spin-\(s\) little-group representation.
Although these carrier spaces are unitarily isomorphic, they are retained as
distinct tagged copies because their translation spectra lie on different
connected mass hyperboloids.

The transformation \(p\mapsto C_{\hat n}p\) maps \(\mathcal O_M^+\)
bijectively onto \(\mathcal O_M^-\) and induces a unitary twisted transport
\begin{equation*}
	T_M:
	\mathcal H_{M,s}^{+}
	\longrightarrow
	\mathcal H_{M,s}^{-}
\end{equation*}
satisfying
\begin{equation}
	\pi_{M,s}^{-}(g)
	=
	T_M\pi_{M,s}^{+}\!\left(\alpha_{C_{\hat n}}(g)\right)T_M^{-1},
	\qquad
	g\in\mathcal P_0.
	\label{eq:uir-massive-twisted-transport}
\end{equation}

The extended massive Hilbert space is
\begin{equation*}
	\mathcal H_{M,s}^{\rm ext}
	=
	\mathcal H_{M,s}^{+}
	\oplus
	\mathcal H_{M,s}^{-}.
\end{equation*}
For
\begin{equation*}
	\psi_+\in\mathcal H_{M,s}^{+},
	\qquad
	\psi_-\in\mathcal H_{M,s}^{-},
\end{equation*}
the connected Poincare group acts diagonally:
\begin{equation*}
	\Pi_{M,s}^{\rm ext}(g)
	\begin{pmatrix}\psi_+\\\psi_-\end{pmatrix}
	=
	\begin{pmatrix}
		\pi_{M,s}^{+}(g)\psi_+\\
		\pi_{M,s}^{-}(g)\psi_-
	\end{pmatrix}.
\end{equation*}
The matrix \(C_{\hat n}\) is represented by
\begin{equation}
	S_C^{(M)}
	=
	\begin{pmatrix}
		0&T_M^{-1}\\
		T_M&0
	\end{pmatrix}.
	\label{eq:uir-massive-exchange-operator}
\end{equation}
As in \eqref{eq:uir-case-A-discrete-exchange-operator}, the operator
\eqref{eq:uir-massive-exchange-operator} exchanges the two sector subspaces,
now using \(T_M\) and \(T_M^{-1}\) to identify their carriers.

After identifying the two carriers through \(T_M\), they become two tagged
copies of a common model Hilbert space \(\mathcal H_{M,s}\):
\begin{equation}
	\mathcal H_{M,s}^{\rm ext}
	\simeq
	\mathcal H_{M,s}\otimes\mathbb C^2,
	\qquad
	S_C^{(M)}
	\simeq
	I_{\mathcal H_{M,s}}\otimes\sigma_x.
	\label{eq:uir-massive-common-carrier}
\end{equation}
where
\[
\sigma_x
=
\begin{pmatrix}
	0&1\\
	1&0
\end{pmatrix}
\]
is the Pauli exchange matrix acting on the two-dimensional sector-label
space \(\mathbb C^2\). The two copies carry the inequivalent positive- and negative-energy Wigner
representations \(\pi_{M,s}^{+}\) and \(\pi_{M,s}^{-}\).  In the physical
terminology adopted in this work, these will be referred to as the particle
and antiparticle sectors.  Strictly speaking, the Clifford--Mackey
construction establishes the pairing of the two energy branches; their
particle--antiparticle interpretation is an additional physical
identification.

The point little group is not enlarged by \(C_{\hat n}\) because of \eqref{eq:uir-Cn-massive-branch-exchange}.
Therefore \(C_{\hat n}\) does not belong to the stabilizer of the single
momentum \(p_+\).  It stabilizes the unordered pair $\{p_+,p_-\}$
not the point \(p_+\) itself.  Hence the massive little group remains $SO(3)$

\subsubsection{Massless sector}
\label{subsubsec:uir-massless-sector}

Consider a future-directed null representative
\begin{equation*}
	k_+
	=
	\omega(1,\hat r)^T,
	\qquad
	|\hat r|=1,
	\qquad
	\omega>0.
\end{equation*}
It satisfies
\begin{equation*}
	k_+^T\eta k_+
	=
	0,
	\qquad
	k_+^0>0.
\end{equation*}
The little group is
\begin{equation*}
	H_{k_+}
	\simeq
	ISO(2).
\end{equation*}
An ordinary helicity representation is
\begin{equation*}
	\pi_{0,h}^{+}
	=
	\operatorname{Ind}_{\mathbb V\rtimes ISO(2)}^{\mathcal P_0}
	\left(
	\chi_{k_+}\otimes\sigma_h
	\right),
\end{equation*}
where \(\sigma_h\) denotes the helicity representation of \(ISO(2)\).

The action of \(C_{\hat n}\) gives
\begin{equation*}
	C_{\hat n}k_+
	=
	\omega
	\left(
	-1,
	(I_3-2\hat n\hat n^T)\hat r
	\right)^T.
\end{equation*}
This vector is still null:
\begin{equation*}
	(C_{\hat n}k_+)^T\eta(C_{\hat n}k_+)
	=
	0.
\end{equation*}
However,
\begin{equation*}
	(C_{\hat n}k_+)^0
	=
	-\omega
	<
	0.
\end{equation*}
Thus \(C_{\hat n}\) maps the future light cone to the past light cone:
\begin{equation*}
	\mathcal O_0^+
	\longleftrightarrow
	\mathcal O_0^-.
\end{equation*}

In the aligned case,
\begin{equation*}
	\hat r=\hat n,
\end{equation*}
we have
\begin{equation*}
	k_+
	=
	\omega(1,\hat n)^T,
\end{equation*}
and
\begin{equation*}
	C_{\hat n}k_+
	=
	-k_+.
\end{equation*}
Thus \(C_{\hat n}\) stabilizes the unoriented null line $\{k_+,-k_+\}$ but not the point \(k_+\) itself.

At the level of ordinary Wigner representations,
\begin{equation*}
	\left(\pi_{0,h}^{+}\right)^{C_{\hat n}}
	\simeq
	\pi_{0,h_C}^{-},
\end{equation*}
where \(h_C\) denotes the little-group label transformed by conjugation with
\(C_{\hat n}\).  Depending on helicity conventions, \(h_C\) may coincide with
or differ by sign from \(h\).  The essential point is that the energy branch is
exchanged.

Since
\begin{equation*}
	\pi_{0,h}^{+}
	\not\simeq
	\pi_{0,h_C}^{-}
\end{equation*}
as representations of \(\mathcal P_0\), the massless sector is also an
instance of Case A.  The corresponding extended massless UIR is
\begin{equation*}
	\Pi_{0,h}^{\rm ext}
	=
	\operatorname{Ind}_{\mathcal P_0}^{\mathcal P_{\eta,C_{\hat n}}}
	\pi_{0,h}^{+}.
\end{equation*}
Its restriction is
\begin{equation}
	\operatorname{Res}_{\mathcal P_0}^{\mathcal P_{\eta,C_{\hat n}}}
	\Pi_{0,h}^{\rm ext}
	\simeq
	\pi_{0,h}^{+}
	\oplus
	\pi_{0,h_C}^{-}.
	\label{eq:uir-extended-massless-restriction}
\end{equation}
Thus the two-component extension naturally produces a forward/backward
massless doublet.

The future- and past-directed massless Wigner representations may be realized
on
\begin{align*}
	\mathcal H_{0,h}^{+}
	&=
	L^2\!\left(\mathcal O_0^+,d\mu_0;\mathcal V_h\right),
	\\
	\mathcal H_{0,h_C}^{-}
	&=
	L^2\!\left(\mathcal O_0^-,d\mu_0;\mathcal V_{h_C}\right),
\end{align*}
where \(\mathcal V_h\) and \(\mathcal V_{h_C}\) carry the corresponding
little-group representations.  The two carriers are unitarily isomorphic
after transporting the momentum variable and the little-group indices with
\(C_{\hat n}\), but they remain distinct tagged copies because their
translation spectra lie on different connected components of the null cone.

The map \(k\mapsto C_{\hat n}k\) induces a unitary twisted transport
\begin{equation*}
	T_0:
	\mathcal H_{0,h}^{+}
	\longrightarrow
	\mathcal H_{0,h_C}^{-}
\end{equation*}
satisfying
\begin{equation*}
	\pi_{0,h_C}^{-}(g)
	=
	T_0\pi_{0,h}^{+}\!\left(\alpha_{C_{\hat n}}(g)\right)T_0^{-1},
	\qquad
	g\in\mathcal P_0.
\end{equation*}

From here on the discussion goes along the same lines of massive sector equations \eqref{eq:uir-massive-twisted-transport} and \eqref{eq:uir-massive-common-carrier}, the only difference being the two tagged copies that carry the inequivalent connected-Poincare
representations \(\pi_{0,h}^{+}\) and \(\pi_{0,h_C}^{-}\) represent
future- and past-directed massless Wigner sectors instead of particle/anti-particle sectors.

\subsubsection{Tachyonic sector}
\label{subsubsec:uir-tachyonic-sector}

Now take a spacelike representative aligned with the direction \(\hat n\):
\begin{equation*}
	q
	=
	(0,-M\hat n)^T.
\end{equation*}
It satisfies
\begin{equation}
	q^T\eta q
	=
	+M^2.
	\label{eq:uir-tachyonic-invariant}
\end{equation}
The ordinary Poincare little group is
\begin{equation*}
	H_q
	\simeq
	SO(2,1).
\end{equation*}
A tachyonic Wigner UIR is therefore
\begin{equation}
	\pi_{M,\sigma}^{\rm tach}
	=
	\operatorname{Ind}_{\mathbb V\rtimes SO(2,1)}^{\mathcal P_0}
	\left(
	\chi_q\otimes\sigma
	\right),
	\label{eq:uir-tachyonic-wigner-rep}
\end{equation}
where \(\sigma\) is a unitary irreducible representation of \(SO(2,1)\).
Standard treatments of the unitary representation theory of the Lorentz
group and of harmonic analysis on the associated noncompact homogeneous
spaces may be found in
\cite{Bargmann1947,Naimark1964,VilenkinKlimyk1991,Helgason1984,Camporesi1990}.

For the aligned representative, one has
\begin{equation*}
	C_{\hat n}q
	=
	-q.
\end{equation*}
Thus \(C_{\hat n}\) maps the spacelike momentum to its opposite.  However, the spacelike mass shell
\eqref{eq:uir-tachyonic-invariant}
is connected under \(SO^+(1,3)\).  Therefore \(q\) and \(-q\) lie on the same
connected orbit:
\begin{equation*}
	q\sim_{SO^+(1,3)} -q.
\end{equation*}
The tachyonic sector differs from the massive and massless sectors.  Since the
spacelike mass shell is connected under \(SO^+(1,3)\), the momenta \(q\) and
\(-q\) belong to the same connected orbit.  Therefore \(C_{\hat n}\) does not
necessarily produce an inequivalent branch.  It acts internally on the
tachyonic orbit and on the \(SO(2,1)\) little-group representation.

If
\begin{equation*}
	\sigma^{C_{\hat n}}\not\simeq\sigma,
\end{equation*}
then the tachyonic sector is again of Case A and the extended representation
restricts to a doublet:
\begin{equation}
	\operatorname{Res}_{\mathcal P_0}^{\mathcal P_{\eta,C_{\hat n}}}
	\Pi_{M,\sigma}^{\rm tach,ext}
	\simeq
	\pi_{M,\sigma}^{\rm tach}
	\oplus
	\pi_{M,\sigma^{C_{\hat n}}}^{\rm tach}.
	\label{eq:uir-tachyonic-summary-doublet}
\end{equation}
If instead
\begin{equation*}
	\sigma^{C_{\hat n}}\simeq\sigma,
\end{equation*}
then the tachyonic Wigner UIR is invariant under the \(C_{\hat n}\)-action and
extends to two inequivalent UIRs of the full extended group:
\begin{equation}
	\Pi_{M,\sigma}^{\rm tach,+},
	\qquad
	\Pi_{M,\sigma}^{\rm tach,-}.
	\label{eq:uir-tachyonic-summary-two-extensions}
\end{equation}
These two extensions are distinguished by the genuine discrete sign
\begin{equation*}
	\varepsilon=\pm1.
\end{equation*}
Thus the tachyonic sector is the only one among the three standard Wigner
classes in which the \(C_{\hat n}\)-extension may act internally on a single
connected orbit rather than necessarily pairing two disconnected energy
branches.

\subsubsection{Summary of extended Poincare group UIRs classification}
\label{subsubsec:uir-summary-local-classification}

The UIRs of the extended group \eqref{eq:srf-extended-poincare-group}
follow the Clifford--Mackey alternative, using the conjugate representation
\eqref{eq:uir-Cn-action-on-representations}.  In Case~A, induction and
restriction are given by \eqref{eq:uir-case-A-induced} and
\eqref{eq:uir-case-A-restriction}; in Case~B, the two extensions are
specified by \eqref{eq:uir-case-B-two-extensions}.

For massive and massless sectors, the extension exchanges the two energy
branches and gives the restrictions
\eqref{eq:uir-extended-massive-restriction} and
\eqref{eq:uir-extended-massless-restriction}.  Both are instances of Case~A,
with the discrete element represented by a sector-exchange operator.

For the tachyonic sector, the spacelike shell
\eqref{eq:uir-tachyonic-invariant} is connected, and opposite momenta belong
to the same orbit.  The action on the little-group representation determines
whether the restriction is the doublet
\eqref{eq:uir-tachyonic-summary-doublet}, or whether the two inequivalent
extensions \eqref{eq:uir-tachyonic-summary-two-extensions} occur, distinguished
by a sign.

Thus the point little groups remain those of
\eqref{eq:uir-massive-orbit-type}, \eqref{eq:uir-massless-orbit-type} and
\eqref{eq:uir-tachyonic-orbit-type}.  The extension changes the organization
of ordinary Wigner sectors, pairing disconnected energy branches or acting
within a connected spacelike orbit.

\subsection{Infinite-Speed Matrices and Isomorphic Poincare Representations}
\label{subsec:infinite-speed-isomorphisms-poincare-representations}

We now examine the role of the two infinite-speed matrices
\begin{equation*}
	\Lambda_{+}(\hat n),
	\qquad
	\Lambda_{-}(\hat n)
	=
	\Lambda_{+}(-\hat n),
\end{equation*}
defined in
\eqref{eq:srf-lambda-plus-matrix}
and
\eqref{eq:srf-lambda-minus-matrix}.
At this stage, we use only their linear and group-theoretical properties.
Their interpretation within the Poincare groupoid will be introduced in
Section~\ref{sec:minkowski-pointed-poincare-groupoid}.

Let
\begin{equation*}
	[\hat n]
	:=
	\{\hat n,-\hat n\}
\end{equation*}
denote the unoriented spatial axis determined by \(\hat n\).
The two infinite-speed matrices define the same transported metric:
\begin{equation}
	\eta_{[\hat n]}
	:=
	\Lambda_{+}(\hat n)^{-T}
	\eta
	\Lambda_{+}(\hat n)^{-1}
	=
	\Lambda_{-}(\hat n)^{-T}
	\eta
	\Lambda_{-}(\hat n)^{-1}.
	\label{eq:srf-directional-metric-two-signs}
\end{equation}
Consequently,
\begin{equation}
	\Lambda_{\varepsilon}(\hat n)^{T}
	\eta_{[\hat n]}
	\Lambda_{\varepsilon}(\hat n)
	=
	\eta,
	\qquad
	\varepsilon\in\{+,-\}.
	\label{eq:srf-infinite-speed-metric-isomorphism}
\end{equation}

Thus each \(\Lambda_{\varepsilon}(\hat n)\) is a metric isomorphism between
the two isomorphic metric spaces
\begin{equation*}
	(\mathbb V,\eta)
	\qquad\text{and}\qquad
	(\mathbb V,\eta_{[\hat n]}).
\end{equation*}
where $\mathbb V$, as in (\ref{definition-translation-group}), is the translation group.
In particular, \(\Lambda_{\varepsilon}(\hat n)\) is not a Lorentz
transformation of the Minkowski metric \(\eta\).  It relates two different
matrix realizations of metric spaces having the same signature.

The corresponding connected Lorentz groups are
\begin{equation*}
	SO^{+}(\eta)
	=
	\left\{
	L\in GL(\mathbb V):
	L^{T}\eta L=\eta
	\right\}
\end{equation*}
and
\begin{equation*}
	SO^{+}(\eta_{[\hat n]})
	=
	\left\{
	L'\in GL(\mathbb V):
	L'^{T}\eta_{[\hat n]}L'=\eta_{[\hat n]}
	\right\}.
\end{equation*}
For either sign, conjugation by the corresponding infinite-speed matrix gives
\begin{equation}
	SO^{+}(\eta_{[\hat n]})
	=
	\Lambda_{\varepsilon}(\hat n)
	SO^{+}(\eta)
	\Lambda_{\varepsilon}(\hat n)^{-1}.
	\label{eq:srf-conjugate-lorentz-groups}
\end{equation}
Indeed, if \(L^{T}\eta L=\eta\) and
\begin{equation*}
	L_{[\hat n]}^{(\varepsilon)}
	:=
	\Lambda_{\varepsilon}(\hat n)
	L
	\Lambda_{\varepsilon}(\hat n)^{-1},
\end{equation*}
then
\begin{equation*}
	\left(L_{[\hat n]}^{(\varepsilon)}\right)^{T}
	\eta_{[\hat n]}
	L_{[\hat n]}^{(\varepsilon)}
	=
	\eta_{[\hat n]}.
\end{equation*}

The associated connected Poincare groups are
\begin{equation}
	\mathcal P_{\eta}^{0}
	=
	\mathbb V\rtimes SO^{+}(\eta),
	\qquad
	\mathcal P_{[\hat n]}^{0}
	=
	\mathbb V\rtimes SO^{+}(\eta_{[\hat n]}).
	\label{eq:srf-two-poincare-groups}
\end{equation}
Each infinite-speed matrix induces a group isomorphism
\begin{equation*}
	\Phi_{\varepsilon\hat n}:
	\mathcal P_{\eta}^{0}
	\longrightarrow
	\mathcal P_{[\hat n]}^{0}
\end{equation*}
defined by
\begin{equation}
	\Phi_{\varepsilon\hat n}(a,L)
	=
	\left(
	\Lambda_{\varepsilon}(\hat n)a,
	\Lambda_{\varepsilon}(\hat n)
	L
	\Lambda_{\varepsilon}(\hat n)^{-1}
	\right).
	\label{eq:srf-poincare-group-isomorphism}
\end{equation}
This map preserves the Poincare group product because
\begin{align*}
	\Phi_{\varepsilon\hat n}
	\bigl((a,L)(a',L')\bigr)
	&=
	\Phi_{\varepsilon\hat n}
	(a+La',LL')
	\\
	&=
	\Phi_{\varepsilon\hat n}(a,L)
	\Phi_{\varepsilon\hat n}(a',L').
\end{align*}
Therefore the Poincare groups associated with
\(\eta\) and \(\eta_{[\hat n]}\) are isomorphic.

This group isomorphism also establishes a one-to-one correspondence between
their unitary representations.  If
\begin{equation*}
	\Pi_{\eta}:
	\mathcal P_{\eta}^{0}
	\longrightarrow
	U(\mathcal H)
\end{equation*}
is a unitary representation, define
\begin{equation}
	\Pi_{[\hat n]}^{(\varepsilon)}
	:=
	\Pi_{\eta}
	\circ
	\Phi_{\varepsilon\hat n}^{-1}.
	\label{eq:srf-transported-poincare-representation}
\end{equation}
Explicitly,
\begin{equation*}
	\Pi_{[\hat n]}^{(\varepsilon)}(g')
	=
	\Pi_{\eta}
	\left(
	\Phi_{\varepsilon\hat n}^{-1}(g')
	\right),
	\qquad
	g'\in\mathcal P_{[\hat n]}^{0}.
\end{equation*}
Equivalently, the two representations satisfy
\begin{equation*}
	\Pi_{[\hat n]}^{(\varepsilon)}
	\circ
	\Phi_{\varepsilon\hat n}
	=
	\Pi_{\eta}.
\end{equation*}
Since \(\Phi_{\varepsilon\hat n}\) is a group isomorphism, this transport
preserves unitarity, irreducibility and unitary equivalence classes.

The two matrices
\(\Lambda_{+}(\hat n)\) and \(\Lambda_{-}(\hat n)\) therefore provide two
isomorphisms between the same pair of Poincare groups and two corresponding
identifications of their unitary representations.  The groups
\(\mathcal P_{\eta}^{0}\) and \(\mathcal P_{[\hat n]}^{0}\) may be regarded
as the symmetry groups of the two isomorphic metric spaces
\((\mathbb V,\eta)\) and \((\mathbb V,\eta_{[\hat n]})\), respectively.

In Section~\ref{sec:minkowski-pointed-poincare-groupoid}, these symmetry
groups will be identified as the isotropy groups of the corresponding metric
objects, and the geometrical relation between the two infinite-speed
isomorphisms will be formulated within the groupoid structure.

\section{Minkowski-Pointed Poincare Groupoid}
\label{sec:minkowski-pointed-poincare-groupoid}

We now show that isomorphims structure of infinite speed boosts and their product highlighted in the previous section corresponds mathematically to a single Poincare groupoid.  Our use of objects, arrows, composition and isotropy follows the standard categorical and Lie-groupoid terminology
\cite{MacLane,Mackenzie,MoerdijkMrcun}.  For the related operator-algebraic
viewpoint on locally compact groupoids, see \cite{Renault}; the objectwise
observable algebras introduced below should not be confused with the
groupoid \(C^*\)-algebra constructed in that setting.

The central distinction is the following.  The matrices
\(\Lambda_{\pm\infty}(\hat n)\) are not internal Lorentz symmetries of a single
Minkowski metric object but metric isomorphisms between different
objects.  Their closed products, instead, are isotropy loops.  Thus the
geometrical structure is naturally groupoidal: objects carry metric structures
and extended Poincare isotropy groups, while infinite-speed boosts are arrows
between objects.

The physical interpretation is obtained by choosing a distinguished object,
namely the Minkowski laboratory object.  We call the resulting structure a
Minkowski-pointed Poincare groupoid.  The unpointed groupoid encodes the
covariance between metric objects; the pointed object supplies the laboratory
time--space partition through which spectral data take physical meaning.

\subsection{The Poincare Groupoid}
\label{subsec:poincare-groupoid}

Let
\begin{equation*}
	\mathcal D_{\mathrm{pr}}
	\subseteq
	\mathbb{RP}^{2}
\end{equation*}
be a set of unoriented spatial axes.  The Poincare groupoid associated with
\(\mathcal D_{\mathrm{pr}}\) will be denoted by
\begin{equation*}
	\mathfrak P_{\mathcal D_{\mathrm{pr}}}.
\end{equation*}
Its objects are metric space-time structures.  Its arrows are affine metric
isomorphisms between them.  Its isotropy groups are the corresponding extended
Poincare groups.

The pointed structure is the pair
\begin{equation}
	\left(
	\mathfrak P_{\mathcal D_{\mathrm{pr}}},
	\mathcal E_\eta
	\right),
	\label{eq:gpd-pointed-groupoid}
\end{equation}
where \(\mathcal E_\eta\) is the distinguished Minkowski laboratory object.
Its role is operational:
it supplies the laboratory time--space partition used to classify spectral data of momentum and Lorentz generators operators.

\subsubsection{Metric Objects}
\label{subsubsec:gpd-metric-objects}

The reference object is the Minkowski metric object
\begin{equation}
	\mathcal E_\eta
	:=
	(\mathbb V,\eta).
	\label{eq:gpd-object-eta}
\end{equation}

Let
\begin{equation}
	\rho:S^2\longrightarrow\mathbb{RP}^2,
	\qquad
	\rho(\hat n)=[\hat n]:=\{\hat n,-\hat n\},
	\label{eq:gpd-projective-cover}
\end{equation}
be the canonical twofold covering map, and let
\begin{equation*}
	\widetilde{\mathcal D}
	:=
	\rho^{-1}(\mathcal D_{\mathrm{pr}})
	\subseteq S^2
\end{equation*}
be the corresponding set of oriented lifts.  Thus every
\([\hat n]\in\mathcal D_{\mathrm{pr}}\) has precisely two lifts,
\(\hat n\) and \(-\hat n\), in \(\widetilde{\mathcal D}\).

For an oriented lift \(\hat n\in\widetilde{\mathcal D}\), write
\(\eta_{\hat n}\) for the metric defined in
\eqref{eq:srf-directional-metric-two-signs}.  Here \(J_{\hat n}\) and
\(J_{-\hat n}\) denote the oriented arrows represented by
\(\Lambda_+(\hat n)\) and \(\Lambda_-(\hat n)\), respectively.

Using the explicit block form of the infinite-speed matrix, one obtains
\begin{equation}
	\eta_{\hat n}
	=
	\begin{pmatrix}
		1&0\\[2pt]
		0&I_3-2\hat n\hat n^T
	\end{pmatrix}.
	\label{eq:gpd-directional-metric-block-form}
\end{equation}
The projector \(\hat n\hat n^T\) is unchanged by orientation reversal.
Thus the transported metric depends only on the projective axis, consistently
with \eqref{eq:srf-directional-metric-two-signs}; we use
\(\eta_{[\hat n]}\), \(\eta_{\hat n}\) and \(\eta_{-\hat n}\) for this
same metric.

The corresponding directional metric object is
\begin{equation}
	\mathcal E_{[\hat n]}
	:=
	(\mathbb V,\eta_{[\hat n]}),
	\qquad
	[\hat n]\in\mathcal D_{\mathrm{pr}}.
	\label{eq:gpd-projective-directional-object}
\end{equation}
Thus the object set is
\begin{equation}
	\operatorname{Obj}
	(\mathfrak P_{\mathcal D_{\mathrm{pr}}})
	=
	\{\mathcal E_\eta\}
	\cup
	\left\{
	\mathcal E_{[\hat n]}:
	[\hat n]\in\mathcal D_{\mathrm{pr}}
	\right\}.
	\label{eq:gpd-projective-object-set}
\end{equation}

For either oriented lift, the metric-isomorphism identity
\eqref{eq:srf-infinite-speed-metric-isomorphism} holds.  Hence the two lifts
determine metric isomorphisms between the same pair of metric objects.

Accordingly, \(\mathbb{RP}^2\) is the parameter space of directional metric
objects, whereas \(S^2\) parametrizes their oriented lifts.  A point
\([\hat n]\in\mathbb{RP}^2\) is an unoriented spatial axis, not a point of
physical space.  The object \(\mathcal E_{[\hat n]}\) remembers this axis but
contains no choice between the two orientations \(\hat n\) and \(-\hat n\).

This definition also makes clear that the discrete loops are not part of the primitive
data of the object.  The primitive object data are
\begin{equation*}
	\mathcal E_\alpha
	=
	(\mathbb V,\eta_\alpha).
\end{equation*}
with \(\alpha=\eta\) or \(\alpha=[\hat n]\).  The oriented maps
\(J_{\hat n}\) and \(J_{-\hat n}\), as well as the discrete involutions
obtained from their closed compositions, belong instead to the arrow and
isotropy structures of the groupoid.

\subsubsection{Arrows}
\label{subsubsec:gpd-arrows}

An arrow of the Poincare groupoid is a typed affine metric isomorphism
between two metric objects \cite{MacLane}.  More precisely, an arrow is a datum
\begin{equation}
	\gamma
	=
	\left(
	\mathcal E_\alpha
	\xrightarrow{(a,A)}
	\mathcal E_\beta
	\right),
	\label{eq:gpd-typed-affine-arrow}
\end{equation}
where \(a\in\mathbb V\) is a translation and
\(A:\mathbb V\to\mathbb V\) is an invertible linear map satisfying
\begin{equation}
	A^T\eta_\beta A
	=
	\eta_\alpha.
	\label{eq:gpd-arrow-metric-condition}
\end{equation}
The corresponding affine action on coordinates is
\begin{equation}
	x
	\longmapsto
	Ax+a.
	\label{eq:gpd-affine-action}
\end{equation}

The source and target objects are part of the arrow data.  Consequently, the
same affine map may represent different categorical arrows when equipped with
different source and target objects.  This distinction will be important
below, since the infinite-speed matrices are involutive and may therefore
represent both a given arrow and the linear part of its inverse, although the
two categorical arrows have opposite source and target.

If
\begin{equation*}
	(a,A):
	\mathcal E_\alpha
	\longrightarrow
	\mathcal E_\beta
\end{equation*}
and
\begin{equation*}
	(b,B):
	\mathcal E_\beta
	\longrightarrow
	\mathcal E_\gamma,
\end{equation*}
their composition is defined by
\begin{equation}
	(b,B)\circ(a,A)
	=
	(b+Ba,BA):
	\mathcal E_\alpha
	\longrightarrow
	\mathcal E_\gamma.
	\label{eq:gpd-arrow-composition}
\end{equation}
Composition is therefore defined only when the target of the first arrow
coincides with the source of the second.

The identity arrow at \(\mathcal E_\alpha\) is
\begin{equation*}
	\operatorname{id}_{\mathcal E_\alpha}
	=
	(0,I):
	\mathcal E_\alpha
	\longrightarrow
	\mathcal E_\alpha,
\end{equation*}
and the inverse of an arrow is
\begin{equation}
	(a,A)^{-1}
	=
	(-A^{-1}a,A^{-1}):
	\mathcal E_\beta
	\longrightarrow
	\mathcal E_\alpha.
	\label{eq:gpd-inverse-arrow}
\end{equation}
Hence every arrow is invertible.

For a projective axis \([\hat n]\in\mathcal D_{\mathrm{pr}}\), both
oriented representatives satisfy
\eqref{eq:srf-infinite-speed-metric-isomorphism}.

They therefore define two arrows with the same source and the same target:
\begin{equation}
	J_{\hat n},
	J_{-\hat n}:
	\mathcal E_\eta
	\rightrightarrows
	\mathcal E_{[\hat n]}.
	\label{eq:gpd-parallel-oriented-arrows}
\end{equation}

The arrows in \eqref{eq:gpd-parallel-oriented-arrows} are parallel in the
categorical sense \cite{MacLane}: they belong to the same hom-set,
\begin{equation*}
	J_{\hat n},
	J_{-\hat n}
	\in
	\operatorname{Hom}_{\mathfrak P_{\mathcal D_{\mathrm{pr}}}}
	\left(
	\mathcal E_\eta,
	\mathcal E_{[\hat n]}
	\right).
\end{equation*}

The unordered pair
\begin{equation*}
	\left\{
	J_{\hat n},
	J_{-\hat n}
	\right\}
\end{equation*}
depends only on the projective axis \([\hat n]\).  Replacing the representative
\(\hat n\) by \(-\hat n\) merely exchanges the two elements of the pair.

The inverse arrows form a parallel pair in the opposite direction:
\begin{equation}
	J_{\hat n}^{-1},
	J_{-\hat n}^{-1}:
	\mathcal E_{[\hat n]}
	\rightrightarrows
	\mathcal E_\eta.
	\label{eq:gpd-parallel-inverse-arrows}
\end{equation}
At the matrix level the transformations are involutive by
\eqref{eq:srf-lambda-plus-minus-involutions}.  This does not identify a typed
arrow with its categorical inverse: the arrows in
\eqref{eq:gpd-parallel-oriented-arrows} and
\eqref{eq:gpd-parallel-inverse-arrows} have opposite sources and targets.
Their inverse identities are
\begin{equation*}
 J_{\varepsilon\hat n}^{-1}\circ J_{\varepsilon\hat n}
 =\operatorname{id}_{\mathcal E_\eta},\qquad
 J_{\varepsilon\hat n}\circ J_{\varepsilon\hat n}^{-1}
 =\operatorname{id}_{\mathcal E_{[\hat n]}},\quad \varepsilon\in\{+1,-1\}.
\end{equation*}

In particular, although the underlying matrices satisfy
\(J_{\pm\hat n}^{2}=I\), the self-composition of the typed arrow
\(J_{\pm\hat n}:\mathcal E_\eta\to\mathcal E_{[\hat n]}\) is not defined.
Only the compositions with the oppositely typed inverse arrows are
categorically meaningful \cite{MacLane}.

There are also distinguished arrows between two directional metric objects.
Let
\begin{equation*}
	[\hat n],[\hat m]
	\in
	\mathcal D_{\mathrm{pr}},
\end{equation*}
and choose oriented representatives
\begin{equation*}
	\epsilon\hat n
	\in
	\rho^{-1}([\hat n]),
	\qquad
	\delta\hat m
	\in
	\rho^{-1}([\hat m]),
	\qquad
	\epsilon,\delta\in\{+1,-1\}.
\end{equation*}
The corresponding lift-dependent arrow is
\begin{equation}
	J_{\delta\hat m\leftarrow\epsilon\hat n}
	:=
	J_{\delta\hat m}
	\circ
	J_{\epsilon\hat n}^{-1}:
	\mathcal E_{[\hat n]}
	\longrightarrow
	\mathcal E_{[\hat m]}.
	\label{eq:gpd-lift-dependent-directional-arrow}
\end{equation}
Its linear part satisfies
\begin{equation}
	\left(
	J_{\delta\hat m}
	J_{\epsilon\hat n}^{-1}
	\right)^T
	\eta_{[\hat m]}
	\left(
	J_{\delta\hat m}
	J_{\epsilon\hat n}^{-1}
	\right)
	=
	\eta_{[\hat n]}.
	\label{eq:gpd-directional-arrow-metric-condition}
\end{equation}

Thus there is no unique canonical arrow between two projective directional
objects unless oriented representatives are specified.  For fixed
\([\hat n]\) and \([\hat m]\), the choices
\(\epsilon,\delta\in\{+1,-1\}\) determine a family of distinguished
directional-to-directional arrows.  These arrows are related by internal
isotropy transformations.

When the same oriented representative is used at an intermediate object, the
distinguished arrows compose by cancellation:
\begin{equation}
	J_{\kappa\hat\ell\leftarrow\delta\hat m}
	\circ
	J_{\delta\hat m\leftarrow\epsilon\hat n}
	=
	J_{\kappa\hat\ell\leftarrow\epsilon\hat n}.
	\label{eq:gpd-lift-dependent-arrow-composition}
\end{equation}
If opposite representatives of the intermediate projective object are used,
the middle factors do not cancel.  Their residual closed composition is an
isotropy loop. Detailed mathematical and physical investigation of arrows between directional metric objects will be the argument of forthcoming publications. Here are mentioned only for mathematical completeness. In the next sections only arrows from the Minkowski object to a directional metric object are explored in the context of the physical equations they imply.

The groupoid
\(\mathfrak P_{\mathcal D_{\mathrm{pr}}}\) considered here and in what follows is the
subgroupoid of the affine metric groupoid generated by the connected Poincare
isotropy arrows at each object, the oriented infinite-speed arrows introduced
above, and their inverses and compositions.

\subsubsection{Isotropy Loops and Extended Poincare Groups}
\label{subsubsec:gpd-isotropy-loops-extended-poincare}

The previous section constructed two parallel arrows in \eqref{eq:gpd-parallel-oriented-arrows}
corresponding to the two oriented lifts \(+\hat n\) and \(-\hat n\) of the
same projective direction \([\hat n]\).  It also introduced their
closed compositions
\begin{equation}
	C_{[\hat n]}^\eta
	=
	J_{+\hat n}^{-1}\circ J_{-\hat n},
	\label{eq:gpd-C-eta-recall}
\end{equation}
and
\begin{equation}
	C_{[\hat n]}^{\mathrm{dir}}
	=
	J_{-\hat n}\circ J_{+\hat n}^{-1}.
	\label{eq:gpd-C-dir-recall}
\end{equation}

Let
\begin{equation*}
	s,t:
	\operatorname{Mor}(\mathfrak P_{\mathcal D_{\mathrm{pr}}})
	\longrightarrow
	\operatorname{Obj}(\mathfrak P_{\mathcal D_{\mathrm{pr}}})
\end{equation*}
denote the source and target maps.  An arrow
\[
\gamma\in
\operatorname{Mor}(\mathfrak P_{\mathcal D_{\mathrm{pr}}})
\]
is an isotropy loop when
\begin{equation*}
	s(\gamma)=t(\gamma).
\end{equation*}
The isotropy group of an object \(\mathcal E_\alpha\) is therefore
\begin{equation}
	\operatorname{Iso}(\mathcal E_\alpha)
	=
	\operatorname{Hom}_{\mathfrak P_{\mathcal D_{\mathrm{pr}}}}
	(\mathcal E_\alpha,\mathcal E_\alpha).
	\label{eq:gpd-isotropy-definition}
\end{equation}
Because every arrow of a groupoid is invertible,
\(\operatorname{Iso}(\mathcal E_\alpha)\) is a group under categorical
composition, with identity
\[
1_{\mathcal E_\alpha}
\]
and inverse inherited from the groupoid \cite{MacLane}.

For the connected isotropy one has
\begin{equation}
	\operatorname{Iso}_0(\mathcal E_\alpha)
	=
	\mathcal P_\alpha^0
	=
	\mathbb V\rtimes SO^+(\eta_\alpha).
	\label{eq:gpd-connected-isotropy}
\end{equation}
At the two endpoints, \eqref{eq:gpd-connected-isotropy} recovers the
connected Poincare groups of \eqref{eq:srf-two-poincare-groups}.

The loop at the Minkowski object satisfies
\begin{equation*}
	s\!\left(C_{[\hat n]}^\eta\right)
	=
	t\!\left(C_{[\hat n]}^\eta\right)
	=
	\mathcal E_\eta.
\end{equation*}
Thus
\begin{equation*}
	C_{[\hat n]}^\eta
	\in
	\operatorname{Iso}(\mathcal E_\eta).
\end{equation*}
Categorically, it is the closed path
\begin{equation*}
	\mathcal E_\eta
	\xrightarrow{\ J_{-\hat n}\ }
	\mathcal E_{[\hat n]}
	\xrightarrow{\ J_{+\hat n}^{-1}\ }
	\mathcal E_\eta.
\end{equation*}

The loop at the directional object instead satisfies
\begin{equation*}
	s\!\left(C_{[\hat n]}^{\mathrm{dir}}\right)
	=
	t\!\left(C_{[\hat n]}^{\mathrm{dir}}\right)
	=
	\mathcal E_{[\hat n]}.
\end{equation*}
Hence
\begin{equation*}
	C_{[\hat n]}^{\mathrm{dir}}
	\in
	\operatorname{Iso}(\mathcal E_{[\hat n]}).
\end{equation*}
It is the closed path
\begin{equation*}
	\mathcal E_{[\hat n]}
	\xrightarrow{\ J_{+\hat n}^{-1}\ }
	\mathcal E_\eta
	\xrightarrow{\ J_{-\hat n}\ }
	\mathcal E_{[\hat n]}.
\end{equation*}

Therefore
\begin{equation}
	C_{[\hat n]}^\eta
	\neq
	C_{[\hat n]}^{\mathrm{dir}}
	\label{eq:gpd-two-loops-not-equal-as-arrows}
\end{equation}
as groupoid arrows.  They belong to different isotropy groups and have
different sources and targets.  They are related by transport along either
oriented infinite-speed arrow:
\begin{equation}
	C_{[\hat n]}^{\mathrm{dir}}
	=
	J_{\varepsilon\hat n}
	\circ
	C_{[\hat n]}^\eta
	\circ
	J_{\varepsilon\hat n}^{-1},
	\qquad
	\varepsilon\in\{+1,-1\}.
	\label{eq:gpd-two-loops-conjugation}
\end{equation}

Equivalently, the two parallel arrows satisfy
\begin{equation}
	J_{-\hat n}
	=
	J_{+\hat n}
	\circ
	C_{[\hat n]}^\eta
	=
	C_{[\hat n]}^{\mathrm{dir}}
	\circ
	J_{+\hat n}.
	\label{eq:gpd-parallel-arrows-relative-loops}
\end{equation}

\subsection{Representations of Arrows and Loops}
\label{subsec:gpd-arrow-loop-representations}

We now introduce two different realizations of the arrows of
\(\mathfrak P_{\mathcal D_{\mathrm{pr}}}\).
First, arrows and loops are represented by matrices acting on space-time
coordinates.  Second, they are represented by unitary operators acting
between Hilbert spaces associated with the metric objects.

These two notions must be distinguished.  The matrices
\(\Lambda_{\pm}(\hat n)\) act on the space-time vector space
\(\mathbb V\), whereas the corresponding unitary operators act on quantum
states.

\subsubsection{Matrix Representations of Arrows}
\label{subsubsec:gpd-arrow-matrix-representations}

Given an affine metric arrow defined in \eqref{eq:gpd-typed-affine-arrow}  Its coordinate action is defined by \eqref{eq:gpd-arrow-metric-condition} and \eqref{eq:gpd-affine-action}.
The complete affine transformation can be represented in homogeneous
coordinates by the matrix
\begin{equation}
	[\gamma]_{\beta\leftarrow\alpha}
	:=
	\begin{pmatrix}
		A&a\\
		0&1
	\end{pmatrix}.
	\label{eq:gpd-homogeneous-arrow-matrix}
\end{equation}
The subscripts \(\beta\leftarrow\alpha\) record the source and target of the
represented arrow.

If
\begin{equation*}
	\gamma_1:
	\mathcal E_\alpha
	\longrightarrow
	\mathcal E_\beta,
	\qquad
	\gamma_2:
	\mathcal E_\beta
	\longrightarrow
	\mathcal E_\gamma,
\end{equation*}
then
\begin{equation}
	[\gamma_2\circ\gamma_1]_{\gamma\leftarrow\alpha}
	=
	[\gamma_2]_{\gamma\leftarrow\beta}
	[\gamma_1]_{\beta\leftarrow\alpha}.
	\label{eq:gpd-matrix-representation-composition}
\end{equation}
Similarly,
\begin{equation*}
	[\gamma^{-1}]_{\alpha\leftarrow\beta}
	=
	[\gamma]_{\beta\leftarrow\alpha}^{-1},
\end{equation*}
and
\begin{equation*}
	[\operatorname{id}_{\mathcal E_\alpha}]_{\alpha\leftarrow\alpha}
	=
	I_5.
\end{equation*}

For zero-translation arrows, the relevant information is entirely contained
in the \(4\times4\) linear matrix \(A\).  In particular, the two parallel
infinite-speed arrows in \eqref{eq:gpd-parallel-oriented-arrows}
have matrix representations
\begin{equation}
	[J_{\hat n}]_{[\hat n]\leftarrow\eta}
	=
	\Lambda_+(\hat n),
	\qquad
	[J_{-\hat n}]_{[\hat n]\leftarrow\eta}
	=
	\Lambda_-(\hat n)
	=
	\Lambda_+(-\hat n).
	\label{eq:gpd-two-infinite-speed-arrow-matrices}
\end{equation}

From \eqref{eq:srf-infinite-speed-metric-isomorphism} they represent two distinct metric isomorphisms between the same pair of objects.

The inverse arrows are represented by the inverse matrices:
\begin{equation*}
	[J_{\hat n}^{-1}]_{\eta\leftarrow[\hat n]}
	=
	\Lambda_+(\hat n)^{-1},
	\qquad
	[J_{-\hat n}^{-1}]_{\eta\leftarrow[\hat n]}
	=
	\Lambda_-(\hat n)^{-1}.
\end{equation*}
By \eqref{eq:srf-lambda-plus-minus-involutions}, an arrow and its inverse
have the same numerical \(4\times4\) matrix.  Their typing remains opposite,
as recorded in \eqref{eq:gpd-parallel-oriented-arrows} and
\eqref{eq:gpd-parallel-inverse-arrows}.

Thus a numerical matrix does not, by itself, determine a groupoid arrow.
Its source and target must also be specified.

The directional arrow defined in
\eqref{eq:gpd-lift-dependent-directional-arrow}
is represented by
\begin{equation*}
	[J_{\delta\hat m\leftarrow\varepsilon\hat n}]
	_{[\hat m]\leftarrow[\hat n]}
	=
	\Lambda_\delta(\hat m)
	\Lambda_\varepsilon(\hat n)^{-1}.
\end{equation*}
Its metric compatibility is precisely
\eqref{eq:gpd-directional-arrow-metric-condition}.

The same matrix representation also describes the transport of internal
Poincare transformations.  Let
\begin{equation*}
	g=(a,L)
	\in
	\operatorname{Iso}_0(\mathcal E_\eta)
	=
	\mathcal P_\eta^0.
\end{equation*}
Transport along \(J_{\varepsilon\hat n}\) gives the loop
\begin{equation}
	\Phi_{\varepsilon\hat n}(g)
	:=
	J_{\varepsilon\hat n}
	\circ g
	\circ J_{\varepsilon\hat n}^{-1}
	\in
	\operatorname{Iso}_0(\mathcal E_{[\hat n]}).
	\label{eq:gpd-isotropy-transport-by-arrow}
\end{equation}
Its affine matrix is
\begin{equation*}
	[\Phi_{\varepsilon\hat n}(a,L)]
	=
	\begin{pmatrix}
		\Lambda_\varepsilon(\hat n)
		L
		\Lambda_\varepsilon(\hat n)^{-1}
		&
		\Lambda_\varepsilon(\hat n)a
		\\
		0&1
	\end{pmatrix}.
\end{equation*}
The affine pair is exactly \eqref{eq:srf-poincare-group-isomorphism}.
Thus this matrix realizes the Poincare-group isomorphism introduced in
Subsection~\ref{subsec:infinite-speed-isomorphisms-poincare-representations}.

\subsubsection{Matrix Representations of Loops}
\label{subsubsec:gpd-loop-matrix-representations}

The two parallel arrows associated with the axis \([\hat n]\) determine the
typed isotropy loops in \eqref{eq:gpd-C-dir-recall}, \eqref{eq:gpd-C-eta-recall}

Their matrix representations are respectively
\begin{align*}
	[C_{[\hat n]}^\eta]_{\eta\leftarrow\eta}
	&=
	\Lambda_+(\hat n)^{-1}
	\Lambda_-(\hat n),
	\\
	[C_{[\hat n]}^{\mathrm{dir}}]
	_{[\hat n]\leftarrow[\hat n]}
	&=
	\Lambda_-(\hat n)
	\Lambda_+(\hat n)^{-1}.
\end{align*}

For the explicit infinite-speed matrices, the two products coincide
numerically:
\begin{equation}
	\Lambda_+(\hat n)^{-1}\Lambda_-(\hat n)
	=
	\Lambda_-(\hat n)\Lambda_+(\hat n)^{-1}
	=
	\mathbf C_{[\hat n]},
	\label{eq:gpd-common-C-matrix-representation}
\end{equation}
Here \(\mathbf C_{[\hat n]}\) is the numerical matrix
\(C_{\hat n}\) already given in \eqref{eq:srf-Cn-block-form}.

This matrix depends only on the projective axis because
\begin{equation*}
	\mathbf C_{[-\hat n]}
	=
	\mathbf C_{[\hat n]}.
\end{equation*}

Although the two loops have the same numerical matrix, their different
typing was established in \eqref{eq:gpd-two-loops-not-equal-as-arrows}.
The matrix representation of an arrow must therefore always be supplied with information regarding target and source to be categorically meaningful.

The common matrix is involutive by \eqref{eq:srf-Cn-involution}
and preserves both metrics:
\begin{equation*}
	\mathbf C_{[\hat n]}^T
	\eta
	\mathbf C_{[\hat n]}
	=
	\eta,
	\qquad
	\mathbf C_{[\hat n]}^T
	\eta_{[\hat n]}
	\mathbf C_{[\hat n]}
	=
	\eta_{[\hat n]}.
\end{equation*}
At the level of typed arrows, however, the corresponding involution
identities are
\begin{equation}
	\left(C_{[\hat n]}^\eta\right)^2
	=
	\operatorname{id}_{\mathcal E_\eta},
	\qquad
	\left(C_{[\hat n]}^{\mathrm{dir}}\right)^2
	=
	\operatorname{id}_{\mathcal E_{[\hat n]}}.
	\label{eq:gpd-C-typed-involution-identities}
\end{equation}
The two identity arrows appearing on the right-hand side are distinct,
although both are represented by the identity matrix \(I_4\).

The source and target loops obey the typed conjugation law
\eqref{eq:gpd-two-loops-conjugation}.

Their matrix representations consequently satisfy
\begin{equation}
	\mathbf C_{[\hat n]}
	=
	\Lambda_\varepsilon(\hat n)
	\mathbf C_{[\hat n]}
	\Lambda_\varepsilon(\hat n)^{-1}.
	\label{eq:gpd-loop-matrix-conjugation}
\end{equation}

The equality in
\eqref{eq:gpd-loop-matrix-conjugation}
is a special property of the present matrix realization.  The intrinsic
groupoid statement is the typed conjugation relation
\eqref{eq:gpd-two-loops-conjugation}.

\subsubsection{Unitary Representations}
\label{subsubsec:gpd-unitary-representations}

A unitary representation of the Poincare groupoid assigns a Hilbert space $\mathcal H_\alpha$
to every metric object 	$\mathcal E_\alpha$, 
$\in\operatorname{Obj}(\mathfrak P_{\mathcal D_{\mathrm{pr}}})$ 
and assigns to every arrow between a pair of objects
\begin{equation*}
	\gamma:
	\mathcal E_\alpha
	\longrightarrow
	\mathcal E_\beta
\end{equation*}
a unitary linear map
\begin{equation}
	\mathcal U(\gamma):
	\mathcal H_\alpha
	\longrightarrow
	\mathcal H_\beta.
	\label{eq:gpd-unitary-arrow-map}
\end{equation}
where
\begin{equation*}
	\left\langle
	\mathcal U(\gamma)\psi,
	\mathcal U(\gamma)\phi
	\right\rangle_{\mathcal H_\beta}
	=
	\langle\psi,\phi\rangle_{\mathcal H_\alpha}
\end{equation*}
for all
\(\psi,\phi\in\mathcal H_\alpha\).

In the language of category theory, the unitary representation of the Poincaré groupoid discussed herein is strictly equivalent to a covariant functor
\[\mathcal U: \mathfrak P_{\mathcal D_{\mathrm{pr}}} \rightarrow \mathbf{Hilb}_U \]
from the Poincaré groupoid to the category of complex Hilbert spaces and unitary operators. 
Under this functorial assignment, each metric object \(\mathcal{E}_{\alpha }\) is mapped to an object \(\mathcal{H}_{\alpha }\) in \(\mathbf{Hilb}_{U}\), and each affine arrow \(\gamma \) is mapped to a morphism \(\mathcal U(\gamma)\).

The assignment preserves identities, compositions and inverses:
\begin{align}
	\mathcal U(\operatorname{id}_{\mathcal E_\alpha})
	&=
	I_{\mathcal H_\alpha},
	\notag\\
	\mathcal U(\gamma_2\circ\gamma_1)
	&=
	\mathcal U(\gamma_2)\mathcal U(\gamma_1),
	\label{eq:gpd-unitary-composition}
	\\
	\mathcal U(\gamma^{-1})
	&=
	\mathcal U(\gamma)^{-1}
	=
	\mathcal U(\gamma)^*.
\notag\end{align}

Restriction to the isotropy group of the Minkowski object gives one of the representations derived in section \ref{subsec:uirs-extended-poincare-group} while restriction to any other directional object gives representations isomorphic to those:
\begin{equation}
	\Pi_\alpha
	:=
	\left.
	\mathcal U
	\right|_{\operatorname{Iso}(\mathcal E_\alpha)}
	:
	\operatorname{Iso}(\mathcal E_\alpha)
	\longrightarrow
	U(\mathcal H_\alpha).
	\label{eq:gpd-isotropy-unitary-representation}
\end{equation}

For each projective axis \([\hat n]\), the two parallel arrows are represented
by unitary maps
\begin{equation}
	\mathcal U(J_{\hat n}),
	\mathcal U(J_{-\hat n})
	:
	\mathcal H_\eta
	\longrightarrow
	\mathcal H_{[\hat n]}.
	\label{eq:gpd-two-unitary-infinite-speed-arrows}
\end{equation}
Their inverses act in the opposite direction:
\begin{equation*}
	\mathcal U(J_{\hat n})^{-1},
	\mathcal U(J_{-\hat n})^{-1}
	:
	\mathcal H_{[\hat n]}
	\longrightarrow
	\mathcal H_\eta.
\end{equation*}

The two typed loops are represented by unitary operators acting within the
corresponding Hilbert spaces:
\begin{align}
	\mathcal U(C_{[\hat n]}^\eta)
	&=
	\mathcal U(J_{\hat n})^{-1}
	\mathcal U(J_{-\hat n})
	:
	\mathcal H_\eta
	\longrightarrow
	\mathcal H_\eta,
	\label{eq:gpd-unitary-C-eta}
	\\
	\mathcal U(C_{[\hat n]}^{\mathrm{dir}})
	&=
	\mathcal U(J_{-\hat n})
	\mathcal U(J_{\hat n})^{-1}
	:
	\mathcal H_{[\hat n]}
	\longrightarrow
	\mathcal H_{[\hat n]}.
	\label{eq:gpd-unitary-C-dir}
\end{align}
Consequently,
\begin{equation*}
	\mathcal U(C_{[\hat n]}^{\mathrm{dir}})
	=
	\mathcal U(J_{\hat n})
	\mathcal U(C_{[\hat n]}^\eta)
	\mathcal U(J_{\hat n})^{-1}.
\end{equation*}
Equivalently, the two unitary arrow representations satisfy
\begin{equation*}
	\mathcal U(J_{-\hat n})
	=
	\mathcal U(J_{\hat n})
	\mathcal U(C_{[\hat n]}^\eta)
	=
	\mathcal U(C_{[\hat n]}^{\mathrm{dir}})
	\mathcal U(J_{\hat n}).
\end{equation*}

Since the typed loops are involutions, their unitary representatives satisfy
\begin{equation*}
	\mathcal U(C_{[\hat n]}^\eta)^2
	=
	I_{\mathcal H_\eta},
	\qquad
	\mathcal U(C_{[\hat n]}^{\mathrm{dir}})^2
	=
	I_{\mathcal H_{[\hat n]}}.
\end{equation*}
They are therefore unitary self-adjoint operators:
\begin{equation*}
	\mathcal U(C_{[\hat n]}^\alpha)^*
	=
	\mathcal U(C_{[\hat n]}^\alpha),
	\qquad
	\alpha\in\{\eta,\mathrm{dir}\},
\end{equation*}
with possible eigenvalues
\begin{equation*}
	+1,
	\qquad
	-1.
\end{equation*}

Finally, the unitary arrow \(\mathcal U(J_{\varepsilon\hat n})\)
intertwines the isotropy representations at its source and target.  For
\(g\in\operatorname{Iso}(\mathcal E_\eta)\), let
\(\Phi_{\varepsilon\hat n}(g)\) be its transport as defined in
\eqref{eq:gpd-isotropy-transport-by-arrow};
one has
\begin{equation}
	\Pi_{[\hat n]}
	\left(
	\Phi_{\varepsilon\hat n}(g)
	\right)
	=
	\mathcal U(J_{\varepsilon\hat n})
	\Pi_\eta(g)
	\mathcal U(J_{\varepsilon\hat n})^{-1}.
	\label{eq:gpd-unitary-isotropy-intertwining}
\end{equation}
Solving \eqref{eq:gpd-unitary-isotropy-intertwining} for the target
representation expresses it as the pullback along
\(\Phi_{\varepsilon\hat n}^{-1}\), conjugated by the unitary arrow.

Thus the groupoid representation does not require the Hilbert spaces
\(\mathcal H_\eta\) and \(\mathcal H_{[\hat n]}\) to be literally the same
space.  It requires them to be unitarily isomorphic through the operators
representing  arrows. 

\subsection{Groupoid Algebra}
\label{subsec:gpd-poincare-algebra-wavefunctions}

The unitary groupoid representation introduced in
Subsection~\ref{subsubsec:gpd-unitary-representations} assigns to every metric
object \(\mathcal E_\alpha\) a Hilbert space \(\mathcal H_\alpha\) and a
unitary representation of its Poincare isotropy group.  We now differentiate
these representations, construct the corresponding infinitesimal generators
and Casimir operators, and study how their spectral data are transported by
the inter-object arrows.

\subsubsection{Infinitesimal Unitary Representations and Poincare Algebra}
\label{subsubsec:gpd-infinitesimal-unitary-poincare-algebra}

Let
\begin{equation*}
	\Pi_\alpha^0:
	\mathcal P_\alpha^0
	=
	\mathbb V\rtimes SO^+(\eta_\alpha)
	\longrightarrow
	U(\mathcal H_\alpha)
\end{equation*}
be the restriction of a unitary groupoid representation to the connected
isotropy group of the object
\begin{equation*}
	\mathcal E_\alpha=(\mathbb V,\eta_\alpha).
\end{equation*}
In a momentum-space realization, its induced action may be written as \cite{Wigner1939}:
\begin{equation}
	\bigl[\Pi_\alpha^0(a,L)\psi\bigr](p)
	=
	e^{i\langle p,a\rangle}
	D_\alpha\!\left(s_\alpha(L,p)\right)
	\psi(L^{-1}p),
	\label{eq:gpd-induced-poincare-action-alpha}
\end{equation}
where \(p\) belongs to a momentum orbit in \(\mathbb V^*\),
\(\langle p,a\rangle\) is the canonical dual pairing, and
\(D_\alpha(s_\alpha(L,p))\) is the unitary little-group action determined by
the chosen orbit representative \cite{Wigner1939}.

For a pure translation, \(L=I\), equation
\eqref{eq:gpd-induced-poincare-action-alpha} reduces to
\begin{equation*}
	\bigl[\Pi_\alpha^0(a,I)\psi\bigr](p)
	=
	e^{i\langle p,a\rangle}\psi(p).
\end{equation*}
Writing
\begin{equation*}
	\Pi_\alpha^0(a,I)
	=
	\exp\!\left(i a^\mu\hat P_\mu^{(\alpha)}\right)
\end{equation*}
and differentiating at \(a=0\), one obtains
\begin{equation}
	\bigl(\hat P_\mu^{(\alpha)}\psi\bigr)(p)
	=
	p_\mu\psi(p).
	\label{eq:gpd-momentum-generator-alpha}
\end{equation}
Thus the commuting translation generators act as multiplication by the
momentum coordinates on the corresponding orbit.

Consider next a one-parameter family of Lorentz transformations near the
identity,
\begin{equation*}
	L(\omega)
	=
	\exp\!\left(
	\frac12\omega^{\mu\nu}m_{\mu\nu}^{(\alpha)}
	\right),
	\qquad
	\omega^{\mu\nu}=-\omega^{\nu\mu}.
\end{equation*}
The unitary representation has the infinitesimal expansion
\begin{equation*}
	\Pi_\alpha^0(0,L(\omega))
	=
	I+
	\frac{i}{2}\omega^{\mu\nu}\hat M_{\mu\nu}^{(\alpha)}
	+O(\omega^2).
\end{equation*}
Differentiation of the little-group factor yields:
\begin{equation*}
	\Sigma_{\mu\nu}^{(\alpha)}(p)
	:=
	i\left.
	\frac{\partial}{\partial\omega^{\mu\nu}}
	D_\alpha\!\left(s_\alpha(L(\omega),p)\right)
	\right|_{\omega=0}.
\end{equation*}
The variation of the momentum argument supplies the orbital part.  Combining
the two contributions gives
\begin{equation}
	\bigl(\hat M_{\mu\nu}^{(\alpha)}\psi\bigr)(p)
	=
	i\left(
	p_\mu\frac{\partial}{\partial p^\nu}
	-
	p_\nu\frac{\partial}{\partial p^\mu}
	\right)\psi(p)
	+
	\Sigma_{\mu\nu}^{(\alpha)}(p)\psi(p).
	\label{eq:gpd-lorentz-generator-alpha}
\end{equation}

The differentiated representation therefore yields the Poincare algebra
\begin{equation}
	\mathfrak p_\alpha
	:=
	\operatorname{Lie}(\mathcal P_\alpha^0)
	=
	\mathbb V\rtimes\mathfrak{so}(\eta_\alpha).
	\label{eq:gpd-poincare-algebra-alpha}
\end{equation}
Its generators satisfy
\begin{align}
	[\hat P_\mu^{(\alpha)},\hat P_\nu^{(\alpha)}]
	&=0,
	\label{eq:gpd-PP-alpha}
	\\
	[\hat M_{\rho\sigma}^{(\alpha)},\hat P_\mu^{(\alpha)}]
	&=
	i\left(
	(\eta_\alpha)_{\mu\rho}\hat P_\sigma^{(\alpha)}
	-
	(\eta_\alpha)_{\mu\sigma}\hat P_\rho^{(\alpha)}
	\right),
	\label{eq:gpd-MP-alpha}
	\\
	[\hat M_{\mu\nu}^{(\alpha)},\hat M_{\rho\sigma}^{(\alpha)}]
	&=
	i\bigl(
	(\eta_\alpha)_{\mu\rho}\hat M_{\nu\sigma}^{(\alpha)}
	-
	(\eta_\alpha)_{\mu\sigma}\hat M_{\nu\rho}^{(\alpha)}
	\notag\\
	&\hspace{2.9cm}
	-
	(\eta_\alpha)_{\nu\rho}\hat M_{\mu\sigma}^{(\alpha)}
	+
	(\eta_\alpha)_{\nu\sigma}\hat M_{\mu\rho}^{(\alpha)}
	\bigr).
	\label{eq:gpd-MM-alpha}
\end{align}
Since all metrics \(\eta_\alpha\) have Lorentzian signature, the algebras
\(\mathfrak p_\alpha\) are mutually isomorphic.  They are different matrix
realizations of the same abstract Poincare Lie algebra:
\begin{equation*}
	\mathfrak p_\alpha
	\simeq
	\mathfrak{iso}(1,3).
\end{equation*}

We now determine the infinitesimal consequence of extending an isotropy group
by the typed loop \(C_{[\hat n]}^\alpha\).  For either endpoint
\(\alpha=\eta\) or \(\alpha=[\hat n]\), the corresponding extended isotropy
group has the form
\begin{equation*}
	\mathcal P_{\alpha,[\hat n]}^{\mathrm{ext}}
	=
	\mathcal P_\alpha^0
	\rtimes_{\operatorname{Ad}_{C_{[\hat n]}^\alpha}}
	\langle C_{[\hat n]}^\alpha\rangle,
	\qquad
	\langle C_{[\hat n]}^\alpha\rangle\simeq\mathbb Z_2.
\end{equation*}
Because the additional factor is discrete, it introduces no new
infinitesimal generator.  Therefore
\begin{equation}
	\operatorname{Lie}
	\left(\mathcal P_{\alpha,[\hat n]}^{\mathrm{ext}}\right)
	=
	\operatorname{Lie}(\mathcal P_\alpha^0)
	=
	\mathfrak p_\alpha.
	\label{eq:gpd-discrete-extension-same-lie-algebra}
\end{equation}
The extension acts instead through the involutive Lie-algebra automorphism
\begin{equation}
	c_{[\hat n]}^\alpha
	:=
	d\!\left(
	\operatorname{Ad}_{C_{[\hat n]}^\alpha}
	\right),
	\qquad
	\left(c_{[\hat n]}^\alpha\right)^2
	=
	\operatorname{id}_{\mathfrak p_\alpha}.
	\label{eq:gpd-loop-lie-algebra-automorphism}
\end{equation}
In coordinate-free form its action is
\begin{align*}
	c_{[\hat n]}^\alpha\bigl(P_\alpha(v)\bigr)
	&=
	P_\alpha(\mathbf C_{[\hat n]}v),
	\\
	c_{[\hat n]}^\alpha\bigl(M_\alpha(u\wedge v)\bigr)
	&=
	M_\alpha
	\left(
	\mathbf C_{[\hat n]}u
	\wedge
	\mathbf C_{[\hat n]}v
	\right).
\end{align*}

Let
\begin{equation*}
	S_{C,\alpha}
	:=
	\mathcal U(C_{[\hat n]}^\alpha)
	:
	\mathcal H_\alpha
	\longrightarrow
	\mathcal H_\alpha
\end{equation*}
be the unitary representative of the loop.  It satisfies
\begin{equation*}
	S_{C,\alpha}^2=I_{\mathcal H_\alpha}
\end{equation*}
and implements the automorphism of the infinitesimal representation:
\begin{equation*}
	S_{C,\alpha}\,
	d\Pi_\alpha(X)\,
	S_{C,\alpha}^{-1}
	=
	d\Pi_\alpha\!\left(c_{[\hat n]}^\alpha(X)\right),
	\qquad
	X\in\mathfrak p_\alpha.
\end{equation*}

In the basis
\begin{equation*}
	\mathcal B_{\hat n}
	=
	(e_0,e_{\hat n},e_1,e_2)
\end{equation*}
adapted to the spatial axis, the common numerical matrix representing the two
typed loops is
\begin{equation*}
	[\mathbf C_{[\hat n]}]_{\mathcal B_{\hat n}}
	=
	\operatorname{diag}(-1,-1,+1,+1).
\end{equation*}
Consequently,
\begin{equation*}
	P_0\mapsto-P_0,
	\qquad
	P_{\hat n}\mapsto-P_{\hat n},
	\qquad
	P_1\mapsto P_1,
	\qquad
	P_2\mapsto P_2,
\end{equation*}
whereas
\begin{equation*}
	M_{\mu\nu}
	\mapsto
	C_\mu{}^\rho C_\nu{}^\sigma M_{\rho\sigma}.
\end{equation*}
If \(R_{\hat n}\) denotes the rotation around \(\hat n\),
\(K_{\hat n}\) the longitudinal boost and \(J_a,K_a\), \(a=1,2\),
the transverse generators, then
\begin{equation*}
	R_{\hat n}\mapsto R_{\hat n},
	\qquad
	K_{\hat n}\mapsto K_{\hat n},
	\qquad
	J_a\mapsto-J_a,
	\qquad
	K_a\mapsto-K_a.
\end{equation*}
The loop therefore constrains and possibly exchanges irreducible sectors of
the connected isotropy representation, as described by the
Clifford--Mackey analysis of
Subsection~\ref{subsec:uirs-extended-poincare-group}, but it leaves the local
Poincare commutation relations unchanged.

\subsubsection{Pauli--Lubanski Vector and Casimir Operators}
\label{subsubsec:gpd-pauli-lubanski-casimirs}

Choose an orientation at \(\mathcal E_\alpha\), and denote the corresponding
Levi--Civita tensor by \(\varepsilon_\alpha^{\mu\nu\rho\sigma}\).  The
Pauli--Lubanski operator at that object is
\begin{equation}
	\hat W_\alpha^\mu
	:=
	\frac12
	\varepsilon_\alpha^{\mu\nu\rho\sigma}
	\hat P_\nu^{(\alpha)}
	\hat M_{\rho\sigma}^{(\alpha)}.
	\label{eq:gpd-pauli-lubanski-alpha}
\end{equation}
The Poincare commutation relations imply
\begin{equation*}
	[\hat W_\alpha^\mu,\hat P_\lambda^{(\alpha)}]=0,
	\qquad
	(\eta_\alpha)_{\mu\nu}
	\hat P^{(\alpha)\mu}\hat W_\alpha^\nu=0.
\end{equation*}
The two intrinsic Casimir operators are therefore
\begin{align}
	\hat C_{1,\alpha}
	&:=
	(\eta_\alpha)_{\mu\nu}
	\hat P^{(\alpha)\mu}\hat P^{(\alpha)\nu},
	\label{eq:gpd-first-intrinsic-casimir}
	\\
	\hat C_{2,\alpha}
	&:=
	(\eta_\alpha)_{\mu\nu}
	\hat W_\alpha^\mu\hat W_\alpha^\nu.
	\label{eq:gpd-second-intrinsic-casimir}
\end{align}
They commute with all generators of \(\mathfrak p_\alpha\).

Substitution of equation~\eqref{eq:gpd-lorentz-generator-alpha} into
\eqref{eq:gpd-pauli-lubanski-alpha} shows that the orbital contribution
vanishes by antisymmetry.  Hence
\begin{equation}
	\hat W_\alpha^\mu
	=
	\frac12
	\varepsilon_\alpha^{\mu\nu\rho\sigma}
	\hat P_\nu^{(\alpha)}
	\Sigma_{\rho\sigma}^{(\alpha)}.
	\label{eq:gpd-pauli-lubanski-spin-part}
\end{equation}

For the Minkowski metric \eqref{eq:srf-minkowski-metric}, choose
\(\varepsilon^{0123}=+1\) and
define
\begin{equation*}
	J_i
	:=
	\frac12\epsilon_{ijk}\Sigma_{jk},
	\qquad
	K_i
	:=
	\Sigma_{i0}.
\end{equation*}
Writing \(P^\mu=(P^0,\mathbf P)\), equation
\eqref{eq:gpd-pauli-lubanski-spin-part} becomes
\begin{equation}
	W^0
	=
	\mathbf P\cdot\mathbf J,
	\qquad
	\mathbf W
	=
	P^0\mathbf J-\mathbf P\times\mathbf K.
	\label{eq:gpd-pauli-lubanski-vector-form}
\end{equation}
These formulas will be evaluated below on the spectral representatives obtained by transporting spectral data by means of groupoids arrows.

The loop automorphism preserves both Casimir operators:
\begin{equation}
	c_{[\hat n]}^\alpha(\hat C_{1,\alpha})
	=
	\hat C_{1,\alpha},
	\qquad
	c_{[\hat n]}^\alpha(\hat C_{2,\alpha})
	=
	\hat C_{2,\alpha}.
	\label{eq:gpd-loop-preserves-casimirs}
\end{equation}
It may nevertheless exchange distinct connected components or distinct
Wigner sectors having the same Casimir eigenvalues.  The Casimirs classify
the infinitesimal representation, whereas the unitary operator
\(S_{C,\alpha}\) supplies additional discrete extension data.

\subsubsection{\texorpdfstring{\(*\)-Isomorphism of Poincare Observable
		Algebras }{Isomorphism of Poincare Observable
		Algebras }}
\label{subsubsec:gpd-transport-generators-spectral-data}

Let \(J_{\varepsilon\hat n}\), \(\varepsilon\in\{+1,-1\}\), be one
of the parallel arrows \eqref{eq:gpd-parallel-oriented-arrows}, represented
as in \eqref{eq:gpd-two-infinite-speed-arrow-matrices}.  The source and target
Poincare Lie algebras are the instances \(\alpha=\eta\) and
\(\alpha=[\hat n]\) of \eqref{eq:gpd-poincare-algebra-alpha}.

For
\begin{equation*}
	(a,X),(b,Y)
	\in
	\mathbb V\rtimes\mathfrak{so}(\eta_\alpha),
\end{equation*}
the Lie bracket is
\begin{equation*}
	[(a,X),(b,Y)]
	=
	\left(
	Xb-Ya,
	[X,Y]
	\right).
\end{equation*}

By the metric-isomorphism identity
\eqref{eq:srf-infinite-speed-metric-isomorphism},
conjugation by \(J_{\varepsilon\hat n}\) maps
\(\mathfrak{so}(\eta)\) isomorphically onto
\(\mathfrak{so}(\eta_{[\hat n]})\).  The arrow therefore induces the
Poincare Lie-algebra isomorphism
\begin{equation*}
	\phi_{\varepsilon\hat n}:
	\mathfrak p_\eta
	\longrightarrow
	\mathfrak p_{[\hat n]},
\end{equation*}
defined by
\begin{equation}
	\phi_{\varepsilon\hat n}(a,X)
	=
	\left(
	J_{\varepsilon\hat n}a,
	J_{\varepsilon\hat n}
	X
	J_{\varepsilon\hat n}^{-1}
	\right).
	\label{eq:gpd-poincare-lie-algebra-isomorphism-definition}
\end{equation}
Indeed,
\begin{equation*}
	\phi_{\varepsilon\hat n}
	\left(
	[(a,X),(b,Y)]
	\right)
	=
	\left[
	\phi_{\varepsilon\hat n}(a,X),
	\phi_{\varepsilon\hat n}(b,Y)
	\right].
\end{equation*}

The Lie algebra is sufficient to describe the infinitesimal generators and
their commutation relations.  The structure transported between metric
objects, however, also contains products of generators.  In particular,
the first Poincare Casimir is quadratic in the translation generators,
the Pauli--Lubanski observable contains products of translation and Lorentz
generators, and the second Casimir is quadratic in the Pauli--Lubanski
components.  These products are not themselves elements of the original
Poincare Lie algebra.

To include them in a single associative algebra, we first complexify:
\begin{equation*}
	\mathfrak p_{\alpha,\mathbb C}
	=
	\mathfrak p_\alpha
	\otimes_{\mathbb R}
	\mathbb C.
\end{equation*}
The complexification is natural because the quantum representation spaces
are complex Hilbert spaces and the adjoint operation is conjugate-linear.

For arbitrary
\[
\xi,\zeta\in\mathfrak p_{\alpha,\mathbb C},
\]
let
\[
\mathcal I_\alpha
:=
\left\langle
\xi\otimes\zeta
-
\zeta\otimes\xi
-
[\xi,\zeta]
\;\middle|\;
\xi,\zeta\in\mathfrak p_{\alpha,\mathbb C}
\right\rangle
\]
denote the two-sided ideal generated by the Lie-algebra relations.
The complex universal enveloping algebra is then defined by
\begin{equation}
	U(\mathfrak p_{\alpha,\mathbb C})
	:=
	\frac{
		T(\mathfrak p_{\alpha,\mathbb C})
	}{
		\mathcal I_\alpha
	}.
\end{equation}
where $T(\mathfrak p_{\alpha,\mathbb C})$ denotes the tensor algebra on  $p_{\alpha,\mathbb C}$.
Set-theoretically, the quotient
\[
T(\mathfrak p_{\alpha,\mathbb C})/\mathcal I_\alpha
\]
is the set of equivalence classes
\[
X+\mathcal I_\alpha,
\qquad
X\in T(\mathfrak p_{\alpha,\mathbb C}),
\]
where
\[
X\sim Y
\quad\Longleftrightarrow\quad
X-Y\in\mathcal I_\alpha.
\]
Since \(\mathcal I_\alpha\) is a two-sided ideal \cite{{Humphreys1972}}, the associative
product descends consistently to the quotient:
\[
(X+\mathcal I_\alpha)(Y+\mathcal I_\alpha)
:=
XY+\mathcal I_\alpha.
\]
The quotient is a unital
associative complex algebra.

\(U(\mathfrak p_{\alpha,\mathbb C})\) contains the Poincare generators
and all finite noncommutative polynomials in those generators, subject to the
original Lie-algebra commutation relations.

For example, it contains expressions of the form
\begin{equation*}
	P_{\alpha,\mu}P_{\alpha,\nu},
	\qquad
	P_{\alpha,\mu}M_{\alpha,\rho\sigma},
	\qquad
	M_{\alpha,\mu\nu}M_{\alpha,\rho\sigma},
\end{equation*}
and therefore contains the algebraic representatives of the Poincare
Casimir operators and of the Pauli--Lubanski observable.

We use the Hermitian-generator convention customary in physics.  Let
\[
\iota_\alpha:
\mathfrak p_\alpha
\longrightarrow
U(\mathfrak p_{\alpha,\mathbb C})
\]
denote the canonical embedding of the real Poincare Lie algebra into its
complex universal enveloping algebra.  For \(X\in\mathfrak p_\alpha\), define
the corresponding abstract Hermitian generator by
\[
\widehat X_\alpha
:=
i\,\iota_\alpha(X).
\]
In a unitary representation its image is
\[
\pi_\alpha(\widehat X_\alpha)
=
i\,dU_\alpha(X),
\]
which is the self-adjoint generator of the one-parameter unitary group
associated with \(X\).

We equip \(U(\mathfrak p_{\alpha,\mathbb C})\) with the conjugate-linear
anti-involution determined on the Hermitian generators by
\[
\widehat P_\alpha(a)^*
=
\widehat P_\alpha(a),
\qquad
\widehat M_\alpha(u\wedge v)^*
=
\widehat M_\alpha(u\wedge v),
\]
for real \(a,u,v\), and extended according to
\[
(\lambda A+\mu B)^*
=
\overline{\lambda}A^*
+
\overline{\mu}B^*,
\qquad
(AB)^*
=
B^*A^*,
\qquad
(A^*)^*
=
A.
\]
Since the Hermitian generators generate the enveloping algebra, these
prescriptions determine the involution uniquely.  Their compatibility with
the Poincare commutation relations guarantees that the involution is
well-defined on the quotient defining the universal enveloping algebra.

Equivalently, in terms of the canonical real Lie-algebra generators,
the same involution is written as
\[
\iota_\alpha(X)^*
=
-\iota_\alpha(X).
\]
Thus the minus sign belongs to the skew-adjoint mathematical convention,
whereas the physical generators used below are Hermitian.
We denote the resulting algebraic polynomial \(*\)-algebra by
\begin{equation}
\mathfrak A_\alpha^{\mathrm{alg}}
:=
U(\mathfrak p_{\alpha,\mathbb C}).
\label{eq:gpd-algebraic-observable-algebra}
\end{equation}
Its Hermitian elements, rather than all of its elements, represent algebraic
observables.  No topological or \(C^*\)-completion is assumed.
In every unitary representation
\[
\pi_\alpha:
\mathfrak A_\alpha^{\mathrm{alg}}
\longrightarrow
\operatorname{End}(\mathcal D_\alpha),
\]
this algebraic involution is represented by the Hilbert-space adjoint:
\[
\pi_\alpha(A^*)
=
\pi_\alpha(A)^\dagger.
\]
By the universal property of the enveloping algebra, the Lie-algebra
isomorphism
\(\phi_{\varepsilon\hat n}\) extends uniquely to a unital algebra
isomorphism
\begin{equation}
	\Phi_{\varepsilon\hat n}:
	\mathfrak A_\eta^{\mathrm{alg}}
	\longrightarrow
	\mathfrak A_{[\hat n]}^{\mathrm{alg}}.
	\label{eq:gpd-enveloping-algebra-isomorphism}
\end{equation}
It is characterized by
\begin{equation*}
	\Phi_{\varepsilon\hat n}(1)
	=
	1
\end{equation*}
and by its action on the Poincare generators:
\begin{align}
	\Phi_{\varepsilon\hat n}
	\left(
	P_\eta(a)
	\right)
	&=
	P_{[\hat n]}
	\left(
	J_{\varepsilon\hat n}a
	\right),
	\label{eq:gpd-enveloping-isomorphism-momentum}
	\\
	\Phi_{\varepsilon\hat n}
	\left(
	M_\eta(u\wedge v)
	\right)
	&=
	M_{[\hat n]}
	\left(
	J_{\varepsilon\hat n}u
	\wedge
	J_{\varepsilon\hat n}v
	\right).
	\label{eq:gpd-enveloping-isomorphism-lorentz}
\end{align}
Its action on arbitrary products is then fixed by
\begin{equation}
	\Phi_{\varepsilon\hat n}(AB)
	=
	\Phi_{\varepsilon\hat n}(A)
	\Phi_{\varepsilon\hat n}(B).
	\label{eq:gpd-enveloping-isomorphism-product}
\end{equation}

Since \(J_{\varepsilon\hat n}\) is real, the map
\(\phi_{\varepsilon\hat n}\) sends real infinitesimal generators to real
infinitesimal generators.  It consequently preserves the involution:
\begin{equation}
	\Phi_{\varepsilon\hat n}(A^*)
	=
	\Phi_{\varepsilon\hat n}(A)^*,
	\qquad
	A\in\mathfrak A_\eta^{\mathrm{alg}}.
	\label{eq:gpd-enveloping-isomorphism-star}
\end{equation}
Together, \eqref{eq:gpd-enveloping-algebra-isomorphism} and
\eqref{eq:gpd-enveloping-isomorphism-star} establish that
\(\Phi_{\varepsilon\hat n}\) is a unital \(*\)-isomorphism.

The inverse is induced by the inverse groupoid arrow:
\begin{equation*}
	\Phi_{\varepsilon\hat n}^{-1}
	=
	\Phi_{J_{\varepsilon\hat n}^{-1}}.
\end{equation*}

The groupoid arrow therefore identifies not only the two Poincare Lie
algebras and their commutation relations, but the complete algebraic
structure generated by their infinitesimal observables.  In particular, the
transport extends automatically from the ten Poincare generators to their
noncommutative products, the Pauli--Lubanski observable and the two Poincare
Casimir operators.

Let $\pi_\eta$, $\pi_{[\hat n]}$ be the representations of the two observable algebras.  The unitary arrow in \eqref{eq:gpd-two-unitary-infinite-speed-arrows} implements the \(*\)-isomorphism:
\begin{equation}
		\mathcal U(J_{\varepsilon\hat n})
		\pi_\eta(A)
		\mathcal U(J_{\varepsilon\hat n})^{-1}
		=
		\pi_{[\hat n]}
		\left(
		\Phi_{\varepsilon\hat n}(A)
		\right),
		\qquad
		A\in\mathfrak A_\eta.
	\label{eq:gpd-unitary-implementation-star-isomorphism}
\end{equation}
Equation~\eqref{eq:gpd-unitary-implementation-star-isomorphism} is the
covariance law for the represented observable algebras. Specializing
it to generators gives:
\begin{align}
	\mathcal U(J_{\varepsilon\hat n})
	\hat P_\eta(a)
	\mathcal U(J_{\varepsilon\hat n})^{-1}
	&=
	\hat P_{[\hat n]}
	\left(
	J_{\varepsilon\hat n}a
	\right),
	\label{eq:gpd-arrow-transport-momentum-generators}
	\\
	\mathcal U(J_{\varepsilon\hat n})
	\hat M_\eta(u\wedge v)
	\mathcal U(J_{\varepsilon\hat n})^{-1}
	&=
	\hat M_{[\hat n]}
	\left(
	J_{\varepsilon\hat n}u
	\wedge
	J_{\varepsilon\hat n}v
	\right).
\notag\end{align}
The two equations are therefore consequences of a single
\(*\)-algebraic transport law.

Suppose now that
\(\psi_p\in\mathcal H_\eta\) is a joint generalized eigenstate of the
translation generators:
\begin{equation*}
	\hat P_\eta(a)\psi_p
	=
	\langle p,a\rangle\psi_p
	\qquad
	\text{for every }a\in\mathbb V,
\end{equation*}
where \(p\in\mathbb V^*\) is a momentum covector.  Define the transported
state by
\begin{equation*}
	\psi_{q,\varepsilon}
	:=
	\mathcal U(J_{\varepsilon\hat n})\psi_p
	\in
	\mathcal H_{[\hat n]}.
\end{equation*}
Using
\eqref{eq:gpd-arrow-transport-momentum-generators}, one obtains
\begin{equation*}
	\hat P_{[\hat n]}
	\left(
	J_{\varepsilon\hat n}a
	\right)
	\psi_{q,\varepsilon}
	=
	\langle p,a\rangle
	\psi_{q,\varepsilon}.
\end{equation*}
Writing
\begin{equation*}
	b
	=
	J_{\varepsilon\hat n}a,
	\qquad
	a
	=
	J_{\varepsilon\hat n}^{-1}b,
\end{equation*}
gives
\begin{equation*}
	\hat P_{[\hat n]}(b)
	\psi_{q,\varepsilon}
	=
	\left\langle
	J_{\varepsilon\hat n}^{-T}p,
	b
	\right\rangle
	\psi_{q,\varepsilon}.
\end{equation*}
Therefore the transported momentum covector is
\begin{equation}
		q_{\varepsilon\hat n}
		=
		J_{\varepsilon\hat n}^{-T}p.
\end{equation}

For the symmetric involutive matrices employed in the present construction,
\begin{equation*}
	J_{\varepsilon\hat n}^{-T}
	=
	J_{\varepsilon\hat n},
\end{equation*}
and hence
\begin{equation*}
	q_{\varepsilon\hat n}
	=
	\Lambda_\varepsilon(\hat n)p.
\end{equation*}

Unlike the four mutually commuting translation generators, the six Lorentz
generators do not possess a common set of simultaneous spectral values.
The \(*\)-isomorphism transports the Lorentz generators, their commutation
relations and their representation matrices.  Simultaneous spectral values
may be assigned to central Casimir
operators of Poincarè algebra.

Since momenta are covectors, their intrinsic quadratic form is defined using
the inverse metric:
\begin{equation*}
	\mathcal C_{1,\alpha}
	=
	\eta_\alpha^{-1}
	\left(
	P_\alpha,P_\alpha
	\right).
\end{equation*}
This is the covector form of
\eqref{eq:gpd-first-intrinsic-casimir}; the inverse metric contracts the
lower momentum indices.

The metric-isomorphism identity implies
\begin{equation}
		\Phi_{\varepsilon\hat n}
		\left(
		\mathcal C_{1,\eta}
		\right)
		=
		\mathcal C_{1,[\hat n]}.
\end{equation}
At the spectral level this gives
\begin{equation}
	q_{\varepsilon\hat n}^{T}
	\eta_{[\hat n]}^{-1}
	q_{\varepsilon\hat n}
	=
	p^{T}\eta^{-1}p.
	\label{eq:gpd-intrinsic-first-casimir-transport}
\end{equation}
For the metric matrices used in the present realization,
\begin{equation*}
	\eta^{-1}=\eta,
	\qquad
	\eta_{[\hat n]}^{-1}
	=
	\eta_{[\hat n]},
\end{equation*}
so the inverse metrics in
\eqref{eq:gpd-intrinsic-first-casimir-transport} may be replaced by the
metric matrices themselves.

Let \(W_\alpha\) denote the Pauli--Lubanski observable constructed using the
metric and volume form of \(\mathcal E_\alpha\), and define
\begin{equation*}
	\mathcal C_{2,\alpha}
	=
	\eta_\alpha^{-1}
	\left(
	W_\alpha,W_\alpha
	\right).
\end{equation*}
If the volume form is transported consistently along
\(J_{\varepsilon\hat n}\), then
\begin{equation}
		\Phi_{\varepsilon\hat n}
		\left(
		\mathcal C_{2,\eta}
		\right)
		=
		\mathcal C_{2,[\hat n]}
	\label{eq:gpd-second-casimir-star-transport}
\end{equation}
If a fixed orientation is used instead, the Pauli--Lubanski pseudovector can
acquire an overall sign under an orientation-reversing arrow; this sign
cancels in the quadratic Casimir.

Consequently, if
\begin{equation*}
	\pi_\eta(\mathcal C_{r,\eta})\psi_p
	=
	c_r\psi_p,
	\qquad
	r\in\{1,2\},
\end{equation*}
then
\begin{equation*}
	\pi_{[\hat n]}(\mathcal C_{r,[\hat n]})
	\psi_{q,\varepsilon}
	=
	c_r\psi_{q,\varepsilon}.
\end{equation*}
Groupoid transport therefore preserves the intrinsic Poincare class of the
state: it carries the representation to an isomorphic realization at another
metric object.

\subsection{Physical Interpretation and Common Affine Structure of the Pointed Groupoid}
\label{subsec:gpd-physical-affine-interpretation}

Let $A$ be a common real affine
space of spacetime events, modeled on a four-dimensional translation vector
space $V\simeq\mathbb R^4$. Every metric object of the groupoid is a metric
structure on this same affine event space:
\begin{equation*}
  \mathcal E_\alpha=(A,\eta_\alpha),\qquad
  \mathcal E_\eta=(A,\eta),\qquad
  \mathcal E_{[\hat n]}=(A,\eta_{[\hat n]}).
\end{equation*}
Choosing an affine origin $O\in A$, conventionally $O=(0,0,0,0)$, permits
coordinates on $A$ to be identified with $V$; the choice is not required to
define event separations. Because the translation space $V$ is common, its
dual $V^*=\operatorname{Hom}(V,\mathbb R)$, the vector space of momentum
covectors, is common to all metric objects as well. In particular, no
identification of momentum spaces through a metric is required before
comparing covectors: the metrics enter only when raising indices or
classifying the common covector data.

Let $S,D\in A$ denote a source and a detector event. Their affine separation
is the common translation vector
\begin{equation*}
  a=D-S\in V.
\end{equation*}
In the distinguished laboratory coordinates, if the spatial displacement
$\Delta\mathbf x=\mathbf x_D-\mathbf x_S$ is nonzero, define the oriented
source--detector direction
\begin{equation}
  \hat n_{SD}:=\frac{\Delta\mathbf x}{\|\Delta\mathbf x\|}
  =(\sin\theta\cos\phi,\sin\theta\sin\phi,\cos\theta).
  \label{eq:gpd-source-detector-direction}
\end{equation}
When $S$ is chosen as the origin this is simply the spatial direction of
$D$. The directional metric object is indexed by $[\hat n_{SD}]$, whereas the
ordered event pair and the chosen orientation convention distinguish the
typed arrows $J_{\hat n_{SD}}$ and $J_{-\hat n_{SD}}$ between the same two
metric objects. The separation by itself fixes the direction; selecting the
arrow still requires the declared orientation convention.

An inter-object arrow is an isomorphism between metric descriptions, not a
replacement of the physical events $S$ and $D$. On the translation data its
linear part acts as
\begin{equation*}
  a\longmapsto a_{[\hat n]}=J_{\hat n}a.
\end{equation*}
Its induced action on momentum data must use the corresponding dual map
(with the inverse transpose when treating momentum as a covector). In the
component convention used for the momentum representatives of this paper,
the oriented transport takes the massive Minkowski representative to
\begin{equation}
 p_M=(M,0,0,0)^T
 \quad\longmapsto\quad
 q_{\hat n}=\Lambda_+(\hat n)p_M=(0,-M\hat n)^T,
 \label{eq:gpd-physical-directed-momentum}
\end{equation}
while the oppositely oriented arrow yields
$q_{-\hat n}=(0,+M\hat n)^T$. Here the component notation uses the fixed
laboratory identification of $V$ with $V^*$ where appropriate; invariantly,
quadratic momentum classification uses the inverse metrics acting on
covectors.

The intrinsic and laboratory classifications concern the same transported
momentum datum but use different quadratic forms. With the conventions for
these metric matrices used throughout the paper,
\begin{align}
 \eta_{[\hat n]}^{-1}(q_{\hat n},q_{\hat n})&=-M^2,
 \label{eq:gpd-affine-intrinsic-momentum}\\
 \eta^{-1}(q_{\hat n},q_{\hat n})&=+M^2.
 \label{eq:gpd-affine-pointed-momentum}
\end{align}
The first relation expresses the intrinsic massive class preserved under
the metric-compatible arrow. The second states that the transported datum
belongs to a spacelike orbit when classified relative to the Minkowski
laboratory structure. There is no contradiction: timelike and spacelike
refer here to different metric forms on the common dual momentum space.

The physical readout principle of the Minkowski-pointed groupoid is therefore
the following: intrinsic covariance and algebraic transport are governed by
the metric of each object, but the operational classification of transported
momentum data uses the Lorentz action of the distinguished Minkowski object
on the common dual space $V^*$. Explicitly, after selecting the direction
$\hat n_{SD}$ and transporting $p_M$ to $q_{\hat n}$, one takes the orbit and
stabilizer
\begin{equation*}
 \mathcal O_\eta(q_{\hat n})=SO^+(\eta)\cdot q_{\hat n},\qquad
 H_{q_{\hat n}}^\eta=\operatorname{Stab}_{SO^+(\eta)}(q_{\hat n}).
\end{equation*}
In the nonzero spacelike sector the stabilizer is isomorphic to
$SO^+(2,1)$, up to the appropriate covering-group convention. Thus the
source--detector direction selects a transported orbit representative and an
adapted little-group realization. It does \emph{not}, by itself, select a
unique unitary irreducible representation of that little group. The
Minkowski-pointed wavefunction classification below associates the allowed
induced-representation families to the transported orbit data; a specific
internal representation label $\sigma$ remains additional representation
data, unless a separate dynamical rule is supplied.

This operational classification must not be confused with the intrinsic
isotropy action. The quadratic form $\eta^{-1}(q,q)$ is invariant under the
Minkowski Lorentz action used for readout, but need not be invariant under
the intrinsic Lorentz action associated with a directional metric object.
Consequently the Minkowski-pointed operators introduced below are not
asserted to be central Casimirs of each directional Poincare algebra.

For a laboratory-subluminal process the ordinary Minkowski object is the
natural description. For a process described by an oriented inter-object
arrow, the directional metric supplies intrinsic covariance and the
pointed Minkowski metric supplies the final operational orbit classification.
This separates the two roles of the metrics without identifying distinct
isotropy representations or postulating an isomorphism between their
little groups.

\subsection{Minkowski Pointed Wavefunctions Classification}
\label{subsubsec:gpd-intrinsic-pointed-casimir-classification}

Every metric object \(\mathcal E_\alpha\) carries an isotropy representation
of a Poincare group and therefore a representation of the corresponding
complex universal enveloping algebra
\begin{equation*}
	\pi_\alpha:
	U(\mathfrak p_{\alpha,\mathbb C})
	\longrightarrow
	\operatorname{End}(\mathcal D_\alpha),
\end{equation*}
where \(\mathcal D_\alpha\subseteq\mathcal H_\alpha\) is a common invariant
dense domain for the infinitesimal generators and their polynomial products.

The arrows of the groupoid induce unital \( * \)-isomorphisms
\begin{equation*}
	\Phi_J:
	U(\mathfrak p_{\alpha,\mathbb C})
	\longrightarrow
	U(\mathfrak p_{\beta,\mathbb C}),
\end{equation*}
which transport generators, polynomial operators and their representation
data between metric objects.

The groupoid is Minkowski-pointed: the distinguished object
\eqref{eq:gpd-object-eta} represents the laboratory structure. The common
affine dual and operational readout principle established in
Section~\ref{subsec:gpd-physical-affine-interpretation} supply the
geometric basis for the following classification.

Let
\[
\mathcal A_\alpha
=
U(\mathfrak p_{\alpha,\mathbb C})
\]
be the complex universal enveloping algebra associated with the isotropy
representation of the metric object \(\mathcal E_\alpha\).  Products of
Poincare generators define operator-valued spacetime tensors with
coefficients in \(\mathcal A_\alpha\).  Every groupoid arrow $J$
transports these tensors through the tensorial extension of the induced
unital \( * \)-isomorphism $\Phi_J$

The intrinsic metrics \(\eta_\alpha\) determine the isotropy groups and the
covariance of inter-object transport.  Physical classification is instead
defined by a single fixed family of contraction tensors, namely the
Minkowski metric \(\eta\) and the associated orientation tensor
\(\varepsilon_\eta\) of the distinguished object
\(\mathcal E_\eta\).

Accordingly, every operator-valued tensor transported to any object of the
groupoid is physically classified by contracting its spacetime indices with
\(\eta\) and \(\varepsilon_\eta\).  The groupoid therefore supports a common
Minkowski-pointed tensor calculus on the family of transported universal
enveloping algebras.

The first
Minkowski-pointed classification operator is:
\begin{equation}
	\widehat{\mathcal Q}_{1,\eta}^{(\alpha)}
	:=
	\eta_{\mu\nu}
	\widehat P_\alpha^\mu
	\widehat P_\alpha^\nu.
	\label{eq:gpd-first-pointed-classification-operator}
\end{equation}
While, for the Pauli-Lubanski vector we have:
\begin{equation}
	\widehat{\mathcal W}_{\eta,\alpha}^{\mu}
	:=
	\frac12
	\varepsilon_\eta^{\mu\nu\rho\sigma}
	\widehat P_{\alpha,\nu}
	\widehat M_{\alpha,\rho\sigma},
	\label{eq:gpd-pointed-pauli-lubanski}
\end{equation}
and the second classification operator is:
\begin{equation}
	\widehat{\mathcal Q}_{2,\eta}^{(\alpha)}
	:=
	\eta_{\mu\nu}
	\widehat{\mathcal W}_{\eta,\alpha}^{\mu}
	\widehat{\mathcal W}_{\eta,\alpha}^{\nu}.
	\label{eq:gpd-second-pointed-classification-operator}
\end{equation}

For \(\alpha=\eta\), these are the ordinary central Casimir operators of the
Minkowski Poincare algebra:
\begin{equation*}
	\widehat{\mathcal Q}_{1,\eta}^{(\eta)}
	=
	\widehat C_{1,\eta},
	\qquad
	\widehat{\mathcal Q}_{2,\eta}^{(\eta)}
	=
	\widehat C_{2,\eta}.
\end{equation*}
For \(\alpha\neq\eta\), the operators
\(\widehat{\mathcal Q}_{i,\eta}^{(\alpha)}\) need not belong to the center of
\(U(\mathfrak p_{\alpha,\mathbb C})\), because the intrinsic Lorentz group of
\(\mathcal E_\alpha\) preserves \(\eta_\alpha\), rather than \(\eta\).
They are therefore used to extract readout data relative to the distinguished
Minkowski object, not as central Casimirs of the directional isotropy algebra.
The orbit classification below is made under $SO^+(\eta)$ acting on the
common dual $V^*$, as established in
Section~\ref{subsec:gpd-physical-affine-interpretation}.

Let \(\psi_q\in\mathcal H_\alpha\) be a generalized joint momentum
eigenstate:
\begin{equation*}
	\widehat P_\alpha^\mu\psi_q
	=
	q^\mu\psi_q.
\end{equation*}
Then
\begin{equation*}
	\widehat{\mathcal Q}_{1,\eta}^{(\alpha)}
	\psi_q
	=
	\left(q^T\eta q\right)\psi_q.
\end{equation*}
The sign of this spectral value determines the Minkowski orbit type:
\begin{equation*}
	q^T\eta q
	\begin{cases}
		<0, & \text{timelike},\\
		=0, & \text{null},\\
		>0, & \text{spacelike}.
	\end{cases}
\end{equation*}

The associated Minkowski orbit and little group are
\begin{align*}
	\mathcal O_\eta(q)
	&:=
	\left\{
	Lq:
	L\in SO^+(\eta)
	\right\},
	\\
	H_q^\eta
	&:=
	\left\{
	L\in SO^+(\eta):
	Lq=q
	\right\}.
\end{align*}

The classification operators determine the orbit invariants, but an
irreducible Poincare representation also requires a unitary irreducible
representation \(\sigma\) of the little group \(H_q^\eta\).  We therefore
define the Minkowski-pointed groupoid classification map by
\begin{equation}
	\operatorname{Class}_\eta:
	(\mathcal E_\alpha,q,\sigma)
	\longmapsto
	\left[
	\rho_{q,\sigma}^{\eta}
	\right],
	\label{eq:gpd-minkowski-classification-map}
\end{equation}
where
\begin{equation}
	\rho_{q,\sigma}^{\eta}
	:=
	\operatorname{Ind}_{
		\mathbb V\rtimes H_q^\eta
	}^{
		\mathcal P_\eta^0
	}
	\left(
	\chi_q\otimes\sigma
	\right),
	\qquad
	\chi_q(a)
	=
	e^{i\langle q,a\rangle}.
	\label{eq:gpd-pointed-induced-representation}
\end{equation}
Here $q$ is selected by the directional transport, while $\sigma$ is a
chosen unitary irreducible representation of the pointed little group.
The source--detector direction fixes $q$ and its orbit type, but does not
by itself determine a unique $\sigma$; such a selection would require
additional representation data or a dynamical prescription.

Thus \(\operatorname{Class}_\eta\) is not a homomorphism between universal
enveloping algebras and is not an operator on a single Hilbert space.  It is a
map from the operator and spectral data carried by the groupoid to the
unitary dual of the distinguished Poincare group:
\begin{equation*}
	\operatorname{Class}_\eta:
	\operatorname{RepData}(\mathfrak P)
	\longrightarrow
	\widehat{\mathcal P_\eta^0}.
\end{equation*}

Transport and Minkowski contraction are distinct operations.  In general,
\begin{equation}
	\operatorname{Contr}_{\eta}
	\circ
	\Phi_J
	\neq
	\Phi_J
	\circ
	\operatorname{Contr}_{\eta_\alpha}.
	\label{eq:gpd-transport-contraction-noncommutativity}
\end{equation}
This noncommutativity is responsible for the fact that data belonging
intrinsically to one Poincare class may determine a different Minkowski
orbit type after inter-object transport.

For example, let \(p_M\) denote the massive representative
\eqref{eq:uir-massive-positive-representative}, with invariant
\eqref{eq:uir-massive-positive-invariant}.

The two oriented arrows give
\begin{align}
	q_{\hat n}
	&:=
	\Lambda_+(\hat n)p_M
	=
	(0,-M\hat n)^T,
	\label{eq:gpd-positive-arrow-massive-transport}
	\\
	q_{-\hat n}
	&:=
	\Lambda_-(\hat n)p_M
	=
	(0,+M\hat n)^T.
	\label{eq:gpd-negative-arrow-massive-transport}
\end{align}
Metric covariance of the inter-object arrows gives
\begin{equation}
	q_{\pm\hat n}^T
	\eta_{[\hat n]}
	q_{\pm\hat n}
	=
	p_M^T\eta p_M
	=
	-M^2.
	\label{eq:gpd-directional-intrinsic-massive-value}
\end{equation}
This identity expresses the preservation of the intrinsic metric class by
groupoid transport.

The Minkowski-pointed classification operator instead gives
\begin{equation}
	q_{\pm\hat n}^T
	\eta
	q_{\pm\hat n}
	=
	+M^2.
	\label{eq:gpd-directional-pointed-spacelike-value}
\end{equation}
Consequently,
\begin{equation}
	\operatorname{Class}_\eta
	\left(
	\mathcal E_{[\hat n]},
	q_{\pm\hat n},
	\sigma
	\right)
	=
	\left[
	\operatorname{Ind}_{
		\mathbb V\rtimes SO(2,1)
	}^{
		\mathcal P_\eta^0
	}
	\left(
	\chi_{q_{\pm\hat n}}\otimes\sigma
	\right)
	\right].
	\label{eq:gpd-directional-data-tachyonic-class}
\end{equation}
The transported massive datum therefore determines a spacelike Poincare
orbit and, for each allowed choice of $\sigma$, a tachyonic Wigner class
relative to the distinguished Minkowski object.

\subsubsection{Massive Wavefunctions}
\label{subsubsec:gpd-massive-wavefunctions}

Consider the positive-energy representative
\eqref{eq:uir-massive-positive-representative}, with invariant
\eqref{eq:uir-massive-positive-invariant} and the \(SO(3)\) little group
of \eqref{eq:uir-massive-orbit-type}.

For a spin-\(s\) UIR \(D^s\) of \(SO(3)\), the classification map gives
\begin{equation*}
	\operatorname{Class}_\eta
	\left(
	\mathcal E_\eta,p_+,D^s
	\right)
	=
	\left[
	\operatorname{Ind}_{
		\mathbb V\rtimes SO(3)
	}^{
		\mathcal P_\eta^0
	}
	\left(
	\chi_{p_+}\otimes D^s
	\right)
	\right].
\end{equation*}

For the standard representative,
\begin{equation*}
	W^\mu=(0,M\mathbf J),
\end{equation*}
and therefore
\begin{equation*}
	C_1=-M^2,
	\qquad
	C_2=M^2\mathbf J^2.
\end{equation*}
Since
\begin{equation*}
	\mathbf J^2=s(s+1)I,
\end{equation*}
the corresponding Casimir equations are
\begin{equation}
	\left(
	\widehat P_\eta^2+M^2
	\right)
	\psi_{M,s}=0,
	\qquad
	\left(
	\widehat W_\eta^2-M^2s(s+1)
	\right)
	\psi_{M,s}=0.
	\label{eq:gpd-massive-wave-equations}
\end{equation}

For the negative-energy representative
\begin{equation*}
	p_-^\mu=(-M,0,0,0),
\end{equation*}
one has
\begin{equation*}
	W^\mu=(0,-M\mathbf J),
\end{equation*}
so the two Casimir eigenvalues remain unchanged.  The loop
\(C_{[\hat n]}^\eta\) exchanges the two energy branches,
\begin{equation*}
	p_+
	\longleftrightarrow
	p_-,
\end{equation*}
without modifying their orbit type, little group or infinitesimal Casimir
equations.  Its unitary representative acts on the extended massive Hilbert
space through the sector-exchange operator described in
Subsection~\ref{subsubsec:uir-massive-sector}.

\subsubsection{Tachyonic Wavefunctions}
\label{subsubsec:gpd-tachyonic-wavefunctions}

The tachyonic Wigner classes arise by transporting massive spectral data from
the Minkowski object to a directional object and then applying
\(\operatorname{Class}_\eta\).

The two oriented arrows give the opposite representatives
\eqref{eq:gpd-positive-arrow-massive-transport} and
\eqref{eq:gpd-negative-arrow-massive-transport}.

They belong to the same connected spacelike Minkowski orbit.  It is therefore
sufficient to choose the representative
\begin{equation*}
	q^\mu=(0,M\hat n),
	\qquad
	q^T\eta q=M^2.
\end{equation*}
Its little group is the \(SO(2,1)\) group of
\eqref{eq:uir-tachyonic-orbit-type}.

Complete \(\hat n\) to a right-handed orthonormal spatial frame
\begin{equation*}
	(\hat e_1,\hat e_2,\hat n),
	\qquad
	\hat e_1\times\hat e_2=\hat n.
\end{equation*}
For the spherical parametrization \eqref{eq:srf-unit-vector},

one may choose
\begin{align*}
	\hat e_1
	&=
	(\cos\theta\cos\phi,
	\cos\theta\sin\phi,
	-\sin\theta),
	\\
	\hat e_2
	&=
	(-\sin\phi,\cos\phi,0).
\end{align*}
Define
\begin{equation}
	R_{\hat n}
	:=
	\hat n\cdot\mathbf J,
	\qquad
	B_1
	:=
	\hat e_1\cdot\mathbf K,
	\qquad
	B_2
	:=
	\hat e_2\cdot\mathbf K.
	\label{eq:gpd-tachyonic-little-generators}
\end{equation}
They satisfy
\begin{equation}
	[R_{\hat n},B_1]=iB_2,
	\qquad
	[R_{\hat n},B_2]=-iB_1,
	\qquad
	[B_1,B_2]=-iR_{\hat n},
	\label{eq:gpd-so21-commutation-relations}
\end{equation}
and generate \(\mathfrak{so}(2,1)\).

For \(P^\mu=q^\mu\), the Minkowski Pauli--Lubanski operator has components
\begin{equation*}
	W^0
	=
	MR_{\hat n},
	\qquad
	\mathbf W
	=
	-M\hat n\times\mathbf K
	=
	M
	\left(
	B_2\hat e_1-B_1\hat e_2
	\right).
\end{equation*}
In the adapted basis
\((e_0,e_{\hat n},e_1,e_2)\),
\begin{equation*}
	W^\mu
	=
	M
	(R_{\hat n},0,B_2,-B_1).
\end{equation*}
The two Casimir operators of the resulting Minkowski Poincare representation
therefore have the values
\begin{align*}
	C_1^\eta
	&=
	M^2,
	\\
	C_2^\eta
	&=
	M^2
	\left(
	B_1^2+B_2^2-R_{\hat n}^2
	\right).
\end{align*}
Defining
\begin{equation}
	C_{\mathfrak{so}(2,1)}
	:=
	B_1^2+B_2^2-R_{\hat n}^2,
	\label{eq:gpd-so21-casimir-definition}
\end{equation}
one obtains
\begin{equation}
	C_2^\eta
	=
	M^2
	C_{\mathfrak{so}(2,1)}.
	\label{eq:gpd-poincare-so21-casimir-relation}
\end{equation}

Let \(\sigma\) be a UIR of \(SO(2,1)\), and let
\(\kappa_\sigma\) be the eigenvalue of its quadratic Casimir:
\begin{equation*}
	C_{\mathfrak{so}(2,1)}
	\psi_\sigma
	=
	\kappa_\sigma\psi_\sigma.
\end{equation*}
Denote by \(\rho_{M,\sigma}^{\mathrm{tach}}\) the induced representation
\eqref{eq:uir-tachyonic-wigner-rep}, using the present choice of representative
\(q\).

On its representation space
\(\mathcal H_{M,\sigma}^{\mathrm{tach}}\), the Casimir equations are
\begin{equation}
	\left(
	\widehat P_\eta^2-M^2
	\right)
	\psi_{M,\sigma}^{\mathrm{tach}}
	=
	0,
	\label{eq:gpd-tachyonic-first-wave-equation}
\end{equation}
and
\begin{equation}
	\left(
	\widehat W_\eta^2-M^2\kappa_\sigma
	\right)
	\psi_{M,\sigma}^{\mathrm{tach}}
	=
	0.
	\label{eq:gpd-tachyonic-second-wave-equation}
\end{equation}
The notation \(\kappa_\sigma\) includes the discrete, principal,
complementary and any other admissible unitary series without imposing a
single spin parametrization on all \(SO(2,1)\) representations.

The opposite representative \(-q\) gives
\begin{equation*}
	W^\mu\longmapsto-W^\mu,
\end{equation*}
and therefore leaves both Casimir eigenvalues unchanged.  The loop
\(C_{[\hat n]}^\eta\), or its conjugate directional loop
\(C_{[\hat n]}^{\mathrm{dir}}\), exchanges the two oriented representatives.
Its action on the \(SO(2,1)\) representation space is governed by the
Case~A/Case~B distinction of the Clifford--Mackey analysis.

No intrinsically tachyonic state is postulated at the directional object.
The tachyonic representation is the class of Poincarè group representation assigned using the Minkowski metric to classify spectral data transported by arrows from the Minkowski object to a directional metric object.

\subsubsection{Massless Wavefunctions}
\label{subsubsec:gpd-massless-wavefunctions}

Let
\begin{equation*}
	p_+^\mu
	=
	(\omega,\omega\hat n),
	\qquad
	\omega>0,
\end{equation*}
be a future-directed null representative, and let
\begin{equation*}
	p_-^\mu
	=
	(-\omega,-\omega\hat n)
\end{equation*}
be the corresponding past-directed representative.  Both satisfy
\begin{equation*}
	p_\pm^T\eta p_\pm=0.
\end{equation*}
For the infinite-speed arrows aligned with \(\hat n\),
\begin{equation}
	\Lambda_+(\hat n)p_+=p_-,
	\qquad
	\Lambda_-(\hat n)p_+=p_+.
	\label{eq:gpd-two-arrows-massless-action}
\end{equation}
The two arrows therefore transport the same intrinsic massless class while
selecting the two time orientations of the Minkowski null cone under
\(\operatorname{Class}_\eta\).

The connected little group of a null representative is
\begin{equation*}
	H_{p_\pm}^\eta
	\simeq
	ISO(2).
\end{equation*}
Using the right-handed frame
\((\hat e_1,\hat e_2,\hat n)\), define
\begin{equation}
	J_{\hat n}
	:=
	\hat n\cdot\mathbf J,
	\qquad
	\Pi_a
	:=
	\hat e_a\cdot
	\left(
	\mathbf K+\hat n\times\mathbf J
	\right),
	\qquad
	a=1,2.
	\label{eq:gpd-massless-little-generators}
\end{equation}
These generators satisfy
\begin{equation}
	[J_{\hat n},\Pi_1]=i\Pi_2,
	\qquad
	[J_{\hat n},\Pi_2]=-i\Pi_1,
	\qquad
	[\Pi_1,\Pi_2]=0.
	\label{eq:gpd-iso2-commutation-relations}
\end{equation}

For \(P^\mu=p_+^\mu\), the Pauli--Lubanski vector is
\begin{equation*}
	W^0
	=
	\omega J_{\hat n},
	\qquad
	\mathbf W
	=
	\omega
	\left(
	J_{\hat n}\hat n
	+
	\Pi_2\hat e_1
	-
	\Pi_1\hat e_2
	\right).
\end{equation*}
Consequently,
\begin{equation*}
	C_1^\eta=0,
	\qquad
	C_2^\eta
	=
	\omega^2
	\left(
	\Pi_1^2+\Pi_2^2
	\right).
\end{equation*}

For helicity representations, the translation subgroup of \(ISO(2)\) acts
trivially:
\begin{equation*}
	\Pi_1=\Pi_2=0.
\end{equation*}
It follows that
\begin{equation*}
	W^\mu=hP^\mu,
	\qquad
	C_1^\eta=C_2^\eta=0,
\end{equation*}
where \(h\) is the helicity.  The corresponding equations are
\begin{equation}
	\widehat P_\eta^2\psi_h=0,
	\qquad
	\widehat W_\eta^2\psi_h=0,
	\label{eq:gpd-helicity-wave-equations}
\end{equation}
supplemented by
\begin{equation}
	\widehat W_\eta^\mu\psi_h
	=
	h\widehat P_\eta^\mu\psi_h.
	\label{eq:gpd-helicity-condition}
\end{equation}

For continuous-spin representations, the translation generators act
nontrivially:
\begin{equation*}
	\left(
	\Pi_1^2+\Pi_2^2
	\right)
	\psi_\rho
	=
	\frac{\rho^2}{\omega^2}
	\psi_\rho,
	\qquad
	\rho>0.
\end{equation*}
Equivalently,
\begin{equation}
	\widehat P_\eta^2\psi_\rho=0,
	\qquad
	\left(
	\widehat W_\eta^2-\rho^2
	\right)
	\psi_\rho=0.
	\label{eq:gpd-continuous-spin-wave-equations}
\end{equation}

The discrete loop fixes the compact rotation generator and reverses the
translation generators:
\begin{equation*}
	c_{[\hat n]}^\alpha(J_{\hat n})
	=
	J_{\hat n},
	\qquad
	c_{[\hat n]}^\alpha(\Pi_a)
	=
	-\Pi_a,
	\qquad
	a=1,2.
\end{equation*}
Its unitary representative satisfies the covariance relation
\begin{equation*}
	S_{C,\alpha}
	D_\alpha(s)
	S_{C,\alpha}^{-1}
	=
	D_\alpha
	\left(
	C_{[\hat n]}^\alpha
	s
	(C_{[\hat n]}^\alpha)^{-1}
	\right),
	\qquad
	s\in ISO(2).
\end{equation*}
For helicity representations this condition is trivial on the translation
subgroup.  For continuous-spin representations it requires a nontrivial
unitary reflection of the translation parameters.

The operator \(S_{C,\alpha}\) represents the typed isotropy loop
\(C_{[\hat n]}^\alpha\), rather than an individual inter-object arrow.  The
operators
\begin{equation*}
	\mathcal U(J_{\pm\hat n}):
	\mathcal H_\eta
	\longrightarrow
	\mathcal H_{[\hat n]}
\end{equation*}
transport states and operator data between Hilbert fibers, whereas
\begin{equation*}
	S_{C,\alpha}:
	\mathcal H_\alpha
	\longrightarrow
	\mathcal H_\alpha
\end{equation*}
acts internally and supplies the discrete extension of the isotropy
representation.  The exchange information carried by this involution is not
detected by the infinitesimal Casimirs and forms an additional part of the
full groupoid classification.

The construction may be summarized as follows.  Every object of the Poincare
groupoid carries an isomorphic realization of the abstract Poincare algebra.
The inter-object arrows transport its generators, polynomial operators,
representations and spectral data through the corresponding unital
\( * \)-isomorphisms of universal enveloping algebras.  The distinguished
Minkowski metric supplies a single contraction rule for the physical
classification of all such transported data.

The operators \eqref{eq:gpd-first-pointed-classification-operator} and
\eqref{eq:gpd-second-pointed-classification-operator} determine the Minkowski
orbit invariants.  The classification map
\eqref{eq:gpd-minkowski-classification-map}, through
\eqref{eq:gpd-pointed-induced-representation}, assigns the corresponding
Wigner representation.

In particular, massive spectral data transported to a directional object
are intrinsically massive with respect to \(\eta_{[\hat n]}\), but are
classified as spacelike if we use Minkowski metric.  They
therefore admit the pointed $SO(2,1)$ little-group families of tachyonic
wavefunctions, with their internal UIR labels specified separately.

\section{An Alternative Explanation of Madame Wu Experiment}
\label{sec:wu-experiment}

\subsection{Experimental Description}
\label{subsec:wu-experimental-description}

The experiment performed by Wu and collaborators tested parity conservation
in the beta decay of polarized \(^{60}\mathrm{Co}\) nuclei
\cite{LeeYang1956,Wu1957}.  The relevant decay is predominantly an allowed
Gamow--Teller transition,
\begin{equation}
	{}^{60}\mathrm{Co}(5^+)
	\longrightarrow
	{}^{60}\mathrm{Ni}^{*}(4^+)
	+
	e^-
	+
	\overline{\nu}_e,
	\label{eq:wu-co60-decay}
\end{equation}
followed by the electromagnetic de-excitation of the daughter nucleus.

The cobalt sample was cooled to a very low temperature and subjected to a
magnetic field in order to produce a non-vanishing nuclear polarization.  The
direction relevant to the beta-decay correlation is the direction of the
expectation value of the nuclear angular momentum,
\begin{equation*}
	\hat n
	:=
	\frac{\langle\mathbf J_N\rangle}
	{\left|\langle\mathbf J_N\rangle\right|}.
\end{equation*}
The magnetic field prepares this polarization but should not, in general, be
identified directly with it without taking into account the sign of the
nuclear magnetic moment.  We therefore use \(\hat n\) for the experimentally
determined nuclear-polarization direction.

Let
\begin{equation*}
	\mathbf P_N
	=
	P\hat n,
	\qquad
	0\leq P\leq1,
\end{equation*}
where \(P\) is the degree of nuclear polarization.  If \(\hat k\) denotes the
direction of the detected electron momentum, the physical angle
\(\vartheta\) is defined by
\begin{equation*}
	\cos\vartheta
	=
	\hat n\cdot\hat k.
\end{equation*}

The parity test relies on the different transformation properties of nuclear
angular momentum and linear momentum.  The nuclear angular momentum is an
axial vector and is therefore unchanged by ordinary spatial parity:
\begin{equation}
	\mathsf P:
	\mathbf J_N
	\longmapsto
	\mathbf J_N.
	\label{eq:wu-parity-spin}
\end{equation}
The electron momentum is a polar vector and changes sign:
\begin{equation}
	\mathsf P:
	\mathbf p_e
	\longmapsto
	-\mathbf p_e.
	\label{eq:wu-parity-momentum}
\end{equation}
Consequently,
\begin{equation}
	\mathsf P:
	\mathbf J_N\cdot\mathbf p_e
	\longmapsto
	-\mathbf J_N\cdot\mathbf p_e.
	\label{eq:wu-parity-pseudoscalar}
\end{equation}
The spin--momentum correlation is therefore a pseudoscalar.

If parity were a symmetry of the decay probability, the two configurations
related by
\(\mathbf p_e\mapsto-\mathbf p_e\), at fixed nuclear polarization, would
have equal rates:
\begin{equation*}
	W(E_e,\vartheta)
	=
	W(E_e,\pi-\vartheta).
\end{equation*}
In particular, the rates parallel and antiparallel to the polarization axis
would be equal:
\begin{equation*}
	W(E_e,0)
	=
	W(E_e,\pi).
\end{equation*}

Wu and collaborators instead observed a pronounced preference for beta
emission in the direction opposite to the nuclear polarization:
\begin{equation*}
	W(E_e,\pi)
	>
	W(E_e,0).
\end{equation*}
When the magnetic field, and hence the prepared nuclear polarization, was
reversed, the counting asymmetry was also reversed.

\begin{figure}[htbp]
	\centering
	\includegraphics[width=0.98\textwidth]{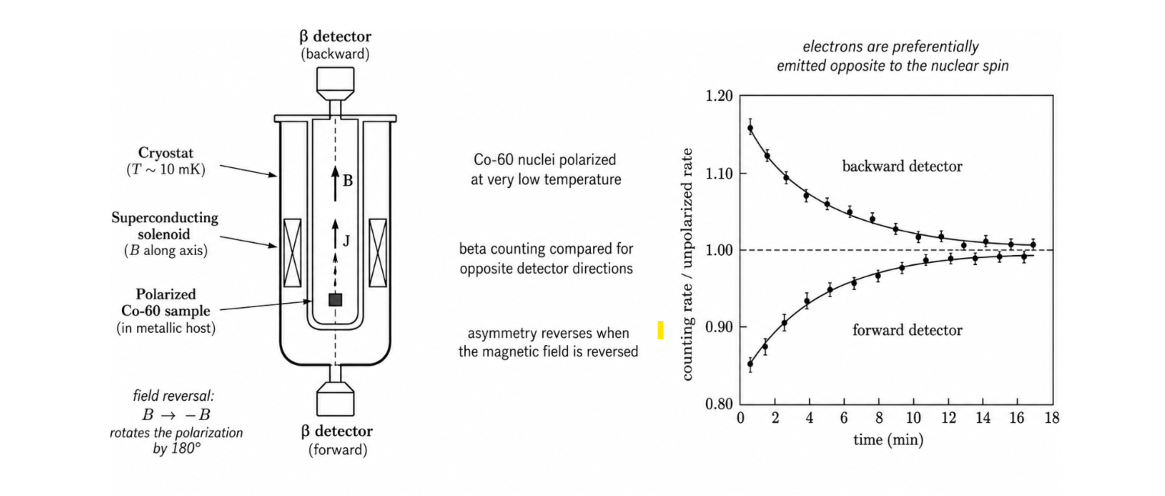}
	\caption{Schematic summary of the Wu experimental arrangement and of the
	qualitative detector-counting asymmetry.  The polarized
	\(^{60}\mathrm{Co}\) source is placed between two opposite beta detectors;
	the larger rate occurs on the side opposite to the nuclear polarization,
	and the asymmetry disappears as the sample depolarizes.  }
	\label{fig:wu-apparatus-counting-asymmetry}
\end{figure}

Figure~\ref{fig:wu-apparatus-counting-asymmetry} summarizes the experimental
geometry and the qualitative behavior of the two counting rates.

The experimental reversal of the magnetic field is not itself the parity
transformation.  It is an operational procedure that exchanges the two
polarized preparations while leaving the detector geometry fixed.  It
therefore permits a comparison of the two opposite spin--momentum
correlations with reduced systematic uncertainty.

\subsection{Standard Electroweak Interpretation}
\label{subsec:wu-standard-electroweak-interpretation}

The standard electroweak interpretation of the Wu experiment is based on two
logically distinct ingredients.  The first is the decomposition of relativistic
spinor fields into the two irreducible Weyl representations of the proper
orthochronous Lorentz group.  The second is the dynamical assumption that the
charged weak interaction couples only to the left-chiral components of those
fields.  The latter assumption is not implied by Lorentz covariance alone; it
is the chiral structure selected by weak-interaction phenomenology and
incorporated into the Standard Model.

\subsubsection{Complexified Lorentz Algebra and Weyl Representations}
\label{subsubsec:wu-complexified-lorentz-weyl}

Let \(J_i\) and \(K_i\) denote the generators of spatial rotations and boosts.
The Lorentz algebra satisfies
\begin{equation*}
	[J_i,J_j]
	=
	i\epsilon_{ijk}J_k,
	\qquad
	[J_i,K_j]
	=
	i\epsilon_{ijk}K_k,
	\qquad
	[K_i,K_j]
	=
	-i\epsilon_{ijk}J_k.
\end{equation*}
After complexification, introduce
\begin{equation}
	A_i
	=
	\frac{1}{2}
	\left(
	J_i+iK_i
	\right),
	\qquad
	B_i
	=
	\frac{1}{2}
	\left(
	J_i-iK_i
	\right).
	\label{eq:wu-sm-two-lorentz-factors}
\end{equation}
These generators obey
\begin{equation*}
	[A_i,A_j]
	=
	i\epsilon_{ijk}A_k,
	\qquad
	[B_i,B_j]
	=
	i\epsilon_{ijk}B_k,
	\qquad
	[A_i,B_j]
	=
	0.
\end{equation*}
Consequently,
\begin{equation}
	\mathfrak{so}(1,3)_{\mathbb C}
	\simeq
	\mathfrak{sl}(2,\mathbb C)_L
	\oplus
	\mathfrak{sl}(2,\mathbb C)_R.
	\label{eq:wu-sm-complex-lorentz-decomposition}
\end{equation}
The two commuting factors are frequently described in the physics literature
as two \(\mathfrak{su}(2)\)-type algebras. This notation refers to their
commutation relations after complexification.

Finite-dimensional complex representations of the proper orthochronous
Lorentz group, or equivalently of its double cover
\begin{equation*}
	\operatorname{Spin}^{+}(1,3)
	\simeq
	SL(2,\mathbb C),
\end{equation*}
are thus labelled by two half-integers
\begin{equation*}
	(j_L,j_R).
\end{equation*}
The two fundamental spinorial representations are
\begin{equation*}
	V_L
	=
	V_{\left(\frac12,0\right)},
	\qquad
	V_R
	=
	V_{\left(0,\frac12\right)}.
\end{equation*}
Where a left-handed Weyl spinor belongs to
\(V_L\) and a right-handed Weyl spinor belongs to \(V_R\):
\begin{equation*}
	\psi_L\in V_L,
	\qquad
	\psi_R\in V_R.
\end{equation*}
As complex representations of
\(\operatorname{Spin}^{+}(1,3)\), they are irreducible and inequivalent:
\begin{equation*}
	\left(\frac12,0\right)
	\not\simeq
	\left(0,\frac12\right).
\end{equation*}

These Weyl representations should not be confused with Wigner UIRs of the
Poincare group.  Since the Lorentz group is noncompact, its nontrivial
finite-dimensional spinorial representations are not unitary.  They describe
the transformation law of local fields.  The one-particle Hilbert spaces are
instead classified by unitary irreducible representations of the Poincare
group \cite{Weinberg}.

\subsubsection{Dirac Spinors and Chiral Projectors}
\label{subsubsec:wu-dirac-spinors-projectors}

A Dirac spinor belongs to the direct-sum space
\begin{equation}
	V_D
	=
	V_L\oplus V_R
	=
	V_{\left(\frac12,0\right)}
	\oplus
	V_{\left(0,\frac12\right)}
	\label{eq:wu-sm-dirac-direct-sum-space}
\end{equation}
and may be written, in the chiral basis, as
\begin{equation*}
	\Psi
	=
	\begin{pmatrix}
		\psi_L\\
		\psi_R
	\end{pmatrix}.
\end{equation*}
The Dirac representation is therefore reducible, over the complex numbers,
when restricted to the proper orthochronous Lorentz group:
\begin{equation}
	D_D(\Lambda)
	=
	\begin{pmatrix}
		D_L(\Lambda)&0\\
		0&D_R(\Lambda)
	\end{pmatrix},
	\qquad
	\Lambda\in SO^{+}(1,3).
	\label{eq:wu-sm-dirac-connected-action}
\end{equation}

The operator that distinguishes the two irreducible subspaces is
\begin{equation*}
	\gamma^5
	=
	i\gamma^0\gamma^1\gamma^2\gamma^3,
\end{equation*}
which satisfies
\begin{equation*}
	(\gamma^5)^2=I,
	\qquad
	\{\gamma^5,\gamma^\mu\}=0.
\end{equation*}
The corresponding chiral projectors are
\begin{equation}
	P_L
	=
	\frac{I-\gamma^5}{2},
	\qquad
	P_R
	=
	\frac{I+\gamma^5}{2}.
	\label{eq:wu-sm-chiral-projectors}
\end{equation}
They satisfy
\begin{equation*}
	P_L^2=P_L,
	\qquad
	P_R^2=P_R,
	\qquad
	P_LP_R=0,
	\qquad
	P_L+P_R=I,
\end{equation*}
and define
\begin{equation*}
	\psi_L=P_L\Psi,
	\qquad
	\psi_R=P_R\Psi.
\end{equation*}
In the chiral basis,
\begin{equation*}
	\gamma^5
	=
	\begin{pmatrix}
		-I_2&0\\
		0&I_2
	\end{pmatrix},
	\qquad
	P_L
	=
	\begin{pmatrix}
		I_2&0\\
		0&0
	\end{pmatrix},
	\qquad
	P_R
	=
	\begin{pmatrix}
		0&0\\
		0&I_2
	\end{pmatrix}.
\end{equation*}

\subsubsection{Parity as an Exchange of Weyl Sectors}
\label{subsubsec:wu-parity-exchange-weyl}

Under spatial parity, rotations remain unchanged whereas boosts change sign:
\begin{equation*}
	\mathsf P J_i\mathsf P^{-1}
	=
	J_i,
	\qquad
	\mathsf P K_i\mathsf P^{-1}
	=
	-K_i.
\end{equation*}
It follows from equation
\eqref{eq:wu-sm-two-lorentz-factors} that parity exchanges the two commuting
factors:
\begin{equation*}
	\mathsf P A_i\mathsf P^{-1}
	=
	B_i,
	\qquad
	\mathsf P B_i\mathsf P^{-1}
	=
	A_i.
\end{equation*}
Consequently,
\begin{equation*}
	\mathsf P:
	\left(\frac12,0\right)
	\longleftrightarrow
	\left(0,\frac12\right),
\end{equation*}
or, equivalently,
\begin{equation*}
	\mathsf P:
	\psi_L
	\longleftrightarrow
	\psi_R.
\end{equation*}

A single Weyl space is therefore not closed under parity.  The minimal
spinorial carrier space on which both the connected Lorentz group and parity
can act is the direct sum
\begin{equation*}
	V_L\oplus V_R.
\end{equation*}
Up to an intrinsic-parity phase, the parity operator is represented in the
chiral basis by
\begin{equation}
	D_D(\mathsf P)
	=
	\gamma^0
	=
	\begin{pmatrix}
		0&I_2\\
		I_2&0
	\end{pmatrix}.
	\label{eq:wu-sm-parity-chiral-matrix}
\end{equation}
Accordingly,
\begin{equation*}
	D_D(\mathsf P)
	\begin{pmatrix}
		\psi_L\\
		\psi_R
	\end{pmatrix}
	=
	\begin{pmatrix}
		\psi_R\\
		\psi_L
	\end{pmatrix}.
\end{equation*}
The chiral projectors are exchanged as well:
\begin{equation*}
	D_D(\mathsf P)P_LD_D(\mathsf P)^{-1}
	=
	P_R,
	\qquad
	D_D(\mathsf P)P_RD_D(\mathsf P)^{-1}
	=
	P_L.
\end{equation*}

This construction has the same index-two exchange structure as Case~A of the
Clifford--Mackey alternative discussed in
Section~\ref{subsec:uirs-extended-poincare-group} and
Appendix~\ref{app:clifford-mackey-index-two}.  Let
\begin{equation*}
	G_0
	=
	\operatorname{Spin}^{+}(1,3)
\end{equation*}
and let \(G_{\mathsf P}\) denote the extension obtained by adjoining spatial
parity.  The parity-conjugate representation of \(D_L\) is
\begin{equation*}
	D_L^{\mathsf P}(g)
	=
	D_L\left(
	\mathsf P^{-1}g\mathsf P
	\right),
	\qquad
	g\in G_0.
\end{equation*}
One has
\begin{equation*}
	D_L^{\mathsf P}
	\simeq
	D_R,
	\qquad
	D_L^{\mathsf P}
	\not\simeq
	D_L.
\end{equation*}
Thus the connected subgroup acts diagonally as in
\eqref{eq:wu-sm-dirac-connected-action}, while the parity matrix
\eqref{eq:wu-sm-parity-chiral-matrix} exchanges the two inequivalent sectors.
These are the two structural ingredients of Case~A.

The analogy concerns the extension and exchange structure.  The Weyl
representations appearing here are finite-dimensional Lorentz-field
representations, whereas the representations classified in
Section~\ref{subsec:uirs-extended-poincare-group} are unitary representations
on one-particle Hilbert spaces.

\subsubsection{What the Wu Experiment Established}
\label{subsubsec:wu-what-experiment-established}

If parity were a symmetry of the weak interaction Hamiltonian, one would have
\begin{equation*}
	\mathsf P H_{\mathrm W}\mathsf P^{-1}
	=
	H_{\mathrm W}.
\end{equation*}
The transition amplitude of a process would then equal, up to conventional
phases, the amplitude of its parity-transformed process.  In particular, for
a nucleus polarized along the unit vector \(\hat n\), parity conservation
would imply
\begin{equation*}
	W(\hat n,\mathbf p_e)
	=
	W(\hat n,-\mathbf p_e).
\end{equation*}
The axial/polar transformation laws are
\eqref{eq:wu-parity-spin} and \eqref{eq:wu-parity-momentum}; the resulting
parity-odd correlation is given in \eqref{eq:wu-parity-pseudoscalar}.

Lee and Yang emphasized that parity conservation had not been experimentally
established for weak interactions and proposed direct tests
\cite{LeeYang1956}.  Wu and collaborators subsequently observed an asymmetric
electron distribution in the beta decay of polarized
\(^{60}\mathrm{Co}\) nuclei \cite{Wu1957}:
\begin{equation*}
	W(\hat n,\mathbf p_e)
	\neq
	W(\hat n,-\mathbf p_e).
\end{equation*}
The direct conclusion of the experiment is therefore
\begin{equation*}
	\mathsf P H_{\mathrm W}\mathsf P^{-1}
	\neq
	H_{\mathrm W}.
\end{equation*}

This result does not mean that the right-handed Weyl representation ceases to
exist, nor that the parity-conjugate spinor space is mathematically or
physically absent.  It means that the weak interaction does not act
symmetrically on the two sectors exchanged by parity.  The more specific
\(V-A\) form was inferred from the combined phenomenology of several weak
processes, rather than from the Wu experiment alone
\cite{MarshakSudarshan1958,FeynmanGellMann1958}.

\subsubsection{Chiral Selection and the \(V-A\) Current}
\label{subsubsec:wu-chiral-selection-va}

Lorentz covariance by itself does not select a unique weak interaction.
Relativistic fermion fields permit scalar, pseudoscalar, vector, axial-vector
and tensor bilinears.  The chiral assumption consists in postulating that the
charged weak current couples to the left-chiral components.

For two Dirac fields \(\psi\) and \(\chi\), one has
\begin{equation*}
	\psi_L=P_L\psi,
	\qquad
	\overline{\psi_L}
	=
	\bar\psi P_R.
\end{equation*}
Using
\begin{equation*}
	P_R\gamma^\mu
	=
	\gamma^\mu P_L,
\end{equation*}
the current formed from the left-chiral components becomes
\begin{equation}
	\begin{aligned}
		2\overline{\psi_L}\gamma^\mu\chi_L
		&=
		2\bar\psi P_R\gamma^\mu P_L\chi
		\\
		&=
		2\bar\psi\gamma^\mu P_L\chi
		\\
		&=
		\bar\psi\gamma^\mu
		\left(
		I-\gamma^5
		\right)
		\chi
		\\
		&=
		\bar\psi\gamma^\mu\chi
		-
		\bar\psi\gamma^\mu\gamma^5\chi.
	\end{aligned}
	\label{eq:wu-sm-left-current-va-derivation}
\end{equation}
The first term is a vector current,
\begin{equation*}
	V^\mu
	=
	\bar\psi\gamma^\mu\chi,
\end{equation*}
whereas the second is an axial-vector current,
\begin{equation*}
	A^\mu
	=
	\bar\psi\gamma^\mu\gamma^5\chi.
\end{equation*}
Therefore the left-chiral current is precisely
\begin{equation*}
	J_L^\mu
	=
	V^\mu-A^\mu.
\end{equation*}
Similarly, projection onto the right-chiral sector would give
\begin{equation*}
	J_R^\mu
	=
	\bar\psi\gamma^\mu
	\left(
	I+\gamma^5
	\right)
	\chi
	=
	V^\mu+A^\mu.
\end{equation*}
Because parity exchanges \(P_L\) and \(P_R\), it also exchanges the two
currents:
\begin{equation*}
	\mathsf P:
	V-A
	\longleftrightarrow
	V+A.
\end{equation*}
An interaction containing \(V-A\) without the corresponding parity-related
\(V+A\) term is consequently not invariant under parity.

In the Standard Model, this chiral selection is implemented by assigning the
left-chiral fermion fields to weak \(SU(2)\) doublets,
\begin{equation*}
	L_e
	=
	\begin{pmatrix}
		\nu_e\\
		e
	\end{pmatrix}_L,
	\qquad
	Q
	=
	\begin{pmatrix}
		u\\
		d
	\end{pmatrix}_L,
\end{equation*}
while the corresponding right-chiral charged fields are weak singlets.  The
charged-current interaction is
\begin{equation}
	\begin{aligned}
		\mathcal L_{\mathrm{CC}}
		=
		-\frac{g}{2\sqrt{2}}\,
		W_\mu^{+}
		\Big[
		&
		\bar\nu_e\gamma^\mu
		\left(
		I-\gamma^5
		\right)e
		\\
		&+
		\bar u\gamma^\mu
		\left(
		I-\gamma^5
		\right)
		V_{ud}d
		\Big]
		+
		\mathrm{h.c.}
	\end{aligned}
	\label{eq:wu-sm-charged-current-lagrangian}
\end{equation}
Thus the charged \(W\) bosons couple to left-chiral fermion fields but not to
a parity-matched right-chiral charged current.

The internal weak gauge group \(SU(2)_{\mathrm W}\) must not be identified
with either factor in the complexified Lorentz algebra
\eqref{eq:wu-sm-complex-lorentz-decomposition}.  The former acts on internal
weak-isospin indices, whereas the latter classifies the Lorentz
transformation properties of spinor fields.

At energies much smaller than the \(W\)-boson mass, the \(W\) field can be
eliminated to obtain an effective four-fermion interaction.  At the nucleon
level it may be written as
\begin{equation}
	\mathcal H_{\mathrm{eff}}
	=
	\frac{G_FV_{ud}}{\sqrt{2}}
	\left[
	\bar p\gamma^\mu
	\left(
	g_V-g_A\gamma^5
	\right)n
	\right]
	\left[
	\bar e\gamma_\mu
	\left(
	I-\gamma^5
	\right)\nu_e
	\right]
	+
	\mathrm{h.c.}
	\label{eq:wu-sm-effective-beta-hamiltonian}
\end{equation}
Here \(G_F\) is the Fermi constant, \(V_{ud}\) is the relevant CKM matrix
element, and \(g_V\) and \(g_A\) are the effective vector and axial nucleon
couplings.  By \eqref{eq:wu-sm-chiral-projectors}, the factor \(I-\gamma^5\) is twice
the left-chiral projector, rather than a separate term inserted independently
into the interaction.

The nuclear transition studied by Wu is \eqref{eq:wu-co60-decay}.

At leading order this is an allowed Gamow--Teller transition.  ``Allowed''
means that the leading lepton partial wave has zero orbital angular momentum
relative to the nucleus.  Since
\begin{equation*}
	\Delta J=1,
	\qquad
	\Delta\pi=0,
\end{equation*}
the transition is dominated by the spatial axial component of the nuclear
current.

The corresponding leading amplitude has the schematic form
\begin{equation*}
	\mathcal M_{\mathrm{GT}}
	\propto
	\left\langle
	{}^{60}\mathrm{Ni}^{*}
	\left|
	\mathbf J_A
	\right|
	{}^{60}\mathrm{Co}
	\right\rangle
	\cdot
	\left[
	\bar u_e\boldsymbol{\gamma}
	\left(
	I-\gamma^5
	\right)v_{\bar\nu_e}
	\right].
\end{equation*}
The chiral projector selects predominantly negative-helicity electrons and
positive-helicity antineutrinos in the relativistic limit.  For a nucleus
polarized along \(+\hat n\), an electron emitted along \(-\hat n\) with
negative helicity has its spin preferentially aligned along \(+\hat n\).
This is compatible with the angular momentum carried away in the
\(5\rightarrow4\) Gamow--Teller transition.  Emission in the opposite
direction has the unfavoured spin orientation.

After summing over the unobserved antineutrino and daughter-nucleus variables,
the electron distribution takes the standard beta-decay correlation form
\cite{JacksonTreimanWyld1957}
\begin{equation}
	\frac{dN}{d\Omega_e}
	=
	N_0(E_e)
	\left[
	1
	+
	A_\beta(E_e)
	P_N
	\frac{|\mathbf p_e|}{E_e}
	\cos\theta
	\right],
	\label{eq:wu-sm-standard-angular-distribution}
\end{equation}
where
\begin{equation*}
	\cos\theta
	=
	\hat n\cdot\hat p_e,
\end{equation*}
\(P_N\) is the nuclear polarization and \(A_\beta\) is the beta-asymmetry
coefficient.  In the ideal pure Gamow--Teller approximation for a
beta-minus transition with
\begin{equation*}
	J_f=J_i-1,
\end{equation*}
one obtains
\begin{equation*}
	A_\beta=-1
\end{equation*}
at leading order \cite{WautersEtAl2010}.  Nuclear, recoil, electromagnetic and radiative effects
produce corrections to this ideal value.

The negative sign implies that the counting rate is maximal near
\begin{equation*}
	\cos\theta=-1,
	\qquad
	\mathbf p_e\parallel-\hat n.
\end{equation*}
Under reversal of the nuclear polarization,
\begin{equation*}
	\hat n
	\longmapsto
	-\hat n,
\end{equation*}
the angular term changes sign and the excess and deficit measured by the two
opposite detectors are exchanged.

The Lorentz representation theory supplies the two Weyl sectors and shows
that parity exchanges them.  The chiral projector \(P_L\), and hence the \(V-A\)
interaction, is an additional input selected by weak-interaction phenomenology with no geometric counterpart. It must also be pointed out that this model implies that weak force is special among the others because the right handed spinor is not removed entirely from the theory, the electron remains a Dirac Field and its mass term couples the two chiral sectors. The Wu asymmetry is therefore interpreted as evidence that only weak dynamics does not realize the exchange symmetry between the two
parity-conjugate Weyl sectors.

\subsection{Groupoid Interpretation}
\label{subsec:wu-groupoid-kinematic-correspondence}

The starting point of the present interpretation is the exceptional
representation-theoretic character of electroweak radiation.  
In the Standard Model, the Dirac direct sum
\eqref{eq:wu-sm-dirac-direct-sum-space} is coupled by the charged weak
interaction through the left-chiral projector
\eqref{eq:wu-sm-chiral-projectors}.

The special character of the interaction is therefore represented by
selecting one of the two inequivalent Weyl sectors contained in the Dirac
representation. The groupoid construction suggests a different possibility.
Rather than describing electroweak radiation by restricting the
chirality content of an ordinary massive representation, it enlarges the
available kinematical classes of wavefunctions.  Massive spectral data
transported to a directional metric object are classified, with respect to
the Minkowski metric, by a spacelike Poincare orbit and
hence by a unitary representation with little group \(SO(2,1)\) (see section \ref{subsubsec:gpd-tachyonic-wavefunctions}). 
The physical ansatz explored below is that the exceptional directional
properties of electroweak radiation may be associated with this new
groupoid generated tachyonic class, rather than being encoded
through a fundamental restriction to the left-chiral sector of a
Dirac spinor.  In this section we will show that Madame Wu experiment does not falsify this ansatz.
The groupoid construction provides a kinematic correspondence between the
oriented nuclear preparation and a spacelike momentum representative.  The
purpose of this subsection is to show that this
correspondence could be responsible for the experimentally determined counting distribution in Madame Wu experiment or, at least, that Madame Wu experiment does not disprove our main ansatz.  In
particular, the groupoid fixes the preferred momentum direction and its
transformation under reversal of the magnetic preparation.
A complete alternative to the
standard electroweak description would additionally require a quantitative
derivation of the decay amplitudes and asymmetry coefficients and should also clarify the role of neutrinos. This will be investigated in a forthcoming publication.

\subsubsection{Oriented Transport and Preferred Momentum Direction}
\label{subsubsec:wu-oriented-transport-preferred-direction}

Use the distinguished Minkowski object \eqref{eq:gpd-object-eta}.
The magnetic preparation selects an unoriented nuclear-polarization axis
\([\hat n]\in\mathbb{RP}^2\), as defined by
\eqref{eq:gpd-projective-cover}, and hence the directional object
\eqref{eq:gpd-projective-directional-object}.  The orientation independence
of the metric in \eqref{eq:srf-directional-metric-two-signs} means that this
object does not distinguish the two possible orientations.

Introduce
\begin{equation*}
	\sigma\in\{+1,-1\},
	\qquad
	\hat b_\sigma
	:=
	\sigma\hat n,
\end{equation*}
where \(\hat b_\sigma\) denotes the oriented nuclear-polarization direction
selected by the magnetic preparation.  The corresponding nuclear angular
momentum is written as
\begin{equation*}
	\mathbf J_N^{(\sigma)}
	=
	|\mathbf J_N|\hat b_\sigma
	=
	\sigma|\mathbf J_N|\hat n.
\end{equation*}

The two oriented preparations select the parallel arrows
\eqref{eq:gpd-parallel-oriented-arrows}, with matrix representatives
\eqref{eq:gpd-two-infinite-speed-arrow-matrices}.

The source--arrow correspondence
\begin{equation}
	\mathbf J_N^{(\sigma)}
	\quad\leadsto\quad
	J_{\sigma\hat n}
	\label{eq:wu-source-arrow-correspondence}
\end{equation}
is the primitive physical ansatz of the present application.  Once this
correspondence is imposed, the orientation of the transported momentum is
fixed by the matrix representation of the selected arrow.

Let \(p_\eta\) denote the standard massive value
\eqref{eq:uir-massive-positive-representative}.  Write \(q_\sigma\) for its
transport along the selected arrow: it is the representative in
\eqref{eq:gpd-positive-arrow-massive-transport} for \(\sigma=+1\), and in
\eqref{eq:gpd-negative-arrow-massive-transport} for \(\sigma=-1\).

Following the groupoid ansatz, this transported spectral value is identified
with the standard momentum representative of the emitted beta radiation.
Its spatial orientation is
\begin{equation}
	\hat q_\sigma
	:=
	\frac{\mathbf q_\sigma}
	{|\mathbf q_\sigma|}
	=
	-\sigma\hat n
	=
	-\hat b_\sigma.
	\label{eq:wu-transported-unit-direction}
\end{equation}
Consequently,
\begin{equation}
	\mathbf J_N^{(\sigma)}
	\cdot
	\mathbf q_\sigma
	=
	-|\mathbf J_N|M
	<
	0.
	\label{eq:wu-groupoid-spin-momentum-anticorrelation}
\end{equation}
Equation~\eqref{eq:wu-transported-unit-direction} expresses the preferred
momentum direction: it is antiparallel to the oriented nuclear polarization.

This statement concerns the statistically preferred axis of the momentum distribution.

Transport preserves the intrinsic massive value
\eqref{eq:gpd-directional-intrinsic-massive-value}, while the pointed readout
gives the spacelike value \eqref{eq:gpd-directional-pointed-spacelike-value}.
The classification map \eqref{eq:gpd-minkowski-classification-map} therefore
assigns the tachyonic class \eqref{eq:gpd-directional-data-tachyonic-class},
with little group \(SO(2,1)\).

Within the proposed correspondence, the emitted state is therefore assigned
to the tachyonic Wigner class constructed from a UIR of \(SO(2,1)\).

\subsubsection{Compatibility with the Observed Angular Distribution}
\label{subsubsec:wu-compatibility-angular-distribution}

Let \(\hat k\) denote a detector direction.  We use the conventional
experimental angle measured from the oriented nuclear-polarization axis:
\begin{equation*}
	\cos\theta_\sigma
	:=
	\hat b_\sigma\cdot\hat k,
	\qquad
	0\leq\theta_\sigma\leq\pi.
\end{equation*}
Since \(\hat q_\sigma=-\hat b_\sigma\), one has
\begin{equation}
	\mu_\sigma
	:=
	\hat q_\sigma\cdot\hat k
	=
	-\cos\theta_\sigma.
	\label{eq:wu-directional-invariant-mu}
\end{equation}

If the prepared state and the detector are covariant under rotations about
the selected axis, the counting probability can depend on the detector
direction only through \(\mu_\sigma\).  Its most general axially symmetric
form is therefore
\begin{equation}
	W_{\sigma,P}(E,\hat k)
	=
	F_{E,P}(\mu_\sigma),
	\label{eq:wu-general-axisymmetric-counting-law}
\end{equation}
where \(P\in[0,1]\) denotes the degree of nuclear polarization.

Equation~\eqref{eq:wu-general-axisymmetric-counting-law} does not determine
the function \(F_{E,P}\).  In particular, the orientations at
\(\theta_\sigma=0\) and \(\theta_\sigma=\pi\) do not by themselves imply a
linear dependence on \(\cos\theta_\sigma\).

Use the experimental angular law
\eqref{eq:wu-sm-standard-angular-distribution}, with
\(E_e=E\), \(\theta=\theta_\sigma\), \(P_N=P\),
\(\beta_e=|\mathbf p_e|/E_e\), and normalization denoted by \(W_0(E)\).

Define
\begin{equation*}
	\kappa(E)
	:=
	-A_\beta(E)\beta_e.
\end{equation*}
Using
\(\mu_\sigma=-\cos\theta_\sigma\), the same experimentally determined law
takes the groupoid-adapted form
\begin{equation}
		W_{\sigma,P}^{\mathrm{exp}}
		(E,\hat k)
		=
		W_0(E)
		\left[
		1
		+
		\kappa(E)P\,
		\hat q_\sigma\cdot\hat k
		\right].
	\label{eq:wu-observed-angular-law-groupoid-form}
\end{equation}
In terms of the polarization angle, insert
\eqref{eq:wu-directional-invariant-mu} into
\eqref{eq:wu-observed-angular-law-groupoid-form}.

The limiting configurations are therefore
\begin{align*}
	\theta_\sigma=0:
	\qquad
	&
	\hat k=\hat b_\sigma,
	\qquad
	\hat q_\sigma\cdot\hat k=-1,
	\\[-2pt]
	&
	W_{\sigma,P}^{\mathrm{exp}}(E,0)
	=
	W_0(E)
	\left[
	1-\kappa(E)P
	\right],
	\\[6pt]
	\theta_\sigma=\pi:
	\qquad
	&
	\hat k=-\hat b_\sigma=\hat q_\sigma,
	\qquad
	\hat q_\sigma\cdot\hat k=+1,
	\\[-2pt]
	&
	W_{\sigma,P}^{\mathrm{exp}}(E,\pi)
	=
	W_0(E)
	\left[
	1+\kappa(E)P
	\right].
\end{align*}
For \(\kappa(E)>0\), the larger counting rate occurs at
\(\theta_\sigma=\pi\), namely along the transported momentum and opposite to
the nuclear polarization.

The same dipolar law \eqref{eq:wu-observed-angular-law-groupoid-form}
applies at intermediate angles through
\eqref{eq:wu-directional-invariant-mu}.  Thus the observed angular dependence
is compatible with the oriented axis selected by groupoid transport.

More generally, an axially symmetric function on the detector sphere admits
the Legendre expansion
\begin{equation*}
	F_{E,P}(\mu)
	=
	\sum_{\ell=0}^{\infty}
	w_\ell(E,P)P_\ell(\mu).
\end{equation*}
The observed leading law in
equation~\eqref{eq:wu-observed-angular-law-groupoid-form} contains the
monopole \(P_0(\mu)=1\) and the dipole \(P_1(\mu)=\mu\).  These are ordinary
angular harmonics on the laboratory detector sphere.  The groupoid
construction does not, by itself, prove that higher multipoles vanish or
determine the coefficient \(\kappa(E)\).

\subsubsection{Magnetic Reversal and Exchange of Detector Counts}
\label{subsubsec:wu-magnetic-reversal-count-exchange}

Consider the physical reversal of the magnetic preparation,
\begin{equation*}
	\sigma
	\longmapsto
	-\sigma.
\end{equation*}
This operation must not be confused with setting
\(\theta_\sigma=\pi\).  The latter changes the detector direction relative to
a fixed preparation, whereas magnetic reversal changes the preparation while
the laboratory detector may remain fixed.

Because
\begin{equation*}
	[\hat n]
	=
	[-\hat n],
\end{equation*}
the projective axis and its directional metric object remain unchanged:
\begin{equation*}
	\mathcal E_{[\hat n]}
	\longmapsto
	\mathcal E_{[\hat n]}.
\end{equation*}
The selected oriented arrow is instead exchanged:
\begin{equation}
	J_{\sigma\hat n}
	\longmapsto
	J_{-\sigma\hat n}.
	\label{eq:wu-arrow-exchange-under-reversal}
\end{equation}
Consequently,
\begin{equation}
	q_{-\sigma}
	=
	-q_\sigma,
	\qquad
	\hat q_{-\sigma}
	=
	-\hat q_\sigma.
	\label{eq:wu-momentum-reversal}
\end{equation}

\begin{figure}[htbp]
	\centering
	\includegraphics[width=0.72\textwidth]{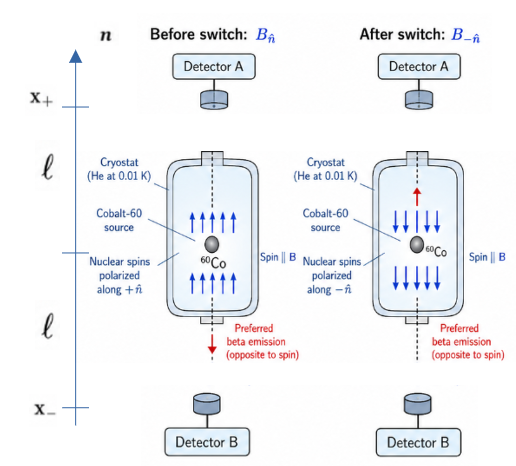}
	\caption{Schematic representation of magnetic reversal at fixed detector
	geometry.  Reversal of the prepared nuclear polarization exchanges the two
	oriented configurations, reverses the preferred beta-emission direction and
	therefore exchanges the excess and deficit recorded by the opposite
	detectors. }
	\label{fig:wu-magnetic-reversal-detector-exchange}
\end{figure}

The exchange is illustrated in
Figure~\ref{fig:wu-magnetic-reversal-detector-exchange}.  In that schematic,
the sign of the nuclear magnetic moment is absorbed into the preparation
convention, so that \(\hat n\) labels the oriented nuclear polarization used
throughout the text rather than an independently defined magnetic-field
direction.

At a fixed detector direction \(\hat k\),
\begin{align*}
	\cos\theta_{-\sigma}
	&=
	\hat b_{-\sigma}\cdot\hat k
	=
	-\hat b_\sigma\cdot\hat k
	=
	-\cos\theta_\sigma,
	\\
	\mu_{-\sigma}
	&=
	\hat q_{-\sigma}\cdot\hat k
	=
	-\hat q_\sigma\cdot\hat k
	=
	-\mu_\sigma.
\end{align*}
Hence
\begin{equation*}
	\theta_{-\sigma}
	=
	\pi-\theta_\sigma.
\end{equation*}

Provided that the reversed preparation is described by the same covariant
function \(F_{E,P}\), one obtains
\begin{align}
	W_{-\sigma,P}(E,\hat k)
	&=
	F_{E,P}
	\left(
	-\hat q_\sigma\cdot\hat k
	\right)
	\notag\\
	&=
	F_{E,P}
	\left(
	\hat q_\sigma\cdot(-\hat k)
	\right)
	\notag\\
	&=
	W_{\sigma,P}(E,-\hat k).
	\label{eq:wu-groupoid-rate-covariance}
\end{align}
Magnetic reversal therefore exchanges the counting distributions of two
antipodal detector directions.

The same transformation can be expressed in the Legendre expansion by using
\begin{equation*}
	P_\ell(-\mu)
	=
	(-1)^\ell P_\ell(\mu).
\end{equation*}
Even angular components remain unchanged under reversal, whereas odd
components reverse sign relative to a fixed laboratory detector basis.  In
particular,
\begin{equation*}
	P_1(-\mu)
	=
	-P_1(\mu),
\end{equation*}
which is precisely the transformation exhibited by the experimentally
observed cosine term.

The exchange of the parallel arrows is described by the typed isotropy
loops \eqref{eq:gpd-C-eta-recall} and \eqref{eq:gpd-C-dir-recall}.
Their numerical matrices coincide by \eqref{eq:gpd-common-C-matrix-representation},
but their typing distinguishes them as in
\eqref{eq:gpd-two-loops-not-equal-as-arrows}.

The directional loop exchanges the transported spectral representatives:
\begin{equation*}
	C^{\mathrm{dir}}_{[\hat n]}q_{\hat n}
	=
	q_{-\hat n}
	=
	-q_{\hat n}.
\end{equation*}
At the unitary level, the corresponding states satisfy
\begin{equation}
	\varrho_{-\hat n}
	=
	\mathcal U
	\left(
	C^{\mathrm{dir}}_{[\hat n]}
	\right)
	\varrho_{\hat n}
	\mathcal U
	\left(
	C^{\mathrm{dir}}_{[\hat n]}
	\right)^{-1}.
	\label{eq:wu-unitary-loop-state-action}
\end{equation}
The exchange of the two detector counting rates is therefore the statistical
counterpart of the exchange of the two parallel groupoid arrows.

\subsubsection{Scope of the Correspondence}
\label{subsubsec:wu-scope-groupoid-correspondence}

The source--arrow correspondence
\eqref{eq:wu-source-arrow-correspondence} and the assignment of the emitted
state to the transported directional sector are the physical inputs of the
ansatz.  Conditional on them, the geometry fixes the preferred direction
\eqref{eq:wu-transported-unit-direction} and its reversal
\eqref{eq:wu-momentum-reversal}.

Axial covariance permits \eqref{eq:wu-general-axisymmetric-counting-law}
but does not determine its function \(F_{E,P}\).  The observed dipolar law
\eqref{eq:wu-observed-angular-law-groupoid-form} has the orientation and
reversal covariance \eqref{eq:wu-groupoid-rate-covariance} required by the
construction.  The Wu experiment therefore does not falsify the proposed
source--arrow correspondence.

The present result should not be interpreted as a complete dynamical
derivation of beta decay.  In particular, the construction developed here
does not yet determine

\begin{enumerate}
	\item the magnitude and energy dependence of \(\kappa(E)\);
	\item the suppression or absence of higher angular multipoles;
	\item the complete beta-decay transition amplitude and total rate;
	\item the detailed relation between the selected \(SO(2,1)\) UIR and the
	source and detector observables.
\end{enumerate}

These questions require a groupoid-covariant interaction theory and explicit
transition matrix elements.  Their analysis is deferred to subsequent work.
The present paper establishes the kinematic and representation-theoretic
correspondence linking the preparation ansatz
\eqref{eq:wu-source-arrow-correspondence}, the preferred direction
\eqref{eq:wu-transported-unit-direction}, arrow exchange
\eqref{eq:wu-arrow-exchange-under-reversal}, and the observed counting law
\eqref{eq:wu-observed-angular-law-groupoid-form}.

No chiral projector is required to obtain the geometrical relation
\(\hat q_\sigma=-\hat b_\sigma\).  The present correspondence shows
that the experimentally observed directional behavior is compatible with 
the new class of groupoid wavefunctions developed in this work.

\section{Conclusions and Outlook}
\label{sec:conclusions-outlook}

This work has developed a Minkowski-pointed Poincare groupoid that extends
the coordinate structure of special relativity to reference frames which are
superluminal with respect to a distinguished rest frame.

The construction began with a real four-dimensional matrix representation of
superluminal Lorentz transformations.  Every superluminal boost was shown to
admit a factorization into an ordinary subluminal boost, an infinite-speed
transformation and, when required, a spatial rotation.  The two oriented
infinite-speed limits
\(\Lambda_{+\infty}(\hat n)\) and
\(\Lambda_{-\infty}(\hat n)\) are metric isomorphisms between the Minkowski
object and a directional metric object.

The metric transported by either infinite-speed limit depends only on the
unoriented spatial axis
\[
[\hat n]=\{\hat n,-\hat n\}\in\mathbb{RP}^{2}.
\]
Consequently, the two orientations do not define two different metric
objects.  They define two parallel arrows
\[
J_{\hat n},J_{-\hat n}:
\mathcal E_\eta
\longrightarrow
\mathcal E_{[\hat n]}
\]
between the same source and target.  Their relative compositions generate
typed isotropy loops at the Minkowski and directional objects.  Although
these loops have the same numerical matrix representative in the present
realization, they remain distinct groupoid arrows because their source and
target objects are different.

The direction-dependent extensions of the Poincare group were thereby
reinterpreted as isotropy groups of the metric objects.  The involution
\(C_{[\hat n]}\) acts internally within an isotropy group, whereas the
infinite-speed matrices transport data between different objects.  The
Clifford--Mackey analysis then determines how the discrete loop either
exchanges two inequivalent Wigner sectors or extends an invariant sector by
a discrete sign.

A Hilbert space and a unitary representation of the corresponding isotropy
group were assigned to every object.  At the infinitesimal level, each object
carries an isomorphic realization of the abstract Poincare algebra.  Passing
to the complex universal enveloping algebras makes it possible to transport
not only the generators, but also their noncommutative polynomial products,
including the elements from which the Casimir operators are constructed.
Every inter-object arrow induces a unital \( * \)-isomorphism between these
algebras, while every isotropy loop induces an internal
\( * \)-automorphism.

The distinguished Minkowski object supplies the common contraction structure
used to classify all transported operator data.  For the Minkowski object
itself, this prescription reproduces the ordinary massive and massless
Wigner classes.  A massive momentum representative transported to a
directional object remains intrinsically massive with respect to the
directional metric.  When the same transported spectral data are classified
with the Minkowski metric, however, the momentum is spacelike and its little
group is \(SO(2,1)\).  The corresponding joint Casimir eigenvalue equations
are therefore the tachyonic Wigner wave equations.

In this construction, tachyonic wavefunctions are not introduced by
postulating a new species of intrinsically tachyonic particle.  They arise
from the Minkowski classification of states transported between isomorphic
metric objects.  The groupoid thus enlarges both the coordinate structures
available to relativistic kinematics and the representation classes that can
be assigned to physical wavefunctions.

The Wu experiment was considered as a first kinematic application.  Under
the working hypothesis that beta radiation belongs to the transported
\(SO(2,1)\) sector, the oriented nuclear preparation selects one of the two
parallel arrows above the same projective object.  Transport of the standard
massive representative then fixes a preferred momentum axis opposite to the
nuclear polarization.  Reversal of the magnetic preparation exchanges the
two arrows and reverses the transported momentum, consistently with the
observed exchange of the detector counts.

The experimentally measured dipolar angular distribution has the orientation
and reversal covariance required by this construction.  The Wu experiment
therefore does not falsify the proposed source--arrow correspondence.
The present paper does not, however, derive the beta-decay amplitude, the
energy-dependent asymmetry coefficient or the absence of higher angular
multipoles.  The result obtained here is a kinematic and
representation-theoretic correspondence, rather than a completed alternative
dynamics of the electroweak interaction.

The first direction for future work is to determine whether electroweak
radiation systematically exhibits the properties predicted for states
associated with directional metric objects.  This requires replacing the
present compatibility argument by explicit transition amplitudes and testing
the construction against the full phenomenology of beta decay and other
electroweak processes.  In particular, a groupoid-covariant calculation
should determine the energy dependence of the beta-asymmetry coefficient,
the angular multipole content, the continuous beta spectrum and the total
transition rates.

This investigation is directly connected with the more radical conjecture
that the neutrino is not an independent physical particle.  According to
this conjecture, phenomena currently represented by an unobserved neutrino
four-momentum would instead be described by \(SO(2,1)\) propagation and by
spacelike Minkowski momenta associated with states that are intrinsically
massive at directional objects.  The neutrino would then be an auxiliary
entity introduced when such processes are forced into a single-object
Minkowski description.

In this restricted methodological sense, the neutrino would play a role
analogous to that of a modern ether: not a material medium added to the
present groupoid construction, but an additional entity introduced to
represent phenomena for which the underlying coordinate structure may have
been incomplete.  This analogy is heuristic and is not a result of the
present paper.  The non-existence conjecture can be regarded as viable only
if the groupoid theory reproduces, within a single quantitative framework,
the experimental phenomena presently attributed to neutrinos, including
missing energy and momentum, scattering events, flavour-dependent
phenomena, helicity measurements and propagation effects.  Failure to
reproduce any of these established observations would falsify the
conjecture.

A second research direction concerns perturbative quantum field theory.
Ordinary Feynman loop integrals already range over off-shell spacelike
four-momenta.  Therefore, improved ultraviolet behaviour cannot follow
merely from allowing spacelike momentum values.  The potentially new
ingredient supplied by the groupoid is that certain spacelike spectral data
are treated as typed physical sectors, connected to timelike data by
inter-object arrows and constrained by their object, arrow and isotropy
structure.

Future work will investigate whether interaction vertices admitting these
typed spacelike sectors lead to modified propagators, integration measures,
composition rules or cancellations.  In particular, it will be necessary to
determine whether groupoid composability and the simultaneous presence of
timelike and spacelike branches restrict the virtual configurations entering
loop amplitudes.  The objective is to establish whether this structure can
provide a mathematically controlled route to finite perturbative amplitudes
without the divergent integrals of the conventional formulation.  Such a
claim would require explicit calculations and could be accepted only if
unitarity, covariance, the relevant Ward identities and the established
low-energy predictions of quantum field theory are preserved.

\appendix

\section{Superluminal Boosts Factorization}
\label{app:superluminal-boost-factorization}

We prove Eq.~\eqref{eq:srf-superluminal-factorization}
using the common directional projector in
\eqref{eq:srf-directional-projector-invariance} and the relation between the
two Lorentz factors in \eqref{eq:srf-gamma-S-gamma-s-relation}.

Let
\begin{equation*}
	\beta:=\frac{|\mathbf v|}{c},
	\qquad
	P_{\hat n}:=\hat n\hat n^T.
\end{equation*}
Since \(\mathbf v\) and \(\mathbf V\) are related by the velocity inversion,
they have the same direction and may be written as
\begin{equation}
	\mathbf v=c\beta\hat n,
	\qquad
	\mathbf V=\frac{c}{\beta}\hat n,
	\qquad
	0<\beta<1.
	\label{eq:app-velocity-parametrization}
\end{equation}
Equation~\eqref{eq:srf-directional-projector-invariance} then becomes
\begin{equation*}
	\frac{\mathbf v\mathbf v^T}{|\mathbf v|^2}
	=
	\frac{\mathbf V\mathbf V^T}{|\mathbf V|^2}
	=
	P_{\hat n}.
\end{equation*}
The projector identities
\begin{equation}
	P_{\hat n}^2=P_{\hat n},
	\qquad
	P_{\hat n}\hat n=\hat n,
	\qquad
	(I_3-P_{\hat n})\hat n=0,
	\qquad
	\hat n^T(I_3-P_{\hat n})=0
	\label{eq:app-projector-identities}
\end{equation}
will eliminate all mixed longitudinal--transverse terms.

In time--space block form, the ordinary boost and the positive
infinite-speed matrix are
\begin{equation*}
	\Lambda_s(\mathbf v)
	=
	\begin{pmatrix}
		\gamma_s & -\gamma_s\beta\hat n^T\\[2pt]
		-\gamma_s\beta\hat n
		& I_3+(\gamma_s-1)P_{\hat n}
	\end{pmatrix},
	\qquad
	\Lambda_+(\hat n)
	=
	\begin{pmatrix}
		0 & -\hat n^T\\[2pt]
		-\hat n & I_3-P_{\hat n}
	\end{pmatrix}.
\end{equation*}
The four time--space blocks of their product are
\begin{align*}
	\bigl[\Lambda_s\Lambda_+\bigr]_{00}
	&=\gamma_s\beta,
	\\
	\bigl[\Lambda_s\Lambda_+\bigr]_{0\mathrm{x}}
	&=-\gamma_s\hat n^T
	-\gamma_s\beta\hat n^T(I_3-P_{\hat n})
	=-\gamma_s\hat n^T,
	\\
	\bigl[\Lambda_s\Lambda_+\bigr]_{\mathrm{x}0}
	&=-\bigl[I_3+(\gamma_s-1)P_{\hat n}\bigr]\hat n
	=-\gamma_s\hat n,
	\\
	\bigl[\Lambda_s\Lambda_+\bigr]_{\mathrm{xx}}
	&=\gamma_s\beta P_{\hat n}
	+\bigl[I_3+(\gamma_s-1)P_{\hat n}\bigr](I_3-P_{\hat n})
	\\
	&=I_3+(\gamma_s\beta-1)P_{\hat n}.
\end{align*}
Here the subscripts \(\mathrm{x}\) denote the spatial blocks.  Hence
\begin{equation}
	\Lambda_s(\mathbf v)\Lambda_+(\hat n)
	=
	\begin{pmatrix}
		\gamma_s\beta & -\gamma_s\hat n^T\\[2pt]
		-\gamma_s\hat n
		& I_3+(\gamma_s\beta-1)P_{\hat n}
	\end{pmatrix},
	\label{eq:app-product-reduced}
\end{equation}
where all reductions follow directly from
\eqref{eq:app-projector-identities}.

By \eqref{eq:srf-gamma-S-gamma-s-relation},
\begin{equation*}
	\gamma_S=\gamma_s\beta.
\end{equation*}
Moreover, \eqref{eq:app-velocity-parametrization} gives
\begin{equation*}
	\gamma_S\frac{\mathbf V}{c}
	=
	\frac{\gamma_S}{\beta}\hat n
	=
	\gamma_s\hat n.
\end{equation*}
Consequently, the superluminal boost itself can be written as
\begin{equation*}
	\Lambda_S(\mathbf V)
	=
	\begin{pmatrix}
		\gamma_S
		& -\gamma_S\dfrac{\mathbf V^T}{c}\\[5pt]
		-\gamma_S\dfrac{\mathbf V}{c}
		& I_3+(\gamma_S-1)P_{\hat n}
	\end{pmatrix}
	=
	\begin{pmatrix}
		\gamma_s\beta & -\gamma_s\hat n^T\\[2pt]
		-\gamma_s\hat n
		& I_3+(\gamma_s\beta-1)P_{\hat n}
	\end{pmatrix}.
\end{equation*}
The last matrix is exactly the product in
\eqref{eq:app-product-reduced}, proving
\eqref{eq:srf-superluminal-factorization}.

The factorization also extends continuously to the infinite-speed limit:
as \(|\mathbf V|\to\infty\), one has \(\beta\to0\),
\(\Lambda_s(\mathbf v)\to I_4\), and therefore
\(\Lambda_S(\mathbf V)\to\Lambda_+(\hat n)\), consistently with
\eqref{eq:srf-lambda-plus-definition}.

\section{Clifford--Mackey Theory of Groups Extensions representations}
\label{app:clifford-mackey-index-two}

Let \(N\) be a closed normal subgroup of a locally compact group \(G\), and
assume that the extension is split and has index two:
\begin{equation}
	G=N\rtimes\langle c\rangle,
	\qquad
	c^2=e,
	\qquad
	G/N\simeq\mathbb Z_2.
	\label{eq:app-cm-index-two-extension}
\end{equation}
Let
\begin{equation*}
	\pi:N\longrightarrow\mathcal U(\mathcal H_\pi)
\end{equation*}
be an irreducible unitary representation of \(N\).  Conjugation by \(c\)
defines another irreducible unitary representation of \(N\):
\begin{equation*}
	\pi^c(n)
	:=
	\pi(c^{-1}nc),
	\qquad n\in N.
\end{equation*}
Because \(c^2=e\), one has
\begin{equation*}
	(\pi^c)^c=\pi.
\end{equation*}

The inertia group of the equivalence class \([\pi]\in\widehat N\) is
\begin{equation*}
	G_\pi
	:=
	\left\{
	x\in G:\pi^x\simeq\pi
	\right\}.
\end{equation*}
Since \(N\subseteq G_\pi\subseteq G\) and \([G:N]=2\), there are only two
possibilities:
\begin{equation*}
	G_\pi=N
	\qquad\text{or}\qquad
	G_\pi=G.
\end{equation*}
Equivalently, the \(G/N\)-orbit of \([\pi]\) in \(\widehat N\) has either two
elements or one element.  These alternatives give the two cases described
below.

\subsection{Case I: a two-element orbit}
\label{app:cm-two-element-orbit}

Assume first that $\pi^c$ and $\pi$ are inequivalent
\begin{equation*}
	\pi^c\not\simeq\pi.
\end{equation*}
Then \(G_\pi=N\), and the orbit of \([\pi]\) is
\begin{equation*}
	\mathcal O_\pi
	=
	\left\{[\pi],[\pi^c]\right\}.
\end{equation*}
The induced representation
\begin{equation*}
	\Pi
	:=
	\operatorname{Ind}_{N}^{G}\pi
\end{equation*}
can be realized, because the quotient has two elements, on
\begin{equation*}
	\mathcal H_\Pi
	=
	\mathcal H_\pi\oplus\mathcal H_{\pi^c}.
\end{equation*}
After a fixed unitary identification of the two Hilbert-space copies, its
action may be written as
\begin{equation*}
	\Pi(n)
	=
	\begin{pmatrix}
		\pi(n)&0\\
		0&\pi^c(n)
	\end{pmatrix},
	\qquad n\in N,
\end{equation*}
and
\begin{equation*}
	\Pi(c)
	=
	\begin{pmatrix}
		0&I\\
		I&0
	\end{pmatrix}.
\end{equation*}
These operators satisfy
\begin{equation*}
	\Pi(c)\Pi(n)\Pi(c)^{-1}
	=
	\Pi(cnc^{-1}),
	\qquad
	\Pi(c)^2=I,
\end{equation*}
and therefore define a unitary representation of \(G\).

This representation is irreducible.  Indeed, let \(B\) be a bounded operator
commuting with \(\Pi(G)\).  Since \(B\) commutes in particular with \(\Pi(N)\),
the inequivalence and irreducibility of \(\pi\) and \(\pi^c\), together with
Schur's lemma, imply
\begin{equation*}
	B
	=
	\begin{pmatrix}
		aI&0\\
		0&bI
	\end{pmatrix}
\end{equation*}
for some \(a,b\in\mathbb C\).  Commutation with \(\Pi(c)\) then gives \(a=b\).
Thus the commutant of \(\Pi(G)\) consists only of scalar multiples of the
identity, proving irreducibility.  Its restriction to \(N\) is
\begin{equation*}
	\operatorname{Res}_{N}^{G}\Pi
	\simeq
	\pi\oplus\pi^c.
\end{equation*}
Consequently, one irreducible representation of the extended group contains
two inequivalent \(N\)-sectors exchanged by the nontrivial element of
\(G/N\).

\subsection{Case II: an invariant representation}
\label{app:cm-invariant-representation}

Assume instead that $\pi^c$ and $\pi$ are equivalent.
\begin{equation*}
	\pi^c\simeq\pi.
\end{equation*}
Then \(G_\pi=G\), and there exists a unitary intertwiner
\begin{equation*}
	A_c:\mathcal H_\pi\longrightarrow\mathcal H_\pi
\end{equation*}
satisfying
\begin{equation*}
	A_c\pi(n)A_c^{-1}
	=
	\pi(cnc^{-1}),
	\qquad n\in N.
\end{equation*}
Applying this relation twice and using \(c^2=e\) gives
\begin{equation*}
	A_c^2\pi(n)A_c^{-2}
	=
	\pi(n).
\end{equation*}
Schur's lemma therefore implies
\begin{equation*}
	A_c^2=\lambda I,
	\qquad
	\lambda\in U(1).
\end{equation*}
Replacing \(A_c\) with a suitable phase multiple, one may impose the
normalization
\begin{equation}
	A_c^2=I.
	\label{eq:app-cm-Ac-normalization}
\end{equation}
The two choices
\begin{equation}
	\Pi_\pm(n)=\pi(n),
	\qquad
	\Pi_\pm(c)=\pm A_c
	\label{eq:app-cm-two-extensions-definition}
\end{equation}
then define two irreducible unitary representations of \(G\) extending
\(\pi\).  They are inequivalent.  If a unitary \(U\) intertwined them, its
restriction to \(N\) would satisfy
\begin{equation*}
	U\pi(n)=\pi(n)U
\end{equation*}
for every \(n\in N\).  Schur's lemma would give \(U=zI\), whereas the
intertwining relation for \(c\) would require
\begin{equation*}
	zA_c=-A_cz,
\end{equation*}
which is impossible.  The two extensions differ precisely by tensoring with
the two characters of the quotient:
\begin{equation*}
	\widehat{G/N}
	\simeq
	\widehat{\mathbb Z_2}
	=
	\left\{\chi_+,\chi_-\right\},
	\qquad
	\chi_\pm(cN)=\pm1.
\end{equation*}

For a general group extension, an invariant representation of the normal
subgroup can carry a nontrivial Mackey obstruction and may initially extend
only projectively.  In the present split index-two case, the explicit phase
normalization in \eqref{eq:app-cm-Ac-normalization} removes this obstruction.
The two remaining choices are exactly the two ordinary characters of
\(\mathbb Z_2\).

%
%

\end{document}